\documentclass[a4paper,11pt]{article}

\pdfoutput=1
\usepackage{jheppub}
\usepackage{caption}
\usepackage{subcaption}

\definecolor{green}{rgb}{0.1,0.8,0.2}

\title{A Pole-Cut Formula for Holographic Extremal Correlators}
 
\author[a]{Jyotirmoy Bhattacharya }
\author[a]{, Amiya Mishra }
\author[a]{, Aditya Sharma }
\author[a]{and Sourav Singha }
\affiliation[a]{Department of Physics, Indian Institute of Technology Kharagpur, Kharagpur 721302, India}

\emailAdd{jyoti@phy.iitkgp.ac.in, mishramiya@gmail.com, adsharma.d1d4@gmail.com, singhasourav000@gmail.com}

\abstract{
We show that the holographic spectral function of a neutral massless scalar on the extremal  Reissner-Nordstr\"om black brane in AdS$_5$ admits an elegant formula in terms of the quasi-normal modes (QNMs) and branch-cut data. In the extremal case, the spectral data comprises the zeros of the boundary source function, i.e. the QNMs, as well as its
phase discontinuity across the branch cut on the negative imaginary axis of the complex frequency plane. This formula neatly factorizes into an infinite product over the QNM poles and a factor encoding the branch cut discontinuity. We derive this pole-cut formula as the zero-temperature limit of the QNM product formula for holographic thermal correlators, with the branch-cut contribution arising from the coalescence of the QNMs on the negative imaginary axis. We also clarify the role of the near-horizon AdS$_2$ region in this context.
Consistency with the OPE of the boundary correlator provides extremal sum-rules involving both the QNMs as well as the phase discontinuity function. We also find that this holographic consistency forces the asymptotic form of the QNMs to be related to the asymptotics of the phase discontinuity function. We verify our formula and these OPE constraints numerically. We also demonstrate that, in the time domain, the pole and cut factors each produce complex-time singularities, whose location is determined by the optical distance between the boundary and the black-brane singularity. We expect the pole-cut factorization structure of spectral functions to hold for more general fluctuations of extremal black branes.
}

\begin{document}

%preprint
%\begin{flushright} 
%\small{preprint} 
%\end{flushright}

%
\maketitle
\flushbottom
%
%%%%%%%%%%%%%%%%%%%%%%%%%%%%%zzz

%****************************************
\section{Introduction}
%****************************************

The holographic dictionary relates boundary correlation functions to bulk fluctuations in AdS, a connection that has proven remarkably successful and insightful in probing strongly coupled large-$N$ CFTs \cite{Witten:1998qj,Gubser:1998bc}. For instance, studying such correlators for black branes in the long-wavelength limit has yielded remarkable insights into strongly coupled relativistic hydrodynamics \cite{Son:2002sd,Kovtun:2004de}. More recently, there has been significant progress in understanding the analytic structure of holographic thermal correlators \cite{David:2026mlz}. In an important development \cite{Dodelson:2023vrw}, it was shown that the structure of the bulk fluctuation equations, together with the expected analytic properties, enforces a simple product formula for the two-sided thermal correlator in terms of the quasi-normal modes (QNMs). This result has since been pursued in several directions \cite{Bhattacharya:2025vyi,Grozdanov:2025ner,Grozdanov:2025ulc,Horowitz:2023ury, Vrbica:2026krk}, including investigations into the validity of the product structure for large-$N$ theories beyond the holographic context \cite{Dodelson:2024atp, Demulder:2025uda}. It is natural to ask whether a similarly simple structure for holographic correlators survives at zero temperature. This has motivated us to explore whether the two-point correlators for linearized fluctuations of extremal black branes can also be expressed directly in terms of the spectral data. In this paper, we show that the holographic spectral function of a neutral massless scalar on the extremal Reissner--Nordstr\"om black brane in AdS$_5$ admits an elegant formula in terms of its QNMs and branch-cut data.

Let us recall the thermal product formula of \cite{Dodelson:2023vrw} where the two-sided correlator is written as an infinite product over poles whose location is determined
by the QNMs. The QNMs on the complex frequency ($\omega$) plane may be classified into two broad classes, those lying on the negative imaginary axis and
those located away from the imaginary axis organized on two branches of a `Christmas tree' (see Fig.\ref{fig:QNMTH}). Due to our interest in the extremal limit, instead of the two-sided function, we shall focus on the spectral function $\rho(\omega)$ which is given by the imaginary part of the retarded correlator (Im $\mathcal G_{R}$) for real $\omega$
{\footnote{In this paper, we shall follow the conventions of \cite{Dodelson:2023vrw,Bhattacharya:2025vyi} while defining the two-point correlators. With this convention, the
spectral function is given by $\rho = 2 \, \text{Im} \, \mathcal G_R$. See appendix A of \cite{Dodelson:2023vrw} for more details.}.
At finite temperature, the spectral function is simply related
to the two-sided correlator and the product formula implies the following schematic form for the spectral function 
\begin{equation}\label{intro:DodZhiform}
\rho(\omega) = \mathcal G_0 \, \sinh \left( \frac{\omega \beta}{2}\right) \,  \underbrace{\prod_{n} \left( 1 - \frac{\omega^2}{\omega_n^2} \right)^{-1}}_{\text{(off-axis poles)}}    \underbrace{\prod_{m} \left( 1 - \frac{\omega^2}{\omega_m^2} \right)^{-1}}_{\text{(on-axis poles)}} \, ,
\end{equation}
where $\omega_n$ are the QNMs and their dependence on $k$ is left implicit. The $\sinh$ factor ensures consistency with KMS condition and accounts for zeros of the spectral function 
at the Matsubara frequencies. In the extremal limit, we find that the zeros at the Matsubara frequencies and the poles at on-axis QNMs combine to produce a branch cut contribution. As a consequence, at
zero temperature, the spectral function takes the elegant form 
\begin{equation} \label{intro:pole-cut}
\rho(\omega)
= \tilde {\mathcal G}_0 ~ \omega \, \exp \left( {H_{ \text {cut} }} (\omega) \right) 
\underbrace{\prod_{n} \left( 1 - \frac{\omega^2}{\omega_n^2} \right)^{-1}}_{\text{off-axis extremal poles}} \, ,
\end{equation}
In the extremal case, the spectral data consists of the zeros of the boundary source function, i.e. the QNMs, as well as its
phase discontinuity $\theta (\sigma)$ of the source function across the branch cut on the negative imaginary axis of the complex frequency plane
\footnote{Here $\sigma \in \left[   0, \infty \right)$ parametrizes negative imaginary axis of the complex frequency plane.}.
In \eqref{intro:pole-cut}, the branch-cut contribution $H_{ \text {cut}}(\omega)$ is explicitly determined in terms of phase discontinuity function $\theta$. 
This branch cut contribution is responsible for the well known non-analyticities of the two point function. Since the spectral function neatly factorizes into contributions from QNM poles 
and that from the branch cut, we refer to \eqref{intro:pole-cut} as the `pole-cut' formula. 

The fact that in the extremal limit, the on-axis QNMs condense into a branch cut, leading to a non-analytic behavior of the spectral function Im $\mathcal G_R \sim \omega^{2 \nu}$ has been 
well studied in the literature, beginning with the seminal  works such as \cite{Faulkner:2009wj,Iqbal:2011in,Edalati:2010hk,Edalati:2010pn}. The exponent has been calculated for the IR-CFT correlators dual to the AdS$_2$ throat, characteristic of an extremal black hole. More recently, \cite{Arnaudo:2024sen} obtained analytic control over the emergence of the zero temperature branch cut as a confluent limit of the Heun equation, in a double-scaling limit where $\omega, \, T \rightarrow 0$ holding $\omega/T$ fixed. In this double scaling limit the factor with Matsubara frequencies is an order one number which did not play any significant role in the branch-cut contribution.
The effect of a R-charge anomaly  on this non-analytic behaviour for extremal correlators in the shear sector, particularly for the exponent $\nu$,  was also studied recently in \cite{Bhattacharya:2025vyi} using a bulk Chern-Simons term. Also, see \cite{Preau:2025rex} where the near-extremal holographic correlators were studied in the hydrodynamic limit. But the relevance and the emergence of the branch cut in the $T \rightarrow 0$, including its exact treatment is one of the main novelties of our work in this paper. We have explained the  previously known relevant results using  \eqref{intro:pole-cut}, particularly clarifying the role of the near-horizon AdS$_2$ region in this context.

After establishing \eqref{intro:pole-cut} and verifying it numerically, we proceed towards an interesting application of the formula. From the boundary point of view, the extremal correlator is a two-point function in a finite density zero temperature CFT state. Such a correlator admits an operator product expansion (OPE) which can be mapped to a large $\omega$ expansion of the retarded correlator. The coefficients of these large $\omega$ expansion are fixed in terms of the boundary conformal data. The same correlator is evaluated from the bulk using  \eqref{intro:pole-cut}. Consistency with the OPE limit imposes non-trivial constrains on the bulk spectral data which goes into \eqref{intro:pole-cut}. One such constraint which we study carefully is an extremal sum-rule where a particular sum over the QNMs added to an integral over the discontinuity function $\theta$ must converge to a simple function of $k$. This remarkable relationship between all the bulk extremal QNMs and the complete branch cut data could be verified numerically. Apart from this, we also find another interesting connection between the QNMs and the discontinuity function. The asymptotic form of these two quantities takes the form 
\begin{equation} \label{intro:asymform}
\omega_n = a (n + q + \dots) ~~\text{with} ~ n \in \mathbb Z_+ \, ,  ~~ \theta(\sigma) = \alpha  \, \sigma + \beta + \dots  \, , 
\end{equation} 
for large $n$ and $\sigma$. The asymptotic QNM parameters $a$ and $q$ can be determined analytically using WKB techniques and are found to be independent of $k$. Consistency of the spectral
function with the OPE limit from the boundary point of view forces the asymptotic parameter $\alpha$ and $\beta$ in the discontinuity function to be determined in terms of the QNM parameters 
$a$ and $q$. Since the QNM parameters are known exactly, this connection leads to  a robust prediction for the asymptotic form of the discontinuity function $\theta$. We are able to verify these predictions numerically.

Another feature of the thermal correlators which has attracted much of recent attention is an interesting feature popularly being referred to as `bouncing singularities' \cite{Fidkowski:2003nf,Festuccia:2005pi,Parisini:2023nbd,Ceplak:2024bja,Buric:2025anb,Buric:2025fye,Barrat:2025twb,Afkhami-Jeddi:2025wra,Dodelson:2025jff,Ceplak:2025dds,Jia:2025jbi,AliAhmad:2026wem,Giombi:2026kdz,Arnaudo:2026der,Grozdanov:2026cut,Jia:2026ryl,Arnaudo:2026tcy,Grozdanov:2026ktq,Buric:2026qsp}.
It is observed that the thermal retarded correlators in the time domain exhibits singularities at complex values of time. The location of such singularities are controlled by the optical distance between the 
AdS boundary and the black brane singularity. The feature is largely a property of large-N holographic CFTs and disappears with the inclusion of $1/N$ corrections \cite{Komargodski:2026oob, Liu:2026luf}.
Nevertheless, this is a significant observation since this maps the curvature singularity in the bulk to a feature of a boundary observable (see also \cite{Chakravarty:2025ncy}). In fact, it has been recently shown in  \cite{Hartnoll:2026vhu} that the Kasner parameters of the near-singularity geometry is also encoded within the boundary correlators (also see  \cite{Ceplak:2026ard,Gao:2026ieh,Li:2026gqf,Xiao:2026pir,Coviello:2026hdc} for more recent work in this direction). In the extremal case, although the temperature vanishes, we continue to have a curvature singularity in the geometry. Hence, it is expected that the extremal retarded correlator continue to exhibit 
complex-time singularities similar to its thermal counterpart. Using our pole-cut formula \eqref{intro:pole-cut} we explain how such singular behaviour arises in the time-domain extremal retarded correlator. 
We find that the residue at the QNM poles and the branch-cut contribution to the spectral function contribute as separate terms to the time-domain retarded correlator. Both these terms individually exhibit the 
anticipated complex-time singularities. This singular behaviour in turn is governed by the asymptotic form of the QNMs as well as that of the discontinuity function in \eqref{intro:asymform} which conspire to ensure that the location of the singularities given by the optical distance between the boundary and the black-brane singularity. 

Our paper is organized as follows. In section \ref{sec:bkgfluc}, we present our computational set-up including the background and the fluctuation equation. In section \ref{sec:pole-cut}, we derive the pole-cut formula \eqref{intro:pole-cut} by taking the extremal limit of its thermal counterpart. In this section, we also examine the role of the near-horizon AdS$_2$ geometry in the branch cut contribution.
Further, we provide substantial numerical evidence in support of our formula without making any simplifying assumptions.
In section \ref{sec:OPElimit}, we present the constraints on the QNM and branch cut data, including the extremal sum rules, which follow from the OPE limit of the extremal correlator.
In section \ref{sec:singularbounce}, we demonstrate the occurrence of complex time singularities in the time-domain version of the extremal retarded correlator using
our pole-cut formula \eqref{intro:pole-cut}.
In section \ref{sec:disco}, we conclude with a brief discussion of interesting perspectives arising out of our work.
Beside the main text, we also have five appendices. Appendix \ref{app:img} and \ref{app:HT} contain minor technical details. In appendix \ref{app:ads2}, we work out the on-axis QNMs using the near extremal AdS$_2$, employing the standard technique of matching solutions in the near horizon and far regions. Here, we use this
result to explain how much of the phase discontinuity function can be reliably captured
by the near horizon AdS$_2$ geometry. In appendix \ref{app:discon}, we explain the numerical procedure to obtain the discontinuity of the source function across the branch cut.
Finally, in appendix \ref{app:asymexact}, we present the WKB procedure which fixes the first two parameters in the asymptotic expansion of the extremal QNMs.

\textit{Note added: As we were preparing to submit this manuscript, having completed the present work, we became aware of the recent work of Emparan, Murthy and Trezzi \cite{Emparan:2026glk}, which has some thematic overlap with our analysis, particularly in the discussion of quasinormal modes and the analytic structure of correlators in the extremal limit.}
%**********************************************************
\section{Background and fluctuation equations}\label{sec:bkgfluc}
%**********************************************************
%
We shall work with the 5D Maxwell-Einstein system with a negative cosmological constant, governed by the following action
\begin{equation} \label{system}
\mathcal S = \int d^5x ~ \sqrt{-g} \left(  \mathcal R + 12 - \frac{1}{4} \ F_{\mu \nu} F^{\mu \nu} \right) \ . 
\end{equation} 
Together with a topological Chern-Simons term, this system is a consistent truncation of $S^5$ compactification of Type-IIB sugra and admits an R-charged black brane solution which we discuss below
{\footnote{In this paper, we shall be agnostic regarding the stability of our extremal background. Our conclusions are robust for the 
extremal saddle point which definitely exists in a UV complete description, irrespective of the fact whether it is a dominant saddle or otherwise.}}. Hence, all our conclusions for the boundary CFT in this paper are directly applicable for 
the corresponding sector in N$=4$ SYM in 4D. 

\subsubsection*{Background}
%--------------------------------------------------

The system \eqref{system} admits the electrically charged AdS-Reissner Nordstr\"om (AdS-RN) black brane solution, whose metric and gauge field are given by 
\begin{equation} \label{backmet}
\begin{split} 
ds^2 &= g_{\mu \nu} \,  dx^\mu dx^\nu = \frac{1}{z^2} \left( - f(z) dt^2 + \frac{dz^2}{f(z)} + d\vec{x}^{\, 2} \right) , \\  \mathcal A & = A_\mu \, dx^\mu  = a(z)  \, dt  \, , 
\end{split} 
\end{equation} 
where 
\begin{equation} \label{backsol}
\begin{split}
f(z) &= 1 - (1+Q^2) z^4 + Q^2 z^6 \, , \\
a(z) &= \sqrt{3} Q (1-z^2) \, . 
\end{split} 
\end{equation} 
The temperature and chemical potential of this black brane are respectively given by
\begin{equation} \label{Tmu}
T = \frac{1}{\pi} \left( 1 - \frac{Q^2}{2} \right) , ~ \mu = \sqrt{3} Q \, . 
\end{equation} 
Here, we are particularly interested in the extremal limit $T \rightarrow 0$, which is achieved by setting the charge parameter $Q = \sqrt{2}$. Hence for the extremal
black brane we have 
\begin{equation} \label{backextsol}
\begin{split}
f(z) &= 1 - 3 z^4 + 2 z^6 \, , \\
a(z) &= \sqrt{6}  (1-z^2) \, . 
\end{split} 
\end{equation} 
For both solutions \eqref{backsol} and \eqref{backextsol} we have chosen the horizon to be at $z=1$ and the boundary is at $z \rightarrow 0$. With horizon at $z=1$, 
the chemical potential of the extremal black brane is fixed to $\mu_e = \sqrt{6}$. 
Note that setting the horizon at $z=1$ sets our choice of units and the chemical potential is an $\mathcal{O}(1)$ number in these units. Rest of the dimensionful quantities in this paper, such as $\omega$ and $T$ can be converted to a dimensionless
number using the chemical potential $\mu_e$. This effectively amounts to a numerical
rescaling of $\omega$ and $T$
{\footnote{Instead of carrying the $\mu_e = \sqrt{6}$ factor everywhere,
we keep this rescaling implicit in this paper and work with the unscaled
parameters instead. Hence, for example, when we compare $\omega$ with an
$\mathcal{O}(1)$ number, we imply the comparison of $\omega / \mu_e$. \label{fn:unit}}}.

\subsubsection*{Fluctuation equations}
%--------------------------------------------------

In this paper, we shall consider the tensor-sector metric perturbations
{\footnote {The tensor sector refers to the traceless components of the metric along the boundary spatial directions, 
i.e. $\left( g_{ij} - \frac{1}{3} g_{kk} \, \delta_{ij} \right)$, where $\{i,j,k\}$ runs over 
the three boundary spatial coordinates $\vec{x}$.}}. 
The linearized fluctuation equation in this sector can be recast into the equation of motion of the minimally coupled uncharged massless scalar field
\begin{equation} \label{Phieq}
\nabla^2 \Phi = 0 \, . 
\end{equation} 
We will use the following ansatz 
\begin{equation} \label{anza}
\Phi = \exp \left( - i \omega v + i k x \right) \phi(z) \, ,
\end{equation} 
where $x$ is one of the boundary spatial directions, while $v$ is the ingoing Eddington-Finkelstein time 
\begin{equation} \label{intime}
v = t - \int \, \frac{dz}{f(z)} \, . 
\end{equation}
Defining the fluctuation frequency with respect to ingoing time $v$ ensures that the ingoing boundary condition at the horizon is equivalent to imposing horizon
regularity for $\phi (z)$. The equation for $\phi(z)$ which follows from $\eqref{Phieq}$, when evaluated on the AdS-RN background, is given by   
\begin{equation}  \label{phizeqnext}
  f \phi''+
  \left(f'-\frac{3f}{z}+2i\omega\right) \phi'
  -\left(k^2+\frac{3i\omega}{z}\right) \phi=0 \, , 
\end{equation}
where $f$ is given by \eqref{backsol} or  \eqref{backextsol} in the finite temperature and extremal cases respectively. We ensure regularity and fix our normalization by imposing 
the following boundary condition at the horizon
\begin{equation} \label{horbdyc}
\phi (z \rightarrow 1) = 1 \, . 
\end{equation} 
Here, the fluctuation equation \eqref{phizeqnext} together with the horizon
boundary condition \eqref{horbdyc} constitutes the main ingredient of our analysis.

%**********************************************************
\section{Extremal limit of the thermal product formula} \label{sec:pole-cut}
%**********************************************************

%**********************************************************
\subsection{Boundary correlators and QNMs}
%**********************************************************
%
In order to obtain the QNMs and the boundary correlators, we have to solve \eqref{phizeqnext} with the boundary condition \eqref{horbdyc}. 
The near-boundary Frobenius expansion for $\phi(z)$, which follows from \eqref{phizeqnext},  is given by 
\begin{equation} \label{asymexp}
\phi(z) = A + \mathfrak {a} \, z + \mathfrak{b} \, z^2 + \mathfrak{c} \, z^3 + \mathfrak{d} \, z^4 \ln z   + B \, z^4 + \dots \, ,
\end{equation} 
where 
\begin{equation} 
\begin{split} 
\mathfrak {a} &= - i A \omega \, , \\ 
\mathfrak {b} &= - \frac{1}{4} A \left( \omega^2 + k ^2\right)\, ,  \\
\mathfrak {c} &= -\frac{1}{12} i A \omega \left( \omega^2 - 3 k ^2\right) \, ,  \\
\mathfrak {d} &= -\frac{1}{16} A \left( \omega^2 - k ^2\right)^2 \, .  \\
\end{split} 
\end{equation} 
This asymptotic expansion has the schematic form
\begin{equation}\label{fulbdyasym}
\phi(z) = A (\omega, k) + \dots + B (\omega, k) z^4 + \dots \, ,
\end{equation} 
where $A(\omega, k)$ and  $B (\omega, k)$ are the source and operator expectation values of the operator dual to $\Phi$.

To find the QNMs we usually set $A (\omega, k)$ to zero as an asymptotic boundary condition. Then, together with the boundary condition \eqref{horbdyc}, we find the allowed values of $\omega$ for which we can obtain a legitimate solution to the fluctuation equation. In other words, QNMs are the zeros of $A(\omega, k)$.

On the other hand, the boundary retarded two point function is given by 
\begin{equation} \label{fullg}
\mathcal G_R (\omega, k) = \frac{B(\omega, k)}{A(\omega, k)} \, .
\end{equation} 
It is interesting to note that despite the appearance of $B(\omega, k)$ in this formula, the imaginary part of this retarded correlator, i.e. the spectral function
can be obtained directly from $A(\omega, k)$ as follows
\begin{equation} \label{img0}
\rho(\omega , k) \equiv 2 \, \text{Im} \, \left[ \mathcal G_R (\omega, k) \right] = \mathcal C \frac{\omega}{\lvert A(\omega, k) \rvert^2 }\, ,  ~~ \text{for} ~ \omega  \in \mathbb R \,  ,
\end{equation} 
where $\mathcal C$ is an overall normalization constant. 
This formula is valid only for real $\omega$ and when we have fixed the normalization of our solution by the horizon boundary condition \eqref{horbdyc}. 
The details of the steps leading to \eqref{img0} have been provided in appendix \ref{app:img}. The real part of $\mathcal G_R$ is related to this imaginary part
through the Kramers-Kronig relations.
Note that the form of the spectral function in \eqref{img0} must obey the unitarity condition
\begin{equation}
 \omega \, \rho(\omega, k) \geq 0 \, , ~~ \text{for} ~\omega \in \mathbb R \, .
\end{equation}
Consequently, we must have $\mathcal C \geq 0$.
Also, since $A^\star(\omega) = A(-\omega)$ for real $\omega$, \eqref{img0} is also consistent with the fact that the spectral function is odd, i.e.
\begin{equation}\label{imgodd}
 \rho(\omega , k ) =  - \rho(- \omega , k ) \, .
\end{equation}
For real $\omega$, this oddness of spectral function directly follows from the reality condition of the retarded correlator
\begin{equation}
 \left( \mathcal G_{R} \left( - \omega^\star \right) \right)^\star  = \mathcal G_{R} \left( \omega \right)
\end{equation}
The spectral function is then analytically continued throughout the complex $\omega$ plane maintaining the oddness property \eqref{imgodd}
{\footnote{Note that $\rho(\omega , k) = 2 \, \text{Im} \, \left[ \mathcal G_R (\omega, k) \right]$ is valid only for real $\omega$. More generally away from the real axis it is
given in terms of the retarded and advanced correlators as $\rho(\omega, k) = \textcolor{black}{-} i \left( \mathcal G_R (\omega, k) - \mathcal G_A (\omega, k) \right)$. Recalling that $\mathcal G_A (\omega, k) = \mathcal G_R (-\omega, -k)$, and our system \eqref{phizeqnext} is invariant under the reversal of momentum, we conclude that the spectral function
must be odd in $\omega$ throughout the complex plane.}}. Performing this analytic continuation in \eqref{img0} we get
\begin{equation} \label{img}
\rho(\omega , k ) = \mathcal C \frac{\omega}{A(\omega, k)  A(-\omega, k) } \,   ,
\end{equation} 
so that on the real axis with $A^\star(\omega) = A(-\omega)$, we recover \eqref{img0}.

In recent papers \cite{Dodelson:2023vrw,Bhattacharya:2025vyi}, 
it has been shown that, for the AdS-black branes at finite temperature, $A(\omega, k)$ 
factorizes in terms of its zeros which are the QNMs, along with poles at the Matsubara frequencies
{\footnote {Up to a normalization, our $A(\omega , k)$ is same as the Jost function $f(\omega)$ referred to in appendix C of \cite{Dodelson:2023vrw}. }}.
This factorization led to a product formula for $\text{Im} \,  \mathcal G_R $ through \eqref{img}. 
In particular, it was shown in \cite{Dodelson:2023vrw} that for finite inverse temperature $\beta$, the spectral function can be expressed in terms
of the QNMs as follows
{\footnote{While considering the contribution of off-axis QNMs in the formula \eqref{DodZhiform} we must consider those $\omega_n$ appearing
on a `single line' among the
two `Christmas-tree' branches, as shown in Fig.\ref{fig:QNMTH}. The QNMs
on the other line are captured by the factors in \eqref{DodZhiform} containing the conjugates ${\omega_n}^\star$. See \cite{Dodelson:2023vrw,Bhattacharya:2025vyi} for further details of this convention.}}
\begin{equation}\label{DodZhiform}
\rho(\omega, k) = \mathcal G_0 \, \sinh \left( \frac{\omega \beta}{2}\right) \,  \underbrace{\prod_{n} \left( 1 - \frac{\omega^2}{\omega_n^2} \right)^{-1} \left( 1 - \frac{\omega^2}{({{\omega_n}^\star})^2} \right)^{-1}}_{\text{(off-axis poles: single line)}}    \underbrace{\prod_{m} \left( 1 - \frac{\omega^2}{\omega_m^2} \right)^{-1}}_{\text{(on-axis poles)}} \, ,
\end{equation}
where $\omega_n$ are the values of QNMs which are classified by whether they appear on the negative imaginary axis or off the imaginary axis on the complex $\omega$ plane, as shown in Fig.\ref{fig:QNMTH}. The $k$ dependence in this formula arising through the QNMs and $\mathcal G_0$ remains implicit. The factor $\sinh \left( \frac{\omega \beta}{2}\right)$ is a factor with zeros at the Matsubara frequencies, whose presence can be traced back to the imposition of KMS condition on the Wightman function (see appendix A of \cite{Dodelson:2023vrw} for a brief summary). Henceforth, we shall refer to this factor as the KMS factor.
Our main interest in this paper is the extremal limit, when the on-axis poles coalesce to form a branch cut. As we shall see in our discussion below, in this $T \rightarrow 0$ limit, the KMS factor plays a crucial role in cancelling divergences arising out of the factor with on-axis QNMs.

The factor of $\mathcal G_0$ in \eqref{DodZhiform} is a normalization factor. This factor can be fixed either by normalizing $\rho(\omega, k)$ at a specific reference frequency $\omega = \mathfrak w$,
%(in this case $g_0 =1$)
or by matching with a direct computation using \eqref{fullg} so that $\rho(\omega = \mathfrak w) = g_0$. In either case, $\mathcal G_0$ will take the following form
\begin{equation}\label{G0}
\mathcal G_0 =  \frac{g_0}{\sinh \left( \frac{\mathfrak w\beta}{2}\right)} \,  
\prod_{n} \left( 1 - \frac{\mathfrak w^2}{\omega_n^2} \right) \left( 1 - \frac{\mathfrak w^2}{({{\omega_n}^\star})^2} \right)
\prod_{m} \left( 1 - \frac{\mathfrak w^2}{\omega_m^2} \right) \, . 
%\underbrace{\prod_{n} \left( 1 - \frac{\omega_c^2}{\omega_n^2} \right)^{-1} \left( 1 - \frac{\omega_c^2}{{\omega^\star}_n^2} \right)^{-1}}
%%_{\text{(off-axis poles: single line)}}   
% \underbrace{\prod_{n} \left( 1 - \frac{\omega_c^2}{\omega_n^2} \right)^{-1}}
% %_{\text{(on-axis poles)}} 
%\frac{\sinh \left( \frac{\omega \beta}{2}\right)}{\sinh \left( \frac{\mathfrak w_0 \beta}{2}\right)}
\end{equation} 
In this form, the dependence of $\mathcal G_0$ on the on-axis poles is manifest. Therefore, this form of $\mathcal G_0$ will be very useful while taking the extremal limit on \eqref{DodZhiform}. 

%********************************************************************************
\subsection{Revisiting the thermal product formula}
%********************************************************************************
In the extremal limit, the off-axis QNMs remain discrete and away from the imaginary axis in the complex $\omega$ plane. On the other hand, the sequence of on-axis QNMs on the
negative imaginary axis become dense and is replaced with a branch cut (see Fig.\ref{fig:QNMEx}). Also, the KMS factor has zeros at the Matsubara frequencies
\begin{equation} \label{Matsu}
\omega =   i \, 2 \pi n T \equiv i \Omega_n \, , ~~ n \in \mathbb Z \, .
\end{equation}
As $T \rightarrow 0$, just like the on-axis QNMs, these Matsubara frequencies also become dense on the imaginary axis, particularly for $n<0$. Hence, in the extremal limit, the KMS factor is also expected to play an important role
in the branch cut contribution. These zeros effectively reduce the pole-density coming from the QNMs in the $T \rightarrow 0$ limit.
So in \eqref{DodZhiform}, the KMS factor together with the factor of on-axis QNMs unites to produce a branch cut factor in the extremal limit. Isolating these factors which can potentially contribute to the branch cut, while keeping track of the contribution coming from $\mathcal G_0$ in \eqref{G0}, let us define the finite temperature quantity
\begin{equation} \label{Hdef}
\exp  H_T  =  \frac{\mathfrak w}{\omega} \, \frac{\sinh \left( \frac{\omega \beta}{2}\right)}{\sinh \left( \frac{\mathfrak w \beta}{2}\right)}   \, \prod_{m} \left( 1 - \frac{\omega^2}{\omega_m^2} \right)^{-1}  \left( 1 - \frac{\mathfrak w^2}{\omega_m^2} \right)\, , 
\end{equation}
where we have additionally divided by the factor $\omega / \mathfrak w$, anticipating the $\omega$ factor in \eqref{img}. 
Let us assume that, at any non-zero temperature, however small, the on-axis poles remain discrete and are denoted by 
\begin{equation} \label{impole}
\omega_m = - i \sigma_m , ~~\text{with} ~~ \sigma_m > 0. 
\end{equation}
Here we will assume that there are no zero-modes in the QNM spectrum. This fact is certainly true for our fluctuation equation {\footnote {For other fluctuations in the shear and sound channels, where we expect a hydrodynamic zero mode i.e. $\omega_n \rightarrow 0$ as $k \rightarrow 0$, this analysis may require special care.}}.
Also, using the infinite product representation of the $\sinh$ function we can write 
\begin{equation} \label{sinh}
\frac{\mathfrak w}{\omega} \, \frac{\sinh \left( \frac{\omega \beta}{2}\right)}{\sinh \left( \frac{\mathfrak w \beta}{2}\right)} = \prod_{n=1}^{\infty}  ~
\frac{\Omega_n^2 + \omega^2}{\Omega_n^2 + \mathfrak w^2} \, ,
\end{equation}
where $\Omega_n$ are the magnitude of the Matsubara frequencies in \eqref{Matsu}. Now using \eqref{impole} and \eqref{sinh} back in \eqref{Hdef}, we get 
\begin{equation} \label{HTdef}
\exp  H_T  =  \prod_{n=1}^{\infty}  ~
\frac{\Omega_n^2 + \omega^2}{\Omega_n^2 + \mathfrak w^2} ~ \prod_{m}  ~
\frac{\sigma_m^2 + \mathfrak w^2}{\sigma_m^2 + \omega^2}
\end{equation} 
Taking a Logarithm on both sides converts the product into a sum and we have 
\begin{equation} \label{HTlogexp}
H_T  = - \left(   \sum_{m}  ~
\ln \left( \frac{\sigma_m^2 + \omega^2}{\sigma_m^2 + \mathfrak w^2} \right)  - \sum_{n=1}^{\infty}  ~
\ln \left( \frac{\Omega_n^2 + \omega^2}{\Omega_n^2 + \mathfrak w^2}  \right) \right) \, . 
\end{equation} 
The sums can now be converted into an integral by introducing delta functions 
\begin{equation} 
H_T  = -    \int_0^{\infty} ~ d\sigma  ~ \left(
\ln \left( \frac{\sigma^2 + \omega^2}{\sigma^2 + \mathfrak w^2} \right)  \sum_m \delta (\sigma - \sigma_m) - ~
\ln \left( \frac{\sigma^2 + \omega^2}{\sigma^2 + \mathfrak w^2}  \right) \sum_{n=1}^{\infty} \delta (\sigma - \Omega_n)  \right) \, . 
\end{equation} 
Further, since the delta function is a derivative of the step function $\Theta$, we can simplify 
\begin{equation} 
H_T  = -    \int_{0}^{\infty} ~ d\sigma  
\ln \left( \frac{\sigma^2 + \omega^2}{\sigma^2 + \mathfrak w^2} \right)  \frac{d}{d\sigma} \left( \sum_m \Theta (\sigma - \sigma_m) - 
\sum_{n=1}^{\infty}  \Theta (\sigma - \Omega_n)  \right) \, .
\end{equation} 
Finally, performing an integration-by-parts, we can write 
\begin{equation} \label{Hasint}
H_T  = 2   \int_{0}^{\infty} ~ \sigma d\sigma  ~ \mathcal N (\sigma) ~\left( \frac{1}{\sigma^2 + \omega^2} - \frac{1}{\sigma^2 + \mathfrak w^2} \right) \, , 
\end{equation} 
where 
\begin{equation} \label{cntfn}
\mathcal N (\sigma) = \sum_m \Theta (\sigma - \sigma_m) - 
\sum_{n=1}^{\infty}  \Theta (\sigma - \Omega_n) \, ,
\end{equation} 
counts the difference between the number of QNMs and  number of Matsubara frequencies on the negative imaginary axis, which occur below a given absolute value of $\sigma$. 
Note that, when we performed the integration-by-parts the boundary terms vanish, since when $\sigma \rightarrow \infty$ the argument of the Log term becomes 1, while
when $\sigma \rightarrow 0$ the function $\mathcal N(\sigma)$ vanishes.

Now, the discontinuity of $H_T$ as a function of $\omega$, about the negative imaginary axis in the complex $\omega$-plane, is given by 
\begin{equation} \label{discHT}
\begin{split} 
\text{Disc} \, H_T (\omega = -i \lambda) =&  \lim_{\epsilon \rightarrow 0} \left[ H_T ( -i \lambda + \epsilon) - H_T ( -i \lambda - \epsilon) \right] \\
=& 2 \pi  i \, \mathcal N(\lambda) \, , ~ \text{for real} ~ \lambda , \, \epsilon > 0 \, .
\end{split}
\end{equation}
This can be derived directly from \eqref{Hasint}. Also, it is insightful to understand how this result follows from \eqref{HTlogexp}. In \eqref{HTlogexp}, we must choose the branch cut of the individual logs
along the negative imaginary axis, since the QNMs and KMS zeros condense along this half line. The discontinuity of the log along the negative imaginary axis is 
\begin{equation} 
\text{Disc} \, \ln (-i x) = \lim_{\epsilon \rightarrow 0} \left[ \ln ( -i x + \epsilon) - \ln ( -i x - \epsilon) \right] = - 2 \pi i , ~ \text{for real} ~ x , \, \epsilon > 0 \, .
\end{equation} 
In order to compute the discontinuity of $H_T$ we can accumulate the discontinuity of the individual logs, after accounting for their signs properly. From this it follows that the discontinuity in $H_T (\omega = -i \lambda)$ cumulatively arising out of the individual logs will be $2\pi i$ times the difference between the number of QNMs and Matsubara frequencies below $\lambda$. This number difference is exactly counted by $\mathcal N(\lambda)$. Hence \eqref{discHT} follows.   

Now, let $\Xi$ denote the product over off-axis QNMs in \eqref{DodZhiform}, i.e. we can write 
\begin{equation} \label{Xidef}
\Xi =  \underbrace{\prod_{n} \left( 1 - \frac{\omega^2}{\omega_n^2} \right)^{-1} \left( 1 - \frac{\omega^2}{({{\omega_n}^\star})^2} \right)^{-1}}_{\text{(off-axis poles: single line)}} \, .
\end{equation} 
Comparing \eqref{img} with \eqref{DodZhiform} and using \eqref{Hdef}, we can write
\begin{equation} \label{Asqform}
 \frac{1}{  A(\omega) A(-\omega) } = \tilde {\mathcal G}_0 \,  \Xi(\omega) \, \exp {H_T(\omega)} \, , 
\end{equation} 
where $\tilde {\mathcal G}_0$ is an overall number independent of $\omega$. 
Note that, at this stage, \eqref{Asqform} is just a rewriting of the product formula of the finite temperature Jost function as put forward in \cite{Dodelson:2023vrw}. But in this form it is clear that
discontinuity along the negative imaginary axis on the complex $\omega$ plane, which exists in $H_T$ on the RHS, must arise from $\ln A(\omega)$ on the LHS {\footnote {Note that the QNMs, by definition, are zeros of $A(\omega)$ rather than being zeros of $A(-\omega)$. Since the QNMs on the negative imaginary axis are responsible for this discontinuity, hence the discontinuity is associated with $\ln A(\omega)$ rather than $\ln A(-\omega)$.}}. Here, $\ln A(-\omega)$ is analytic on the negative imaginary axis, but is discontinuous on the positive imaginary axis.  Note that, at non-zero temperature all the QNMs are discretely separated, consequently there are no branch cuts in \eqref{Asqform}. However, the Logarithm of the thermal product formula \eqref{DodZhiform} does have many branch cuts. But in this argument we have focused on the branch cuts along the negative imaginary axis on which the QNMs will ultimately condense in the extremal limit of our interest.

%********************************************************************************
\subsection{A pole-cut formula for the extremal spectral function} \label{ssec:polecut}
%********************************************************************************

In the extremal limit, $\Xi(\omega)$ smoothly converges to the product contribution of extremal off-axis QNMs
\begin{equation}
 \lim_{T \rightarrow 0} ~\Xi(\omega) = \mathcal F_{\text{poles}}(\omega) \, .
\end{equation}
On the other hand, $H_T(\omega)$ converges to the branch cut contribution as follows
\begin{equation} \label{Hcut}
\lim_{T \rightarrow 0}~ H_{T}(\omega) \equiv H_{\text{cut}} (\omega) = - \frac{2}{\pi} \int_{0}^{\infty} ~ \sigma d\sigma  ~ \theta (\sigma) ~\left( \frac{1}{\sigma^2 + \omega^2} - \frac{1}{\sigma^2 + \mathfrak w^2} \right) \, ,
\end{equation} 
where, in the extremal limit, the function $\mathcal N(\omega)$ weakly converges to the discontinuity of the Log of the Jost function $A(\omega)$ about the branch cut, which we denote by
$\theta(\omega)$. This follows from the discussion following \eqref{Asqform} in the previous subsection. More precisely,
\begin{equation} \label{thetadef}
\lim_{T \rightarrow 0} \,   \mathcal N ( \lambda ) = \lim_{T \rightarrow 0} \frac{\text{Disc} \, H_T (\omega = - i \lambda) }{2 \pi i} = - \frac{\text{Disc} \, \ln A(\omega = - i \lambda)}{2 \pi i} \equiv -\frac{1}{\pi} \, \theta(\lambda).
\end{equation} 
Clearly, $\theta$ here measures the phase-jump of the Jost function $A$ across the
branch cut which physically exists for the extremal correlator
{\footnote{If we consider $A_\pm \equiv A\left( \omega = - i \lambda \pm \epsilon \right) = \lvert A_0 \rvert e^{\pm i \theta}  $, then $\text{Disc} \ln A = \ln \left( A_+ / A_-\right) = 2 i \, \theta$.}}.

Thus, putting everything together we arrive at the pole-cut formula for the spectral function in the extremal limit
\begin{equation} \label{pole-cut}
\rho(\omega, k)
= \tilde {\mathcal G}_0 ~ \omega \, \exp \left( {H_{ \text {cut} }} (\omega) \right) 
\underbrace{\prod_{n} \left( 1 - \frac{\omega^2}{\omega_n^2} \right)^{-1} \left( 1 - \frac{\omega^2}{({{\omega_n}^\star})^2} \right)^{-1}}_{\mathcal F_{\text{poles}}(\omega) \, = \, \text{off-axis extremal poles (single line)}} \, ,
\end{equation}
where $H_{ \text {cut}}$ is given by \eqref{Hcut}. Note that, in this formula both the non-trivial factors are spectral in nature, i.e. the poles are determined by the zeros of the Jost function $A(\omega)$, while the branch-cut contribution is determined by the discontinuity of the phase of the same Jost function. Interestingly, in this formula the Matsubara zeros have dissolved into the branch-cut contribution.
%
%
%
%
%
%
%*********************************************************************************************************
\subsection{Low frequency branch cut contribution from near-horizon AdS$_2$}\label{ssec:AdS2}
%*********************************************************************************************************
%
In the near-extremal black brane an AdS$_2$ region begins to develop close to the horizon which is a characteristic feature of the extremal geometry.
The nature of boundary correlation functions and QNMs using this near horizon AdS$_2$ has been extensively discussed in the literature
\cite{Faulkner:2009wj, Denef:2009yy, Faulkner:2010da, Edalati:2010hk, Edalati:2010pn, Faulkner:2010tq, Arnaudo:2024sen}. 

At low but non-zero temperature, it is possible
to write down approximate analytical expressions for the on-axis QNMs using the fluctuations of the near-horizon AdS$_2$ black hole geometry.
In the $T \rightarrow 0$ limit, these approximate QNMs coalesce on the negative imaginary axis on the complex $\omega$ plane, i.e. on $\omega = - i \lambda$, exactly
reproducing the QNM density near $\lambda \rightarrow 0$. On the other hand, using the near-horizon
AdS$_2$ geometry in the strict extremal limit, it is possible to obtain an analytic formula for the leading order discontinuity in the logarithm of the source function
$A$ in \eqref{thetadef}. In other words, if we expand $\theta (\lambda)= \theta_{\text{IR}} + \mathcal O (\lambda)$ in the small $\lambda$ expansion, then $\theta_{\text{IR}}$
can be obtained reliably from the IR AdS$_2$.

Thus the near-horizon AdS$_2$ provides us with an excellent testing ground, where $H_{\text{cut}}$ in \eqref{Hcut} may be obtained both from the extremal QNMs density
as well as the discontinuity of $(\ln A)$, thus providing an explicit demonstration of our argument leading to the pole-cut formula in \ref{ssec:polecut}.
In appendix \ref{app:ads2}, we have reviewed the derivation of these approximate near-extremal QNMs, as well as the evaluation of $\theta_{\text{IR}}$.
These QNMs are given by {\footnote{ \textcolor{black}{ In deriving these QNMs, we have assumed that $\nu_k$ does not take half-integer values. At half-integer values, $\omega^{2 \nu_k}$ becomes analytic. } } }
\begin{equation} \label{ads2qnm}
\omega_m = - i \sigma_m = - 2 \pi i  \, T ( m + \mathfrak c), ~\text{where}~ \mathfrak c = \frac{1}{2} + \nu_k ~\text{and}~ m = 0,1,2, \dots \, ,
\end{equation} 
where $$\nu_k = \sqrt{\frac{1}{4} + \frac{k^2}{12}} \, ,$$ is related to the IR dimension of the relevant operator (see \eqref{nuk}).
Using this result, it is straightforward to simplify the RHS of \eqref{HTdef} into a closed form.
Firstly let us note that we can rewrite the infinite products in terms of gamma functions as follows
\begin{equation} 
\begin{split} 
P_1 \equiv&   \prod_{n=1}^{\infty}  ~
 \frac{\Omega_n^2 + \omega^2}{\Omega_n^2 + \mathfrak w^2} 
 = \prod_{n=1}^{\infty}  ~
 \frac{1 + \omega^2 / (2 \pi n T)^2}{1 + \mathfrak w^2 / (2 \pi n T)^2}  
 =   \frac{\Gamma \left( 1 + i \mathfrak w / 2 \pi \, T \right) \Gamma \left( 1 - i \mathfrak w / 2 \pi \, T \right)}{\Gamma \left( 1 + i \omega / 2 \pi \, T \right) \Gamma \left( 1 - i \omega  / 2 \pi \, T \right)} \, ,\\
 % \\ =& \frac{\omega - \mathfrak w}{2T} - \ln \frac{\omega}{\mathfrak w} + \mathcal O \left( T \right) ,\\
P_2 \equiv&  \prod_{m}  ~
 \frac{\sigma_m^2 + \mathfrak w^2}{\sigma_m^2 + \omega^2} 
  =  \prod_{m=0}^\infty  ~
 \frac{1 + \mathfrak w^2/(2 \pi \, T (m + \mathfrak c))^2}{1 + \omega^2 /(2 \pi \, T (m + \mathfrak c))^2}  
= \frac{\Gamma \left( \mathfrak c + i \omega / 2 \pi \, T \right) \Gamma \left( \mathfrak c - i \omega / 2 \pi \, T \right)}{\Gamma \left( \mathfrak c + i \mathfrak w / 2 \pi \, T \right) \Gamma \left( \mathfrak c - i \mathfrak w / 2 \pi \, T \right)} \\
% =& - \frac{\omega - \mathfrak w}{2T} + (2 \mathfrak c - 1) \ln \frac{\omega}{\mathfrak w} + \mathcal O \left( T \right) 
\end{split} 
\end{equation} 
Then, taking  Logarithm on these products and applying Stirling's formula in the $T \rightarrow 0$ limit, we get 
\begin{equation} \label{strfor}
\begin{split} 
\ln P_1 =& \frac{\omega - \mathfrak w}{2T} - \ln \frac{\omega}{\mathfrak w} + \mathcal O \left( T \right) \, , \\
\ln P_2 =& - \frac{\omega - \mathfrak w}{2T} + (2 \mathfrak c - 1) \ln \frac{\omega}{\mathfrak w} + \mathcal O \left( T \right) \, .\\ 
\end{split} 
\end{equation} 
Putting these results together to evaluate $H_T$ in \eqref{HTdef}, we get 
\begin{equation}\label{HTcancel}
H_T(\omega) = (2 \mathfrak c - 2) \ln \frac{\omega}{\mathfrak w} + \mathcal O \left( T \right) \, .
\end{equation} 
Interestingly, the divergent terms of \eqref{strfor}
in the $T \rightarrow 0$ limit cancel between
$\ln P_1$ and $\ln P_2$ leading to an expression which has a smooth extremal limit.
Recall that $P_1$ originated from the KMS factor involving the Matsubara frequencies \eqref{Matsu}, while $P_2$ descended from the product over the on-axis QNMs. This implies that the extremal divergences neatly cancel between the Matsubara and QNMs factors to produce a finite well defined branch cut contribution. The $T \rightarrow 0$ limit on \eqref{HTcancel} smoothly gives the branch cut contribution
\begin{equation} \label{ads2hcutqnm}
H_{\text{cut}}(\omega)  = \lim_{T \rightarrow 0} H_T(\omega)  = (2 \mathfrak c - 2) \ln \frac{\omega}{\mathfrak w} = \left( 2 \, \nu_k - 1 \right) \ln \frac{\omega}{\mathfrak w} \, .
\end{equation}
Note that, although this appears to be an exact result, it is essentially the leading order result in a small $\omega$ expansion, i.e. $\omega \ll \mathcal O(1)$
{\footnote{Strictly speaking, here we mean $\omega \ll \mu_e$ (See \ref{fn:unit}).}}
. This assumption is inherited from the use of AdS$_2$ black hole to compute the QNMs, which only gets the low overtones accurate (see appendix \ref{app:ads2} for further discussion on this issue). It is only for $\omega \ll \mathcal O(1)$, the product is dominated by these low overtones which the AdS$_2$ throat approximation
gets right. So the small $\omega$ condition is inherited from the approximate nature of the QNMs.
Note that both the Stirling approximation as well as the AdS$_2$ throat approximation are simultaneously valid in the range $T \ll \omega \ll \mathcal O(1)$.
% Also, it is very interesting to note that both the KMS factor  $\left( P_1 \right)$ as well as the contribution from the on-axis QNMs $\left( P_2 \right)$ had terms which
% were divergent in the $T \rightarrow 0$ limit. Both these divergent terms neatly cancelled between each other to produce a finite well defined branch cut contribution. We believe  this cancellation is universal for holographic correlation functions and perhaps may be relevant even beyond holographic context.

The result \eqref{ads2hcutqnm} can also be obtained directly from the strict $T \rightarrow 0$ limit.
Focusing on the strict extremal limit, as reviewed in appendix \ref{app:ads2}, the function $\theta(\lambda)$, capturing the discontinuity of the
source function $A$ about the branch cut, can be obtained at leading order in small $\lambda$ expansion as follows
\begin{equation}
 \theta (\lambda) = \theta_{\text{IR}} + \mathcal O (\lambda^{2 \nu_k}) \,
\end{equation}
where $\theta_{\text{IR}} = \pi \left( \nu_k - 1/2 \right)$ (see \eqref{thetaexp}). Using this into \eqref{Hcut}, the leading order answer in small $\omega$ can be obtained by taking
$\theta$ to be constant and we have
\begin{equation}
 H_{\text{cut}} (\omega) = - \frac{2}{\pi} \, \theta_{\text{IR}} \, \int_{0}^{\infty} ~ \sigma d\sigma ~\left( \frac{1}{\sigma^2 + \omega^2} - \frac{1}{\sigma^2 + \mathfrak w^2} \right)
 =  \left( 2 \, \nu_k - 1 \right) \ln \frac{\omega}{\mathfrak w} \, ,
\end{equation}
which matches with \eqref{ads2hcutqnm}. This provides a very non-trivial check of the formula \eqref{Hcut} relating the branch cut factor with the discontinuity of $(\ln A)$.

%***************************************************************************************************
\subsection{Numerical verification using full extremal black brane geometry}
%***************************************************************************************************

We now proceed to verify our pole-cut formula \eqref{pole-cut} numerically using the full AdS$_5$ extremal black brane, without any approximations.
We shall perform this verification for real $\omega$ when the spectral function is given in terms of imaginary part of the retarded correlator,
i.e. $\rho = 2 \, \text{Im}  \, \mathcal G_R$ . For this reason, in this section we will directly use $\text{Im}  \, \mathcal G_R$ to perform the numerical tests.
Also, for the purpose of numerics, for most part of this subsection we work with $k=1$. We have also tested the formula \eqref{pole-cut} sampling over other values of $k$ as we report at the end of this subsection.

At first we will work in the extremal geometry \eqref{backextsol} and numerically solve the fluctuation equation \eqref{phizeqnext} with in-going boundary conditions, fixing the normalization with \eqref{horbdyc}. From this numerical solution we shall read off the source $A(\omega,k)$ and the vev $B(\omega,k)$ of the dual operator from the asymptotic form of the solution \eqref{fulbdyasym}. Subsequently, we can calculate the retarded two point function using \eqref{fullg} and extract its imaginary part
\begin{equation}\label{directGr}
 \frac{1}{2} \, \rho(\omega, k) = \text{Im} \left[ \mathcal G_R \left( \omega , k\right)\right] = \text{Im} \left( \frac{B(\omega,k)}{A(\omega,k)} \right), ~~ \omega \in \mathbb R \, .
\end{equation}
After performing this numerical exercise, we plot $\text{Im} \left[ \mathcal G_R \left( \omega \right)\right]$ for $k = 1,2$, in Fig.\ref{fig:numpolecut}.
This has been shown by the hyphenated lines interspaced with open circles ($-\circ-$)
in Fig.\ref{fig:numpolecut}.
For a verification of our pole-cut formula, we now reproduce this plot using the RHS of \eqref{pole-cut}. This formula has two structural components, one arising from the off-axis QNMs and the other capturing the contribution of the spectral branch cut. We will deal with these two contributions separately.

%%%%%%%%%%%%%%%%%%%%%%%%%%%%%%%%%%%%%%%%%%%%
\begin{figure}
\centering
\begin{subfigure}{.5\textwidth}
  \centering
  \includegraphics[width=0.9\textwidth]{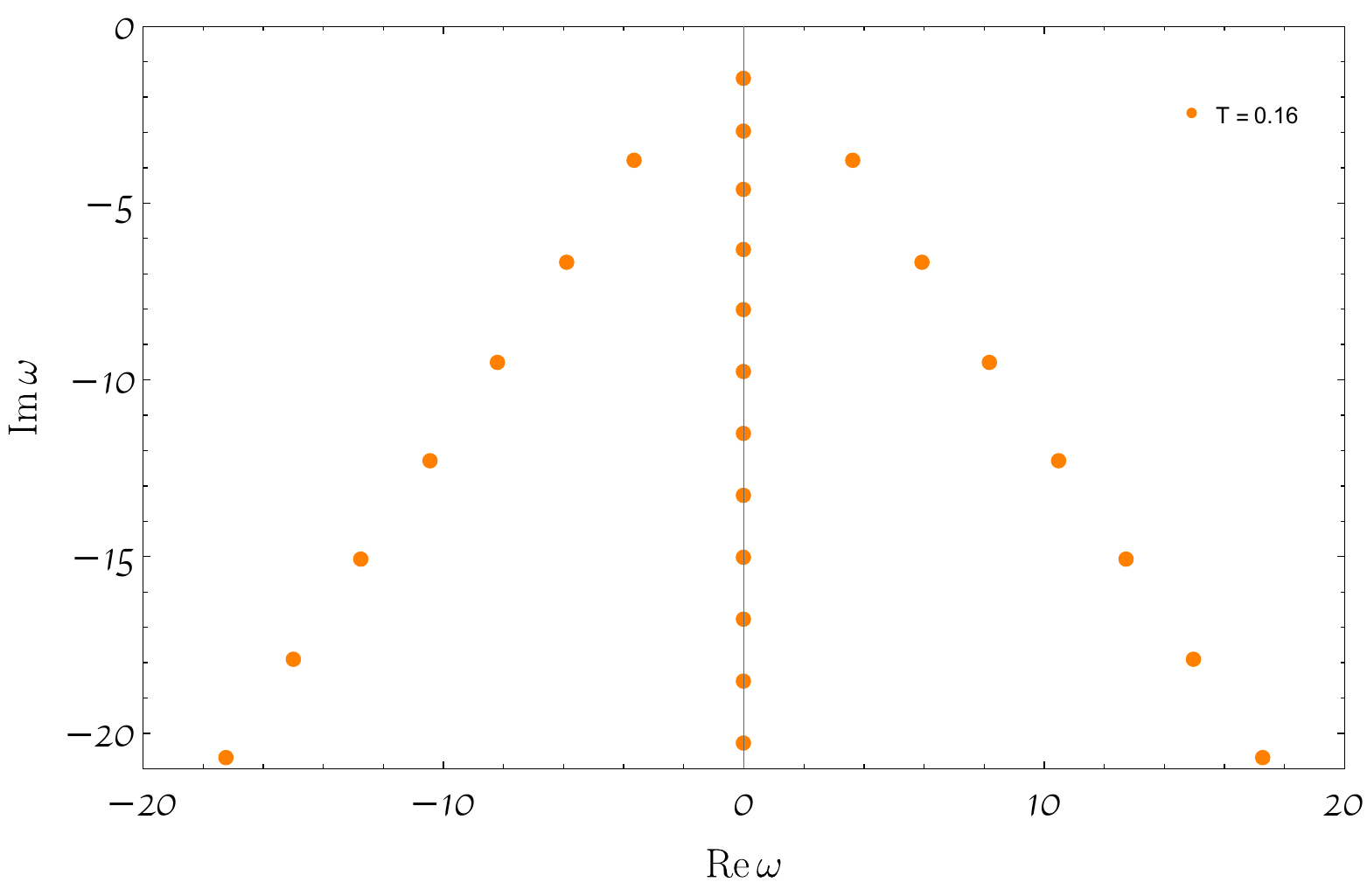}
  \caption{Thermal}
  \label{fig:QNMTH}
\end{subfigure}%
\begin{subfigure}{.5\textwidth}
  \centering
  \includegraphics[width=0.9\textwidth]{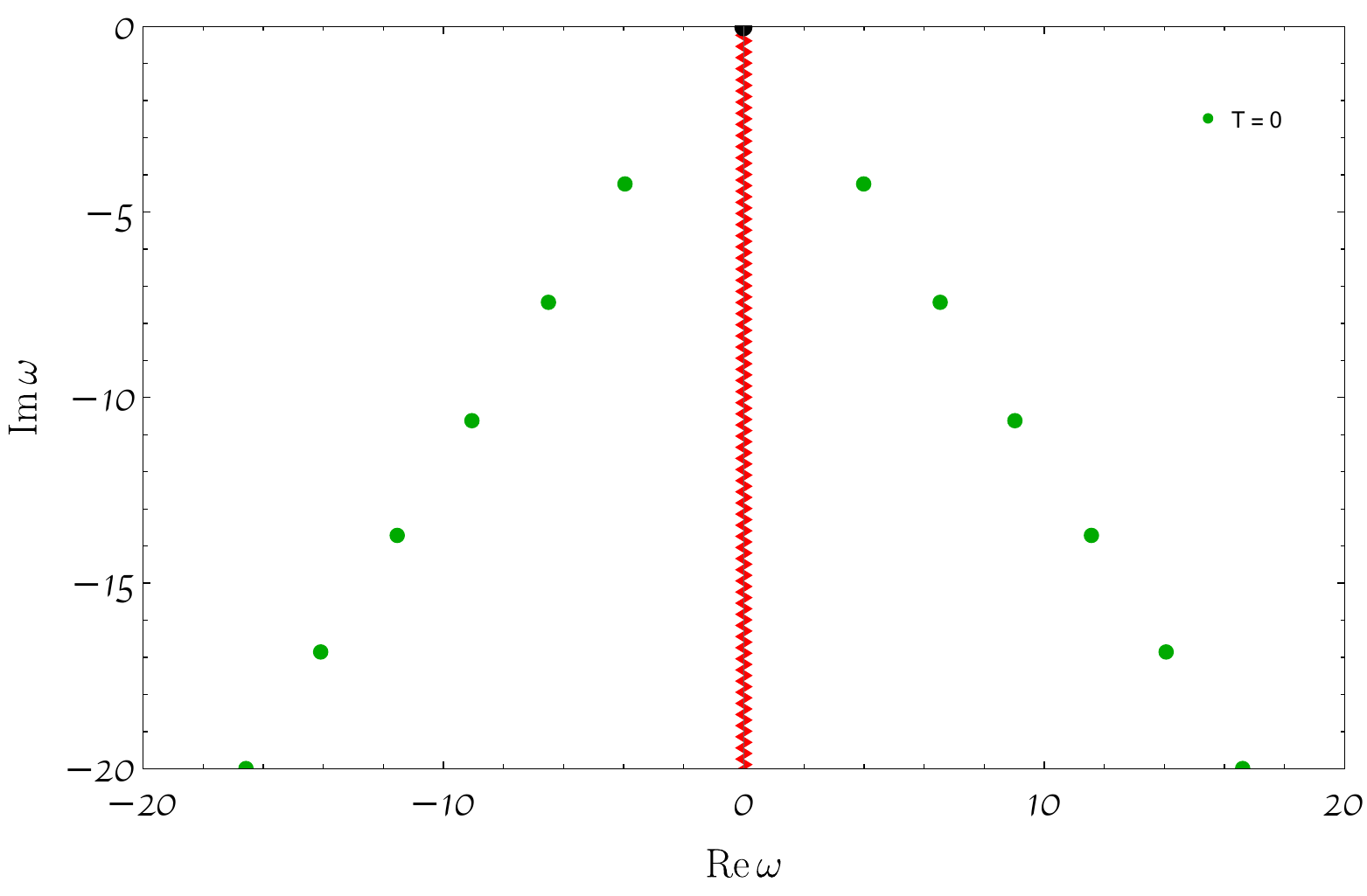}
  \caption{Extremal}
  \label{fig:QNMEx}
\end{subfigure}
\caption{QNM spectrum on complex-$\omega$ plane.}
\label{fig:QNM}
\end{figure}
%%%%%%%%%%%%%%%%%%%%%%%%%%%%%%%%%%%%%%%%%%%%

\vspace{0.40cm}
\noindent
{\bf{Contribution from off-axis QNMs:}}
In \eqref{pole-cut}, the off-axis QNMs appear in a product over poles just like the thermal product formula in \cite{Dodelson:2023vrw,Bhattacharya:2025vyi}.
Hence we will compute this contribution numerically
following the technique used in these previous papers. Using the Mathematica package QNMspectral \cite{Jansen:2017oag} we generated 40 off-axis QNMs some of which have been shown in Fig.\ref{fig:QNMEx}.
The large overtone off-axis QNMs can be approximated by the formula
\begin{equation}\label{qnmasym}
 \omega_n = \omega_{n_0} + r e^{i \vartheta} \, (n - n_0)  \, .
\end{equation}
By fitting the last 10 out of the 40 off-axis QNMs on the right branch of the `Christmas tree'
pattern (see Fig.\ref{fig:QNMEx}), we find
\begin{equation}\label{qnmparavals}
 r \approx 4.00, ~~\vartheta \approx -0.89 , ~~n_0 = 40,
 ~~\omega_{40}  \approx 102.06 - i \, 125.80 \, .
\end{equation}
In appendix \ref{app:asymexact} we have determined these parameters exactly and our numerical results in \eqref{qnmparavals} are completely consistent with \eqref{paravals}. The parameters in \eqref{qnmasym} map to the parameters in
\eqref{qnmasymform} in the following way
\begin{equation}
 a = r e^{i \vartheta} , ~~ q = \frac{\omega_{n_0}}{\, r}  \, e^{- i \vartheta} - \, n_0  .
\end{equation}
Here, the numerical value of $q$ is independent of $n_0$ for large values of
$n_0$.
Approximating the overtones higher than 40 with \eqref{qnmasym}, we can rewrite the infinite product in \eqref{pole-cut} as follows
\begin{equation}\label{Fpoles}
\begin{split}
 \mathcal F_{\text{poles}}(\omega)
% = \prod_{n = 1}^{\infty} \left( 1 - \frac{\omega^2}{\omega_n^2} \right)^{-1} \left( 1 - \frac{\omega^2}{({{\omega_n}^\star})^2} \right)^{-1} \\
= \prod_{n = 1}^{40} \left( 1 - \frac{\omega^2}{\omega_n^2} \right)^{-1} \left( 1 - \frac{\omega^2}{({{\omega_n}^\star})^2} \right)^{-1}
 \times \mathcal F_{\text{tail}} \, ,
\end{split}
\end{equation}
where $\mathcal F_{\text{tail}}$ is a tail correction term which captures
the contribution from all the asymptotic QNMs beyond 40, and is given by
\cite{Dodelson:2023vrw,Bhattacharya:2025vyi}
{\footnotesize{
\begin{equation}
\mathcal F_{\text{tail}} = \frac{ \Gamma \left(1 + \frac{e^{ - i \vartheta }}{r} (\omega_{n_0}-\omega)  \right)
\Gamma \left(1 + \frac{e^{- i \vartheta } }{r} (\omega_{n_0}+\omega)\right)
\Gamma \left(1 + \frac{e^{i \vartheta } }{r}  \left(\omega_{n_0}^*-\omega\right)\right)
\Gamma \left(1 + \frac{e^{i \vartheta } }{r} \left(\omega+\omega_{n_0}^*\right)\right)
}{\left| \Gamma \left(1+ \frac{e^{-i \vartheta }
   \omega_{n_0}}{r}\right)\right| ^4}
\end{equation}
}}
We have plotted the function $\mathcal F_{\text{poles}}(\omega)$ in Fig.\ref{fig:numpolecutFp}, which clearly differs significantly from Im$\, \mathcal G_R$ in Fig.\ref{fig:numpolecut}. Clearly, we need the branch cut contribution for the complete reconstruction of Im$\, \mathcal G_R$.

%%%%%%%%%%%%%%%%%%%%%%%%%%%%%%%%%%%%%%
\begin{figure}
\centering
\begin{subfigure}{.5\textwidth}
  \centering
  \includegraphics[width=0.9\textwidth]{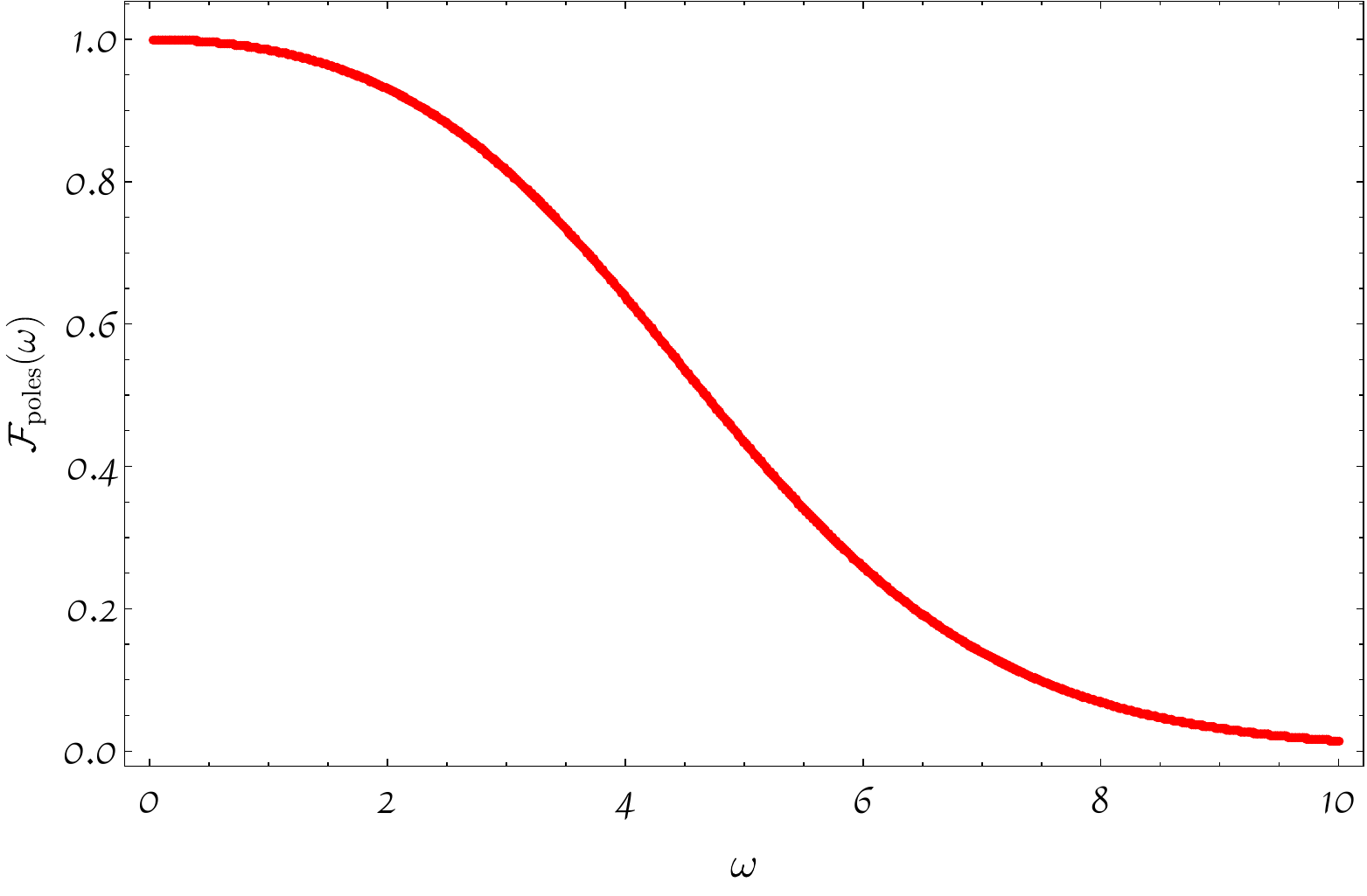}
  \caption{}
  \label{fig:numpolecutFp}
\end{subfigure}%
\begin{subfigure}{.5\textwidth}
  \centering
  \includegraphics[width=0.9\textwidth]{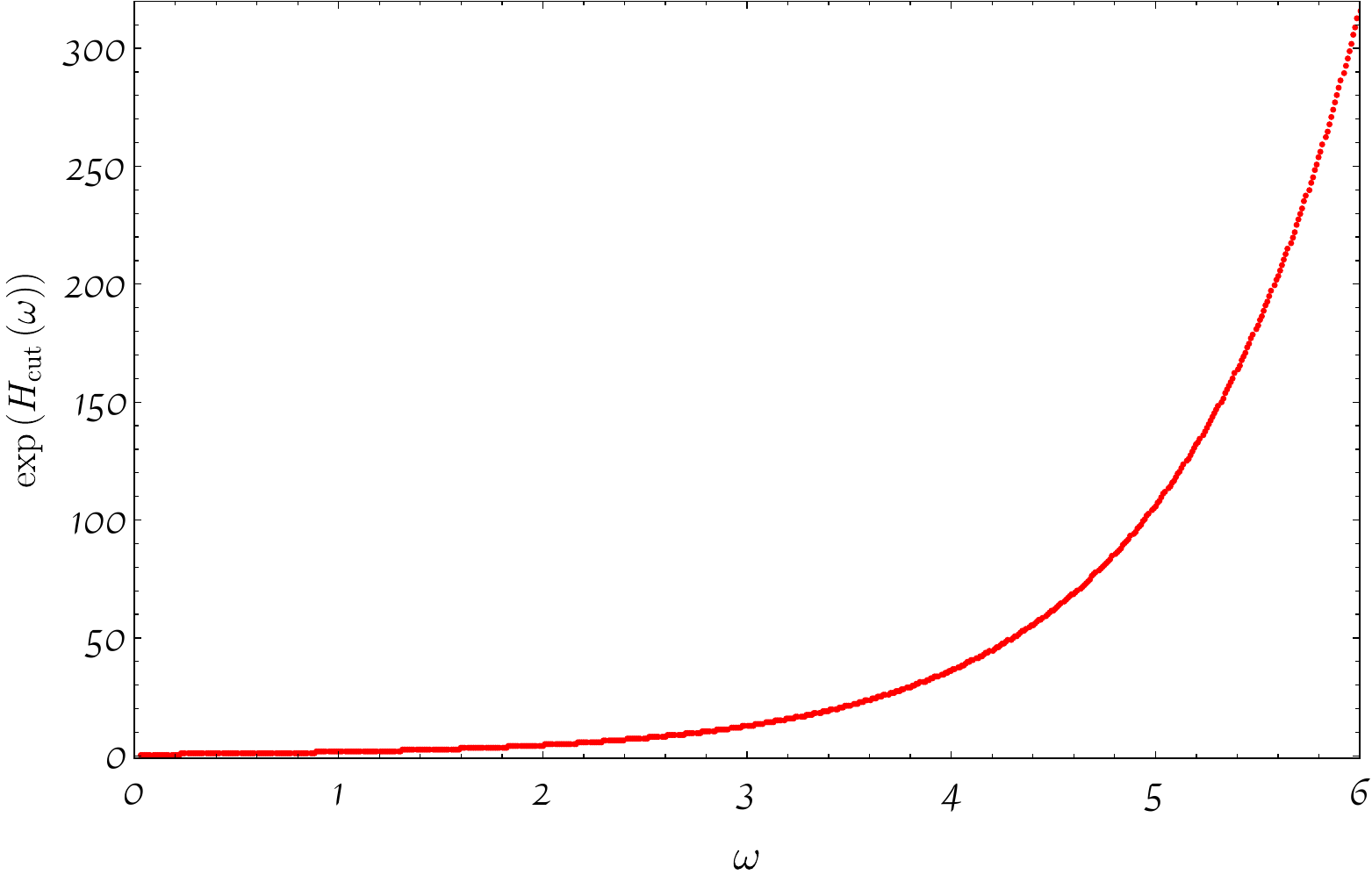}
  \caption{}
  \label{fig:Hcut}
\end{subfigure}
\caption{In (a) we plot the pole-contribution $\mathcal F_{\text{poles}}$ using \eqref{Fpoles}. We have generated 40 QNMs at $k=1$ for this plot. A tail correction formula has
been used to account for the contribution from QNMs beyond 40. In (b) we have plotted the cut-contribution $\exp H_{\text{cut}}$ using \eqref{Hcut}, also for $k=1$.}
\label{fig:PoleCutsep}
\end{figure}
%%%%%%%%%%%%%%%%%%%%%%%%%%%%%%%%%%%%%%

\vspace{0.40cm}
\noindent
{\bf{Contribution from branch cut:}}
Finally, we also numerically compute the branch cut contribution $H_{\text{cut}}$ which is necessary to verify \eqref{pole-cut}. In order to compute
$H_{\text{cut}}$ using \eqref{Hcut} we need to compute $\theta(\lambda)$ in \eqref{thetadef} by evaluating the discontinuity in the source function $A(\omega)$.
The numerical procedure for obtaining this discontinuity has been outlined in appendix \ref{app:discon}. Using this procedure, we evaluated
$\theta(\lambda)$ in the range $\lambda \in \left[ 0.03,50\right]$ for $k=1$ which is plotted using the solid blue line in Fig.\ref{fig:theta}.

%%%%%%%%%%%%%%%%%%%%%%%%%%%%%%%%%%%%%%%%%%%%%%%%%%%%%%%%%%%%%%%%%%%%%%%%%%%%
\begin{figure}[h!tbp]
  \centering
  \includegraphics[width=0.5\textwidth]{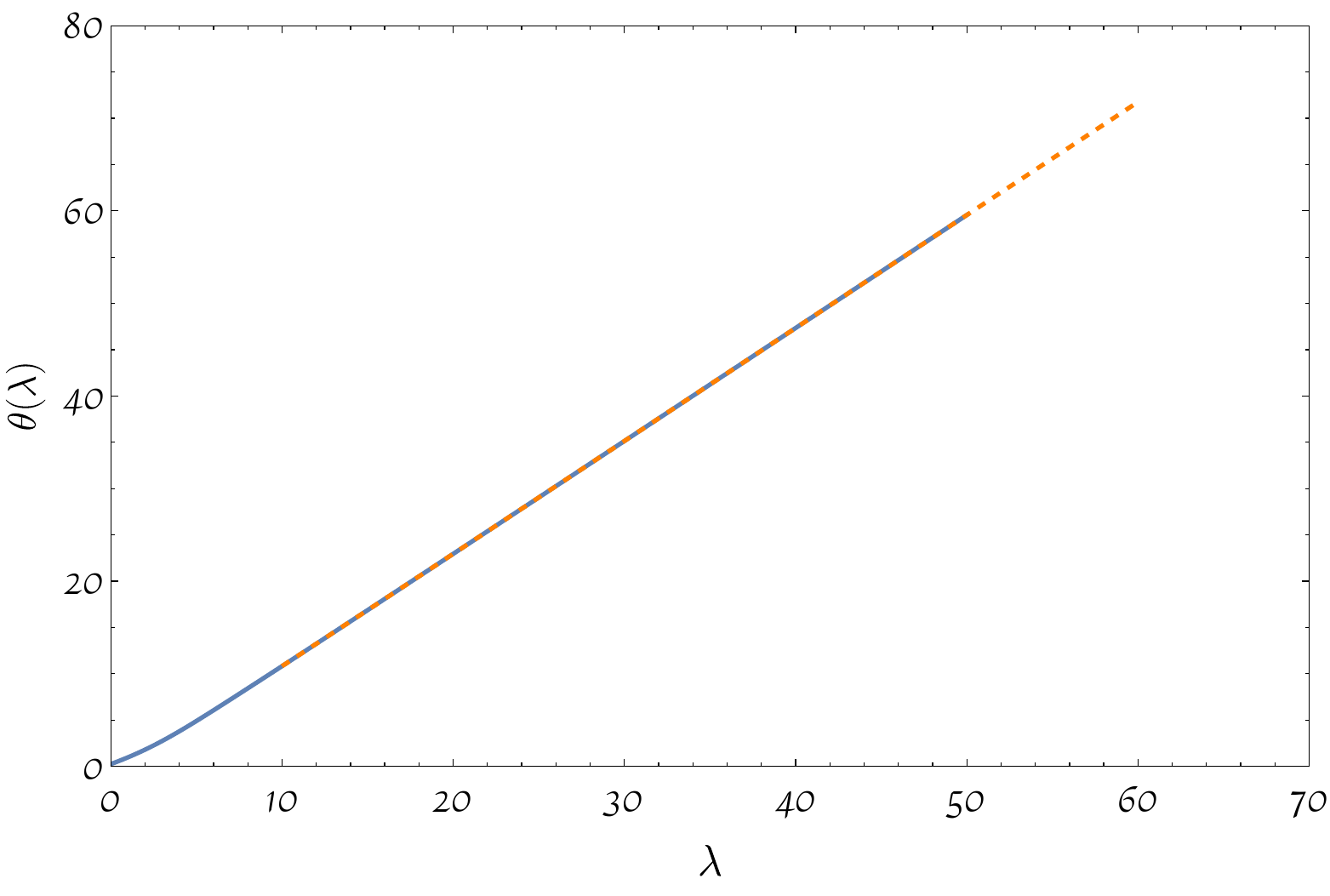}
  \caption{Plot of the discontinuity of $(\ln A)$, i.e. the phase-jump of the source function across the branch cut along the negative imaginary axis
  in the complex $\omega$ plane parameterized by $\omega = - i \lambda$ with $\lambda > 0$. The solid line has been numerically obtained using
  the procedure outlined in appendix \ref{app:discon}. The orange dashed line is a plot of the extrapolation of the numerical function using the ansatz
  \eqref{thetapara}.}
  \label{fig:theta}
\end{figure}
%%%%%%%%%%%%%%%%%%%%%%%%%%%%%%%%%%%%%%%%%%%%%%%%%%%%%%%%%%%%%%%%%%%%%%%%%%%%

For small values of $\lambda$, the behaviour of $\theta(\lambda)$ was anticipated in \eqref{thetaexp}.
Numerical fitting with low values of $\lambda$ gives $\mathcal C_0 \approx 1.16$.

Note that in order to execute the integral in \eqref{Hcut}
we need to know $\theta(\lambda)$ in the entire range $\lambda \in \left[ 0, \infty \right)$. Hence we extrapolate $\theta(\lambda)$ beyond $\lambda = 50$, using the
following ansatz
\begin{equation}\label{thetapara}
 \theta (\lambda)
 %\overset{\lambda \gg 1}{=}
 = \alpha \lambda + \beta + \dfrac{\gamma}{\lambda}, ~~ \text{for}~~ \lambda \gg 1 \, .
\end{equation}
By appropriately fitting the region $\lambda \in \left[ 10,50\right]$ with this ansatz, the parameters were determined to be
\begin{equation}\label{albetvals}
\alpha \approx 1.22 , ~~ \beta \approx - 1.57 , ~~  \gamma \approx  1.74 , ~~ \text{for} ~~ k =1.
\end{equation}
Using these parameters into \eqref{thetapara} provides the behaviour of $\theta(\lambda)$ above $\lambda = 50$ including arbitrary large values of $\lambda$ (see the orange dashed line in Fig.\ref{fig:theta}). In this way, we have numerically obtained $\theta(\lambda)$ over the entire range $\left[ 0, \infty \right)$. This can now be used in \eqref{Hcut}. Performing
a numerical integration in \eqref{Hcut}, we are able to eventually obtain $\exp H_{\text{cut}}(\omega)$, which has been plotted in Fig.\ref{fig:Hcut}.

Combining the numerical results of $\mathcal F_{\text{poles}}(\omega)$ together with $\exp H_{\text{cut}}$
into the RHS of the pole-cut formula \eqref{pole-cut}, we obtain the solid line (for $k=1$)
in Fig.\ref{fig:numpolecut}
{\footnote{In Fig.\ref{fig:numpolecut}, we have plotted $\text{Im} \, \mathcal G_R = \tfrac{1}{2} \rho$ with $\rho$ being given by \eqref{pole-cut}. We have fixed the normalization constant
at the reference frequency $\mathfrak{w} = 0.5$ with the condition $\rho (\omega = 0.5 ) = 2$, which is equivalent to  $\text{Im} \, \mathcal G_R(\omega = 0.5) = 1$. }}
. Clearly, this matches extremely well with hyphenated line interspaced with circles ($-\circ-$), which was obtained from the computation of imaginary part of the retarded correlator using \eqref{directGr}. We have
also found similar agreement for $k = 2$ which we have included in Fig.\ref{fig:numpolecut}. Thus our pole-cut formula is verified numerically over a wide range of $\omega$ and for two distinct values of $k$.

%%%%%%%%%%%%%%%%%%%%%%%%%%%%%%%%%%%%%%
\begin{figure}[h!tbp]
  \centering
  \includegraphics[width=0.7\textwidth]{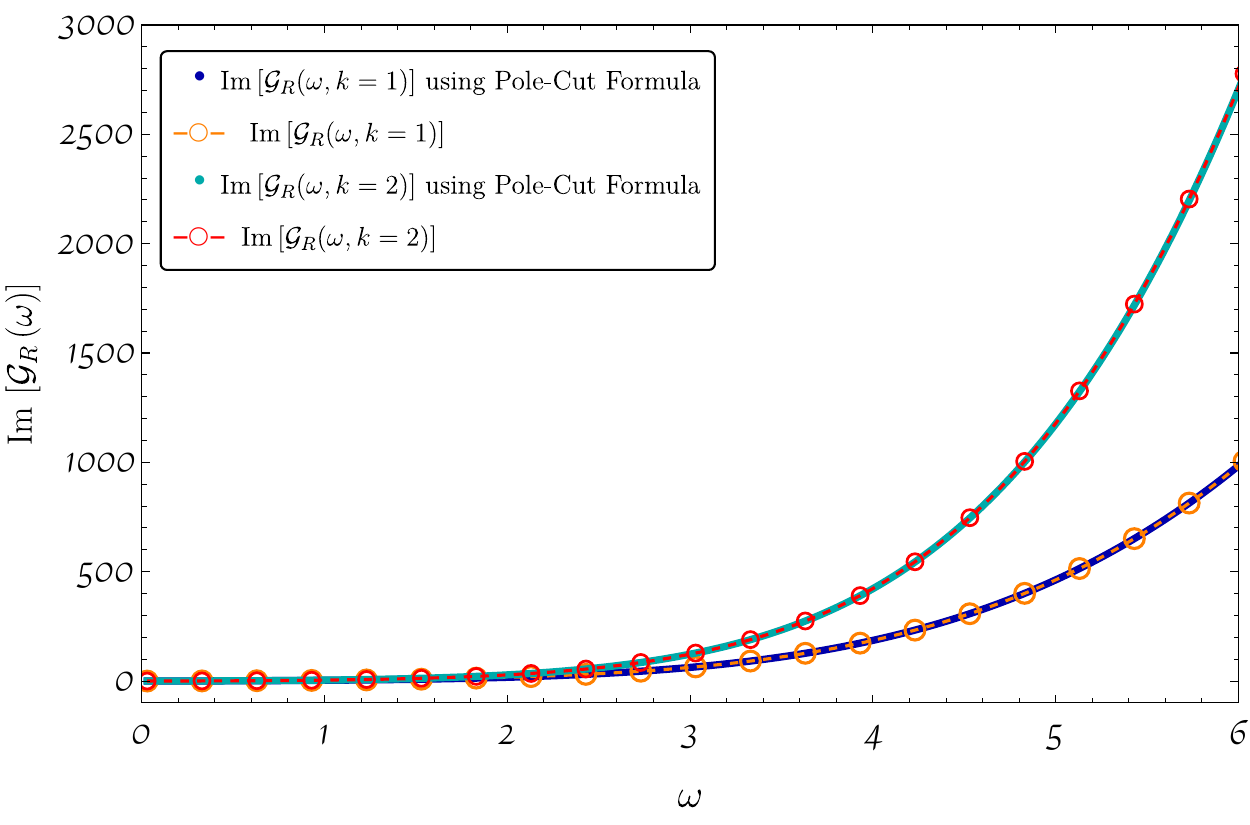}
  \caption{Plot of Im $\mathcal G_R$ and its reconstruction using the pole-cut formula \eqref{pole-cut} for $k=1,2$. The dashed lines interspaced with circles denote the plot of Im $\mathcal G_R$ using the standard holographic dictionary \eqref{directGr}. While the solid lines denote the plot of Im $\mathcal G_R$ using \eqref{pole-cut}, after putting together $\mathcal F_{\text{poles}}$ and $\exp{\left( H_{\text{cut}} \right)}$ computed separately. We have normalized both the plots at reference frequency $\mathfrak{w} = 0.5$ so that $\text{Im} \, \mathcal G_R(\omega = 0.5) = 1$.}
  \label{fig:numpolecut}
\end{figure}
%%%%%%%%%%%%%%%%%%%%%%%%%%%%%%%%%%%%%%

%********************************************************************
\section{OPE limit and Extremal sum-rules}\label{sec:OPElimit}
%********************************************************************

%-----------------------------------------------------------
\subsection{Expectations from the OPE on the CFT side}
%-----------------------------------------------------------
%
The extremal black brane corresponds to a finite density zero temperature state for the boundary CFT. Let us denote this state with $\vert \mu \rangle$. Using the
OPE, we can formally expand any two point function in the following way
\begin{equation}
 \langle \mu \rvert \, \Phi_\Delta  \left( x \right) \Phi_\Delta \left( 0 \right) \, \lvert \mu \rangle
 = \sum_{\mathcal O} \mathcal C_{\Phi \Phi}^{\mathcal O} ~ \lvert x \rvert^{\Delta_{\mathcal O} - 2 \Delta}   ~ \langle \mu \rvert \, \mathcal O \, \lvert \mu \rangle
\end{equation}
where $C_{\Phi \Phi}^{\mathcal O}$ are the relevant OPE coefficients. The leading term in the OPE expansion corresponds to the Identity channel ($\mathcal O  = \mathbf 1 $) which dominates
the UV behaviour of this correlation function. Isolating the Identity channel, we can write
 \begin{equation}\label{OPEfull}
 \langle \mu \rvert \, \Phi_\Delta  \left( x \right) \Phi_\Delta \left( 0 \right) \, \lvert \mu \rangle
 = \underbrace{ \frac{C}{\lvert x \rvert^{ 2 \Delta}} }_{\langle \Phi_\Delta   \Phi_\Delta  \rangle_0 } ~+~
 \sum_{\mathcal O \neq \mathbf 1} \mathcal C_{\Phi \Phi}^{\mathcal O} ~ \lvert x \rvert^{\Delta_{\mathcal O} - 2 \Delta}   ~ \langle \mu \rvert \, \mathcal O \, \lvert \mu \rangle \, ,
\end{equation}
where the term corresponding to the Identity channel is the vacuum two point function of the same operator. Here $C$ is a constant which is usually fixed by the normalization of the
operators. We are interested in the Fourier transform of \eqref{OPEfull}. Let us focus on first term which captures the leading order UV behaviour.
In general, the Fourier transform of the first term in \eqref{OPEfull} in any dimension $d$ is
\begin{equation} \label{2ptFuncFourier}
\mathcal I_{\Delta}(P) = \int_{-\infty}^{\infty} d^{d} x \, \frac{e^{i P . x}}{|x|^{2\Delta}} = \pi^{d/2} 2^{d - 2 \Delta} \frac{\Gamma(\frac{d}{2} - \Delta)}{\Gamma(\Delta)}  (P^2)^{\Delta - d/2} \, .
\end{equation}
Here, the integral is performed with Euclidean signature, using the standard Schwinger parameterization technique.
Now, in this paper, we have focused on the massless probe scalar fluctuations in the bulk 4+1 dimensions. The holographic dual of this scalar field is a $d=4$ scalar operator with dimension $\Delta = 4$. For this value of $\Delta$ and $d$, the numerator of \eqref{2ptFuncFourier} is naively proportional to $\Gamma(-2)$ which is divergent, and must be regulated appropriately.

Let us regulate the integral in \eqref{2ptFuncFourier} by infinitesimally shifting the conformal dimension to $\Delta = 4 - \epsilon$. Subsequently, a small $\epsilon$ expansion yields
\begin{equation} \label{2ptEpsilonExp}
\mathcal I_{4 - \epsilon}(P) = \frac{\pi ^2 P^4}{192 \epsilon }-\frac{1}{576} \pi ^2 P^4 \left(3 \log P^2 +6 \gamma -10-3 \log 4 \right)+O\left(\epsilon ^1\right) .
\end{equation}
where $\gamma$ is the Euler-Mascheroni constant. We can now renormalize this expression by adding counterterms which has support only at $|x| = 0$, i.e. appropriate derivatives on delta function.
In this way, the divergent term can be removed. Note that such a subtraction can also affect terms proportional to $P^4$ in the finite left over piece, thus rendering them scheme dependent.
Thus, after appropriate renormalization, the scheme independent part of the integral is given by
\begin{equation} \label{2ptEpsilonExp}
\mathcal I_{4}^{(\text{ren})}(P) = - \frac{\pi ^2}{192}  (P^2)^2  \log P^2 \, .
\end{equation}
Hence, momentum-space expression for the Euclidean vacuum two point function of the operators is given by
\begin{equation} \label{2ptmomspace}
G^{(0)}_{E}(P) = - C \, \frac{\pi ^2}{192} \, (P^2)^2  \, \log P^2.
\end{equation}
where $C$ is the constant appearing in the first term of \eqref{OPEfull} and the superscript in $G^{(0)}_{E}$ denotes that it is the vacuum correlator.
Note that, since this is the Euclidean correlator, we have $P^2 = \omega_E^2 + \mathbf k^{\, 2}$.
We can obtain the Lorentzian retarded two-point function starting from its Euclidean counterpart \eqref{2ptmomspace}. For this we have to wick rotate $\omega_E$ along with a suitable $i \epsilon$ shift which ensures that we have the right retarded correlator pole prescription. This is achieved by the transformation $\omega_E = - i (\omega + i \epsilon)$. Substituting this in $P^2$ we have
\begin{equation}\label{Plor}
P^2 = \mathbf k^2 - \omega^2 -2 i  \omega \epsilon + O(\epsilon^2) = - s - i \epsilon \, \text{sgn($\omega$)} + O(\epsilon^2) \, ,
\end{equation}
where we defined $s =  \omega^2 - \mathbf k^2$ and absorbed $2 |\omega|$ inside $\epsilon$. Using \eqref{Plor} back in \eqref{2ptmomspace}, we can convert the Euclidean correlator into the following retarded Lorentzian correlator
\begin{equation} \label{GRleadingEpsilon}
G^{(0)}_R(P) =  - \frac{\pi ^2 \, C}{192} ~ s^2 \ln \left( -s - i \epsilon \, \text{sgn}\left( \omega \right) \right) =  - \frac{\pi ^2 \, C}{192} \left( s^2 \, \ln \lvert s \rvert  - i \, \Theta(s) s^2 \pi \, \text{sgn($\omega$)} \right).
\end{equation}
Here, we have retained the terms in leading order in $\epsilon$, and $\Theta(s)$ is the Heaviside step function.
Note that, following standard conventions, we have considered the branch cut of the logarithm along the negative real axis and the imaginary part arises only when $s >0$, which is ensured by
$\Theta(s)$. Redefining the constant $C_0 = - \tfrac{\pi^3 C}{192}$, the imaginary part of the retarded correlator \eqref{GRleadingEpsilon}  is given by
\begin{equation} \label{specDensEqn}
 \text{Im} \left[ G^{(0)}_R(\omega, \mathbf k) \right] =  C_0 \, \text{sgn($\omega$)} (\omega^2 - \mathbf k^2)^2 \, \Theta(\omega^2 - \mathbf k^2),
\end{equation}
Now, let us go back to \eqref{OPEfull}, recalling that the result \eqref{specDensEqn} corresponds only to the first term in \eqref{OPEfull}.
We are interested to obtain the retarded two-point function for our finite density non-thermal state in momentum space $\mathcal G_R \left( \omega, \mathbf k \right)$, which can be compared with
\eqref{pole-cut}. For this, in principle, we can repeat the above steps for all the terms in the expansion \eqref{OPEfull}. However, we are only interested in the large $\omega$ behaviour,
and the leading two terms in \eqref{specDensEqn}, arising from the first term, do not receive corrections from the rest of the terms in \eqref{OPEfull}.
Here is a brief justification for this assertion.

Observe that the first correction to the leading term in \eqref{OPEfull} is proportional to
$\lvert x \rvert^{- 2 \Delta + \Delta_{\mathcal O}}$
corresponding to the operator $\mathcal O$ which has the smallest non-zero value of scaling dimension $\Delta_{\mathcal O}$,
and for which $\langle \mu \rvert \, \mathcal O \, \lvert \mu \rangle \neq 0$.
After performing the Fourier transform of $\lvert x \rvert^{- 2 \Delta + \Delta_{\mathcal O}}$ we see that for such a term to contribute at
$\omega^2$ order, to the imaginary part of the retarded correlator, we must have $\Delta_{\mathcal O} =2$. But in our holographic model with a massless
bulk scalar field coupled to gravity, there are no operators with scaling dimension $\Delta_\mathcal O = 2$ which has a non-zero expectation value in the state $\lvert \mu \rangle$.
In this system, we definitely have the energy momentum tensor which has scaling dimension $4$
{\footnote{As explained in \ref{sec:bkgfluc}, the scalar field with scaling dimension $4$ can also be understood as
the tensor fluctuation of the graviton. This implies the dual boundary operator $\Phi_{\Delta=4}$ can be viewed as a component of the CFT energy momentum tensor.}},
whose contribution to \eqref{OPEfull} can only affect terms of $\mathcal O (\omega^0)$ or higher, in the large $\omega$ expansion. Hence,
 the first two terms proportional
to $\omega^4$ and $\omega^2$ respectively, in the imaginary part of the full retarded correlator, can only arise from the vacuum
two point function as computed in \eqref{specDensEqn}. Hence, we can conclude that
the retarded correlator corresponding to \eqref{OPEfull}, has the following large $\omega$ expansion for $\omega > 0$ and $\omega^2 > \mathbf k^2$
\begin{equation} \label{GRopelim}
 \text{Im} \left[ \mathcal G_R(\omega, \mathbf k) \right] = C_0 \, \text{sgn($\omega$)} \, \omega^4 \, \left(1- 2 \, \frac{\mathbf k^2}{\omega^2} + \dots \right) \, .
\end{equation}
Here, the ratio of the coefficients of the subleading term to that of the leading term is unambiguous and is given by $-2 \mathbf k^2$. In the OPE limit, the structure of
\eqref{GRopelim} can thus be summarized as follows
\begin{equation}\label{schGR}
 \text{Im} \left[ \mathcal G_R(\omega, \mathbf k) \right] = c_1 \, \omega^4 \left( 1 + \frac{c_2}{\omega^2}  + \dots \right) \, , ~~\text{where}~~ c_2  = - 2 \mathbf k^2 \,.
\end{equation}
The parameter $c_1$ is fixed by our normalization convention, in particular, by the choice of our reference frequency $\mathfrak w$.
On the other hand, the parameter $c_2$ is a robust prediction which must be satisfied by our pole-cut formula \eqref{pole-cut}
{\footnote{Unlike the thermal sum-rules of \cite{Dodelson:2023vrw}, here the value of $c_2$ is determined by the dimension of the dual operator rather than
being an OPE coefficient.}. We will now verify this prediction numerically.

%-----------------------------------------------------------
\subsection{Numerical verification using pole-cut formula}
%-----------------------------------------------------------

The first numerical check is somewhat straightforward. We have already performed a numerical reconstruction of $\text{Im} \left[ \mathcal G_R(\omega, \mathbf k) \right]$,
using the pole-cut formula \eqref{pole-cut} as shown in Fig.\ref{fig:numpolecut} for $k=1,2$. For numerical convenience, we consider the combination
\begin{equation}
 \mathbb G \left( \omega \right) = \ln \left( \frac{1}{2 \omega}  \text{Im} \left[ \mathcal G_R(\omega)  \right] \right) \, ,
\end{equation}
so that the CFT prediction \eqref{schGR} implies the following large $\omega$ expansion
\begin{equation}\label{Gfit}
 \mathbb G \left( \omega \right) = \ln \frac{c_1}{2} + 3 \, \ln \omega + \frac{c_2}{\omega^2} + \dots \, .
\end{equation}
We now perform a numerical fit of $\mathbb G \left( \omega \right)$  over the range $\omega \in \left[ 10, 20 \right]$, using the formula
\eqref{Gfit}. The best fit values of these parameters are obtained to be
\begin{equation}
\begin{split}
& k = 1 : c_1 \approx 0.10 \, , ~c_2 \approx - 2.01 \, ,\\
& k = 2 : c_1 \approx 0.10 \, , ~c_2 \approx - 8.01 \, .
\end{split}
\end{equation}
Thus we see that the values of $c_2$ obtained in both the cases are consistent with our prediction $c_2 = - 2 k^2$.

%---------------------------------------------------------------------------------
\subsection{Extremal sum-rule and its numerical verification}\label{ssec:sumrule}
%---------------------------------------------------------------------------------

Just like thermal sum-rules discussed in \cite{Dodelson:2023vrw}, the CFT prediction of the coefficient $c_2$ imposes a nontrivial constraint on the extremal QNM spectrum of the bulk together with the branch cut contribution in \eqref{pole-cut}. This constraint can be expressed as an extremal sum-rule which we now derive and verify numerically.

For this discussion it would be convenient to remove the factor of $\omega$ from \eqref{pole-cut}. So we define the quantity
\begin{equation}\label{Gbyw}
 \mathfrak G (\omega) = \frac{1}{\omega} \, \rho (\omega, k) =
 \tilde {\mathcal G}_0 ~  \exp \left( {H_{ \text {cut} }} (\omega) \right)
\prod_{n} \left( 1 - \frac{\omega^2}{\omega_n^2} \right)^{-1} \left( 1 - \frac{\omega^2}{({{\omega_n}^\star})^2} \right)^{-1}
\end{equation}
In the OPE limit, the expected behaviour of $\mathfrak G$ which follows from \eqref{GRopelim} is given by
\begin{equation}\label{Glargesw}
 \mathfrak G (\omega) = c_1 \, \omega^3 \left( 1 + \frac{c_2}{\omega^2}  + \dots \right) \, , ~~\text{where}~~ c_2  = - 2 \mathbf k^2
\end{equation}
Redefining $\mathfrak G$ in terms of the variable $\mathfrak s = \omega^2$ and subsequently using \eqref{Gbyw} we obtain
\begin{equation}\label{dellnG}
 \partial_{\mathfrak s}  \ln \left( \mathfrak G \left( \mathfrak s \right) \right) = - \sum_n \left( \frac{1}{\mathfrak s - \omega_n^2}
 + \frac{1}{\mathfrak s - \left(\omega_n^\star\right)^2} \right)  + \frac{2}{\pi} \int_0^\infty \, \sigma \, d\sigma \, \frac{\theta(\sigma)}{\left( \mathfrak s + \sigma^2 \right)^2}
\end{equation}
Here, we have used the expression for $H_{\text{cut}}$ in \eqref{Hcut} to obtain the second term.
Now, in the large $\mathfrak s$ limit, we can use \eqref{Glargesw} to evaluate the same quantity as follows
\begin{equation}\label{dellnGLs}
 \partial_{\mathfrak s}  \ln \left( \mathfrak G \left( \mathfrak s \right) \right) =  \frac{3}{2} \, \frac{1}{\mathfrak{s}} - c_2 \, \frac{1}{\mathfrak{s}^2}\ \, + \,  \dots
\end{equation}
Expanding \eqref{dellnG} and comparing the coefficient of $1/\mathfrak s$ and $1/\mathfrak s^2$, we naively get
\begin{subequations}\label{divsum}
\begin{align}
  \frac{3}{2} &\overset{?}{=} - 2 \sum_n \, 1 \, \label{divsumA}\\
  c_2  &\overset{?}{=} 2 \text{Re} \left[ \sum_n \omega_n^2  \right] - \frac{2}{\pi} \int_0^\infty \, \sigma \, d\sigma \theta \left( \sigma \right) \, . \label{divsumB}
\end{align}
\end{subequations}
Clearly, the sums and the integral appearing in these constraints are all divergent. For an accurate comparison with the predictions from the OPE-limit, we must regulate them  appropriately. For the naively divergent sum in \eqref{divsumA} a zeta-function regularization does not give the expected answer. Also, the divergences in the sum and integral
in \eqref{divsumB} should be coordinated appropriately so that we can recover the expected answer. We will now explain the regularization procedure which will allow us to verify
these expectations numerically{\footnote{This regulation scheme was suggested by OpenAI model ChatGPT Astra 6.0.}}.

Let us consider the Laplace transform
\begin{equation}
 \partial_{\mathfrak s}  \ln \left( \mathfrak G \left( \mathfrak s \right) \right) \equiv \int_0^\infty \, d\mathfrak t ~\mathcal K (\mathfrak t) \, e^{- \mathfrak {s t}} \, .
\end{equation}
Then, taking the inverse Laplace transform of \eqref{dellnG} we obtain
\begin{equation}\label{invLKdef}
 \mathcal K \left( \mathfrak{t}\right) = - 2 ~\text{Re} \left[ \sum_n \exp \left( \mathfrak{t} \, \omega_n^2 \right) \right]
 + \frac{2}{\pi} \, \mathfrak{t} \, \int_0^\infty \, \sigma \, d \sigma  \, \theta \left( \sigma \right) \,\exp \left( - \mathfrak{t} \, \sigma^2 \right) \, .
\end{equation}
Note that the large $\mathfrak{s}$ expansion \eqref{dellnGLs} must correspond to the small $\mathfrak{t}$ expansion of \eqref{invLKdef}. Hence, if we take the
inverse Laplace transform of OPE-limit expansion in \eqref{dellnGLs}, we obtain the following prediction for the small  $\mathfrak{t}$ expansion of $ \mathcal K \left( \mathfrak{t}\right)$
\begin{equation}\label{predKt}
 \mathcal K \left( \mathfrak{t}\right) = \frac{3}{2} -  c_2 \, \mathfrak{t} + \mathcal O \left( \mathfrak{t}^2 \right)
\end{equation}

Let us examine the cut-integral and sum over the off-axis QNMs in \eqref{invLKdef} separately. For the cut-integral, a rescaling of the integration variable gives
\begin{equation} \label{Kc}
 \mathcal K_c \left( \mathfrak{t}\right) \equiv   \frac{2}{\pi} \, \mathfrak{t} \, \int_0^\infty \, \sigma \, d \sigma  \, \theta \left( \sigma \right) \,\exp \left( - \mathfrak{t} \, \sigma^2 \right)
 =  \frac{2}{\pi} \, \int_0^\infty \, \sigma \, d \sigma  \, \theta \left( \tfrac{\sigma}{\sqrt{\mathfrak{t}}} \right) \,\exp \left( - \sigma^2 \right)
\end{equation}
This implies that in order to extract the small-$\mathfrak{t}$ behaviour, it would suffice to consider the large $\sigma$ behavior of the cut-discontinuity $\theta (\sigma)$ which
was discussed in \eqref{thetapara}. We have
\begin{equation}\label{thetaexppara}
 \theta \left( \sigma \right) =  \alpha \, \sigma + \beta + \frac{\gamma}{\sigma} + \dots \, ,
\end{equation}
where our numerical analysis suggests that $\alpha$ and $\beta$ are independent of $k$.
Using this form of $\theta (\sigma)$ back in \eqref{Kc}, we obtain the following small $\mathfrak{t}$ expansion
\begin{equation}\label{Kcapprox}
 \mathcal K_c \left( \mathfrak{t}\right) =  \frac{\alpha }{2 \sqrt{\pi}} \, \frac{1}{\sqrt{\mathfrak{t}}} + \frac{\beta}{\pi}  \, + \,  \mathcal O \left( \sqrt{\mathfrak{t}} \right)
\end{equation}
Let us now move our focus to the term with the sum over QNMs in \eqref{invLKdef}
\begin{equation}\label{Kp}
 \mathcal K_p \left( \mathfrak{t}\right) \equiv - 2 ~\text{Re} \left[ \sum_n \exp \left( \mathfrak{t} \, \omega_n^2 \right) \right] \, .
\end{equation}
We notice that if the QNMs have the asymptotic form \eqref{qnmasym}, i.e.
\begin{equation}\label{omexpn}
 \omega_n = a \left( n + q + \frac{p}{n} + \mathcal O(1/n^2) \right),
\end{equation}
then the small $\mathfrak{t}$ behavior of  $\mathfrak{t} \, \omega_n^2$,  and hence that of $\mathcal K_p$ in \eqref{Kp}, is dominated by the coefficients $a$ and $q$. This is because, for
sufficiently small $\mathfrak{t}$,  we have $\mathfrak{t} / n \ll 1$,  even for the lower harmonics of the QNM spectrum. See appendix \ref{app:asymexact}, where we have determined the parameters $a$ and $q$
exactly, which were also verified by our numerical data in \eqref{qnmparavals}. Interestingly, in Appendix\ref{app:asymexact} we have demonstrated that
$a$ and $q$ are independent of $k$.

Using the first two leading terms from \eqref{omexpn}, we can approximate the sum \eqref{Kp} in the following way, ensuring that it reliably captures the $1/\sqrt{\mathfrak{t}}$ and constant terms in the small $\mathfrak{t}$ expansion\footnote{Here we use the Euler-Maclaurin formula
\begin{equation}
\sum_{n=1}^\infty g(n) = \int_1^\infty g(n) \, dn + \frac{1}{2} \, g(1) - \sum_{k \geq 1} \frac{\text{B}_{2k}}{(2 k)!} g^{(2 k -1)}(1) .
\end{equation}
In our case, $g(n) = \exp\left( \mathfrak{t} \, \omega_n^2 \right)$ and the terms retained in \eqref{Kpaprox} capture the first two terms in the small $\mathfrak{t}$ expansion reliably. With this
point of view,
we have also used $g(1) \approx 1 + \mathcal O \left( \mathfrak{t}\right)$.}
\begin{equation}\label{Kpaprox}
\begin{split}
 \mathcal K_p \left( \mathfrak{t}\right)
 &\approx - 2 \, \text {Re} \left( ~\int_1^\infty dn ~\exp \left( \mathfrak{t} \, a^2 \left( n + q \right)^2 \right) + \frac{1}{2} + \,  \mathcal O \left( \sqrt{\mathfrak{t}} \right) \right) \\
& =   \text{Re} \left[ \frac{i  \sqrt{\pi}}{a} \right] \frac{1}{\sqrt{\mathfrak{t}}} + \left( 1 + 2 \, \text{Re} \left[ \, q \, \right] \right) + \,  \mathcal O \left( \sqrt{\mathfrak{t}} \right)  \, .  \\
 \end{split}
\end{equation}
Now, we must have $\mathcal K = \mathcal K_c + \mathcal K_p$, whose small $\mathfrak{t}$ expansion is expected to match \eqref{predKt}. This implies that the
coefficient of $1/\sqrt{\mathfrak{t}}$ must cancel between \eqref{Kpaprox} and \eqref{Kcapprox}, while the constant terms must add to give $3/2${\footnote{Note
that the term proportional to $\sqrt{\mathfrak{t}}$ in $\mathcal K_c$ and $\mathcal K_p$ must also cancel with each other when they are put together in $\mathcal K$ since such a term is not present
in the expected answer \eqref{predKt} from the CFT side. This cancellation relates the parameters $p$ and $\gamma$ in \eqref{omexpn} and \eqref{thetaexppara} respectively. Similarly there
are also terms proportional to $\mathfrak{t} \,\left( \ln \mathfrak{t} \right) $ present in both $\mathcal K_c$ and $\mathcal K_p$, which also must cancel.}}.
In this way, we get the following two constraints
\begin{equation}\label{paramatch}
 \alpha =  - \, \pi \, \text{Re} \left[ \frac{2i}{a} \right], ~~ \beta = - \, \pi \, \left( 2 \, \text{Re} \left[ \, q \, \right] - \frac{1}{2} \right) \, .
\end{equation}
Using the exact expressions of $a$ and $q$ reported in \eqref{paravals}, we find that $\alpha$, $\beta$ evaluates to
\begin{equation}\label{albetexact}
 \alpha = \frac{7 \pi}{18}, ~~\beta = - \frac{\pi}{2} \, .
\end{equation}
These values match remarkably well with our numerical estimates in \eqref{albetvals}.
Similarly, once we proceed to higher orders in the small $\mathfrak{t}$ expansion, the coefficients of the asymptotic expansion of
the QNMs becomes related to those in the asymptotic expansion of the branch-cut phase-jump $\theta(\sigma)$. The coefficient of the
linear term is particularly interesting, since in order to match it with the CFT expectation \eqref{predKt}, we must have the knowledge
of all the QNMs and $\theta(\sigma)$ beyond their asymptotic expansions. Let us focus on this linear term now.

From \eqref{predKt}, it follows that $ \mathcal K '\left( \mathfrak{t}\right) = - c_2 + \mathcal{O} \left( \mathfrak{t} \right)$. However, for numerical convenience
we consider the quantity $\mathcal C \left( \mathfrak{t}\right) \equiv 3/\mathfrak{t} - 2 \mathcal K \left( \mathfrak{t}\right) / \mathfrak{t} + \mathcal K '\left( \mathfrak{t}\right) = c_2 + \mathcal{O} \left( \mathfrak{t}^2 \right)$. Hence
\begin{equation} \label{c2pred}
 c_2 = \lim_{\mathfrak{t} \rightarrow 0} \mathcal C \left( \mathfrak{t}\right) =  \lim_{\mathfrak{t} \rightarrow 0} ~ \left( \frac{3}{\mathfrak{t}} - \frac{2}{\mathfrak{t}} \, \mathcal K \left( \mathfrak{t}\right)  + \mathcal K '\left( \mathfrak{t}\right) \right) \, .
\end{equation}
Note that $\mathcal C \left( \mathfrak{t}\right)$ is better-conditioned for numerical verification compared to $\mathcal K '\left( \mathfrak{t}\right)$,
since for $\mathcal C \left( \mathfrak{t}\right)$ the corrections to
$c_2$ are quadratic in $\mathfrak{t}$ compared to the linear order corrections to the later. Evaluating the RHS of \eqref{c2pred} using \eqref{dellnG}, we obtain the following
constraint for the QNMs and branch cut discontinuity
\begin{equation}\label{c2lim}
 c_2 = \lim_{\mathfrak{t} \rightarrow 0} ~ \left( \frac{3}{\mathfrak{t}} -
  2 ~\text{Re} \left[ \sum_n \left( \omega_n^2 - \frac{2}{\mathfrak{t}}   \right) e^{\mathfrak{t} \, \omega_n^2 }\right]
  - \frac{2}{\pi}  \int_0^\infty \, \sigma \, d \sigma  \, \theta \left( \sigma \right)  \, \left( 1 + \mathfrak{t} \, \sigma^2 \right) e^{- \mathfrak{t} \, \sigma^2 } \right) \,
\end{equation}
For $k=1$, from \eqref{schGR} we expect $c_2 = -2$. Using the numerical data for QNMs and $\theta(\sigma)$ which was used to verify the pole-cut formula in Fig.\ref{fig:numpolecut}, we are able to verify \eqref{c2lim}. In Fig. \ref{fig:c2conv} we have plotted $\mathcal C \left( \mathfrak{t}\right)$, which clearly converges to $-2$ as $\mathfrak{t} \rightarrow 0$. Note that,
in this convergence plot, we were not able to find any straightforward way to incorporate the analogue of the tail-correction which played a crucial role in verifying the pole-cut formula.
Here, we have used only $40$ off-axis QNMs which introduce truncation errors. Also robust numerical data for $\theta(\sigma)$ was only obtained up to $\sigma \approx 50$ which introduced
further errors in numerical evaluation of $\mathcal C \left( \mathfrak{t}\right)$. These limitations put a restriction on the lowest value of $\mathfrak{t}$ up to which we can reliably verify the convergence. In Fig. \ref{fig:c2conv} we have verified the convergence up to $\mathfrak{t} = 0.005$.

\begin{figure}[h!tbp]
  \centering
  \includegraphics[width=0.7\textwidth]{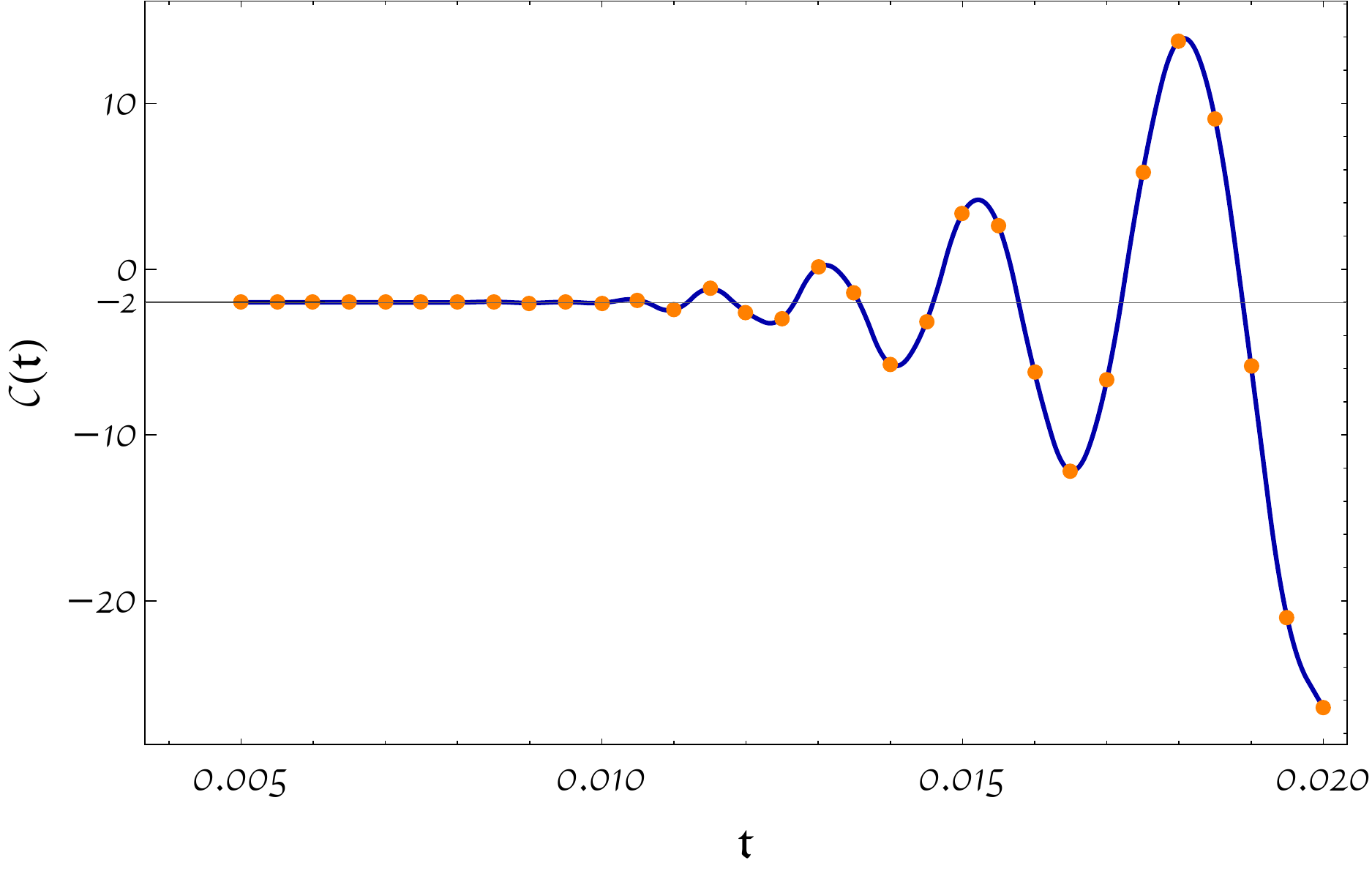}
  \caption{Convergence plot for $\mathcal C \left( \mathfrak{t}\right)$ demonstrating the extremal sum-rule \eqref{c2lim}. Here we have used 40 off-axis QNMs, while
  the numerical extrapolation used for $\theta (\sigma)$ is same as that used in Fig.\ref{fig:numpolecut}. Both QNMs and $\theta(\sigma)$ have been obtained for $k=1$. }
  \label{fig:c2conv}
\end{figure}

Before concluding this section, we would like to re-emphasize one of the most remarkable results in this discussion.
The extremal off-axis QNMs and the branch-cut discontinuity function $\theta(\sigma)$ are a priori distinct spectral data. In other words, there is no straightforward
reason to suspect any relationship between the zeros of the Jost-function $A(\omega)$ and its discontinuity across the branch cut. However,
the fact that there is a dual CFT forces a non-trivial relationship between these two classes of spectral data.
In fact, as we have demonstrated above, the asymptotic data for $\theta(\sigma)$ is determined order by order in terms of coefficients of the large overtone expansion of the QNMs due to
the OPE  expansion of the CFT correlator. Thus the existence of the dual CFT stitches together the QNM and branch-cut spectral data in a very non-trivial fashion.

%***************************************************************************************************
\section{Complex-time singularities of the extremal correlator}\label{sec:singularbounce}
%***************************************************************************************************

Recently, there has been enormous interest on complex time singularities of thermal correlators \cite{Fidkowski:2003nf,Festuccia:2005pi,Parisini:2023nbd,Ceplak:2024bja,Buric:2025anb,Buric:2025fye,Barrat:2025twb,Afkhami-Jeddi:2025wra,Dodelson:2025jff,Ceplak:2025dds,Jia:2025jbi,AliAhmad:2026wem,Giombi:2026kdz,Arnaudo:2026der,Grozdanov:2026cut,Jia:2026ryl,Arnaudo:2026tcy,Grozdanov:2026ktq,Buric:2026qsp}.
The location of such singularities have been shown to be related to the optical distance between the boundary and singularity of AdS black holes.
This singularity has been interpreted as a signature of bulk curvature singularity on a boundary observable \cite{Afkhami-Jeddi:2025wra}.
As anticipated, quantum corrections in the bulk lifts the curvature singularity which is also reflected on the boundary correlators \cite{Komargodski:2026oob,Buric:2026sit}.
Also recently, in \cite{Hartnoll:2026vhu}, it has been shown that more data from the near singularity geometry, i.e. the Kasner coefficients are also encoded
within the holographic thermal correlator. The information of the singularity is essentially present in the high frequency behaviour
of the correlator - the OPE limit. Now, since the thermal correlators can be expressed as an infinite product of QNMs,
the location of the complex time singularities are also related to the asymptotic form of the QNMs. This connection
has been explained thoroughly in the recent work \cite{Arnaudo:2026axe}.

In this section, we follow the lead of \cite{Arnaudo:2026axe} and repeat the analysis for our extremal case. We find that complex time singularities of the extremal correlators
arise separately from both the off-axis pole contribution, as well as the branch cut contribution in our pole-cut formula \eqref{pole-cut}.
The singularity from the pole contribution arises from the asymptotic behaviour of the off-axis poles, very similar to the thermal case.
The novelty of the extremal case is that the branch cut contribution also leads to singular behaviour in complex time. In this case, the location of the
singularity is controlled by the asymptotic behaviour of the phase discontinuity  of the Jost function, i.e. $\theta(\sigma)$ defined in \eqref{thetadef}.
As we have seen in Section \ref{ssec:sumrule} and appendix \ref{app:asymexact}, the asymptotic behaviors of the off-axis QNMs is related to that of $\theta$, both being controlled by the optical distance between the boundary and singularity \eqref{odval}. We will now explain this in more detail.

The retarded correlator in the time domain is given in terms of the spectral function $\rho = 2 \, \text{Im} \, \mathcal G_R$ ($\omega \in \mathbb R$) as follows
\begin{equation}\label{grt}
G_R(t,\mathbf k) = i \,\Theta(t) \int_{-\infty}^{\infty} \frac{d\omega}{\pi}\, e^{-i\omega t}\, \text{Im} \, \mathcal G_R(\omega,\mathbf k) \, .
\end{equation}
Now $\text{Im} \, \mathcal G_R$ is given by our pole-cut formula \eqref{pole-cut}. In order to perform the $\omega$ integral
we must close the contour below real axis as shown in Fig.\ref{fig:freqcont}. This is because $t>0$ because of the $\Theta(t)$ function present in the retarded correlator. As a result, we pick up the contribution from the
poles and the branch cut, all of which lies on the lower half plane of the complex $\omega$-plane. For compactness of notation,
we will keep the dependence on $\mathbf k$ implicit here. Our discussion here applies to arbitrary values of $\mathbf k$.

\begin{figure}[h!tbp]
  \centering
  \includegraphics[width=0.7\textwidth]{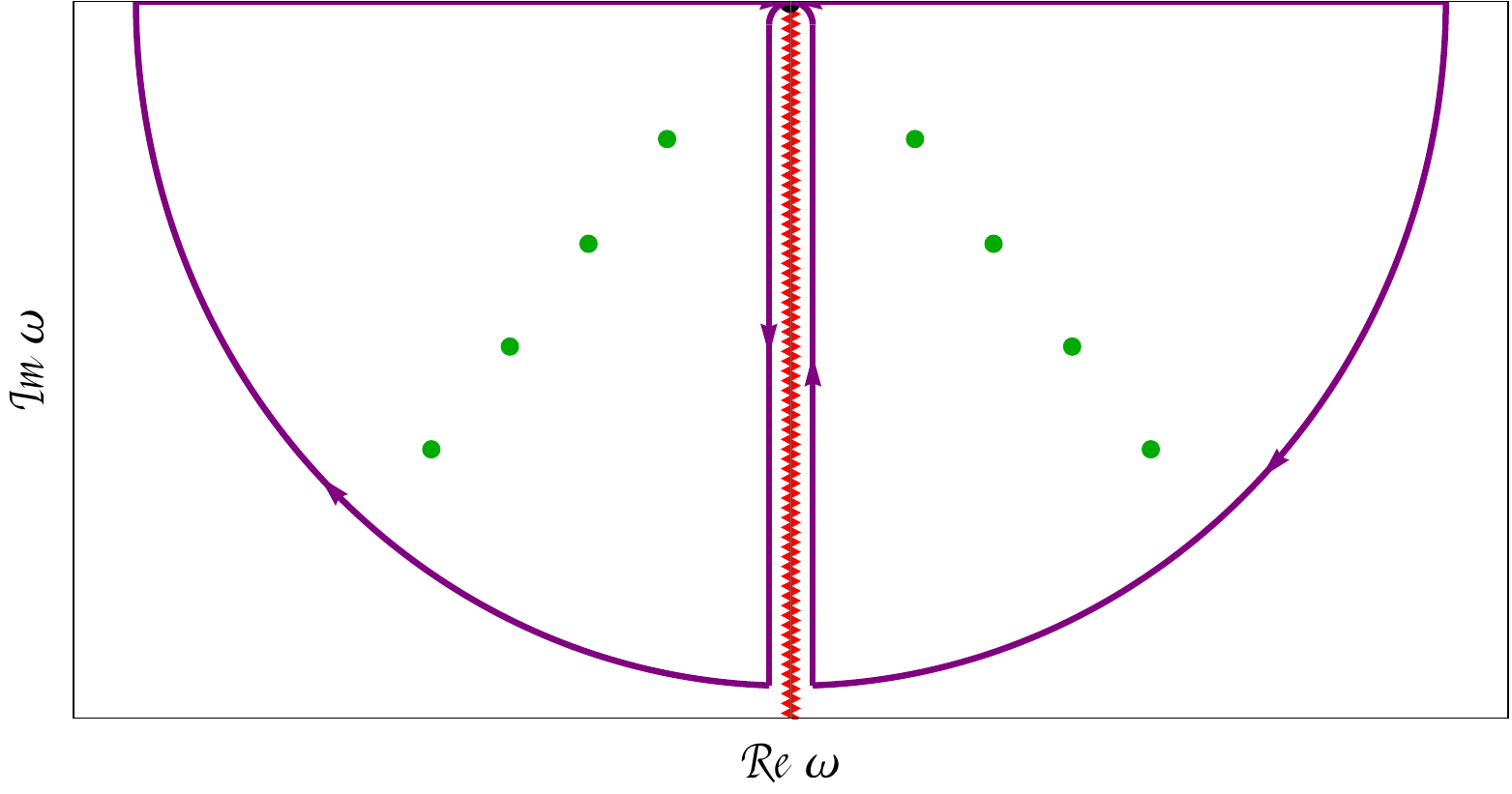}
  \caption{Schematic representation of the contour on the complex $\omega$-plane which is used to evaluate the integral to obtain $G_R(t,\mathbf k)$.
  The green dots denote the location of the off-axis QNMs, while the red jagged line denotes the branch cut on the imaginary axis.}
  \label{fig:freqcont}
\end{figure}

After evaluating the integral \eqref{grt}, the
result for $G_R(t )$ is expressed as a sum of the residue at the poles together with the cut-contribution as follows
\begin{equation}
 G_R(t ) = G^{(p)}_R(t ) + G^{(c)}_R(t ) \, .
\end{equation}
Here the residue from poles corresponding to the off-axis QNMs are collectively given by
\begin{equation}\label{Gpdef}
 G^{(p)}_R(t ) = \textcolor{black}{- \tilde{\mathcal G}_0} \sum_m \, \text{Re} \left[
\omega_m^2 \,
e^{-i\omega_m t} \exp \left[ H_{\text{cut}} \left( \omega_m \right) \right]
 \prod_{n\neq m} \left(1-\frac{\omega_m^2}{\omega_n^2}\right)^{-1}
\prod_l \left(1-\frac{\omega_m^2}{\left({\omega}_l^{^\star}\right)^2}\right)^{-1}
\right] \, .
\end{equation}
where the sum as well as the products run over the QNMs lying on the right branch of the `Christmas-Tree' in Fig.\ref{fig:QNMEx}.
On the other hand, the cut-contribution is given by
\begin{equation} \label{cutcontribGt}
G^{(c)} (t) =  -\frac{i}{2 \pi} \int_{0}^{\infty} d \sigma \, \sigma e^{-\sigma t} \, \text{Disc} \,  \mathfrak G \left( \omega = - i \sigma \right)  \, .
\end{equation}
where $ \mathfrak G (\omega) = \tfrac{2}{\omega} \text{Im} \, \mathcal G_R \left( \omega, k \right)$ was defined in \eqref{Gbyw}.
Both the terms $G^{(p)}$ as well as $G^{(c)}$ exhibit complex time singularities which we now discuss separately.

\vspace{0.5cm}
\noindent
{\bf{Pole Contribution:}}
Since the structural form of $G^{(p)}$ is similar to the thermal case, the procedure to extract the complex time singularities of this function
can be adopted from a similar discussion using the thermal product formula in \cite{Arnaudo:2026axe}.
These complex time singularities for $G^{(p)}$ are entirely captured by the asymptotic QNMs. Note that \eqref{Gpdef} has the following
structure
\begin{equation}\label{sumGp}
 G^{(p)}_R(t ) = \textcolor{black}{- \tilde{\mathcal G}_0} \sum_m  \text{Re} \, \left[ \mathrm C_m \ e^{- i \omega_m t} \right]
\end{equation}
where
\begin{equation}\label{residuefull}
 \mathrm C_m  =
\omega_m^2 \,
 \exp \left[ H_{\text{cut}} \left( \omega_m \right) \right]
 \prod_{n\neq m} \left(1-\frac{\omega_m^2}{\omega_n^2}\right)^{-1}
\prod_l \left(1-\frac{\omega_m^2}{\left({\omega}_l^{^\star}\right)^2}\right)^{-1}
\end{equation}
For our purpose of locating the complex time singularity, it would suffice to
focus on those terms in the sum which corresponds to the high harmonics,
i.e. for asymptotic QNMs $\omega_n = a \left( n + q \right)$, with large $n$.
Here the coefficients $a$ and $q$ have been computed analytically, see \eqref{paravals}.
Let us first consider $H_{\text{cut}} \left( \omega_m \right)$, which is given by \eqref{Hcut}.
In order to capture the large overtone dependence of $H_{\text{cut}}$ it would be
sufficient to compute the integral in \eqref{Hcut} using the asymptotic form of $\theta(\sigma)$.
For large values of its argument, $\theta (\sigma) \approx \alpha \, \sigma + \beta$,
where $\alpha$ and $\beta$ were determined in \eqref{albetvals}. Using this asymptotic form of $\theta (\sigma)$
in to the integral \eqref{Hcut}, we obtain the following result for the large overtones
\begin{equation} \label{Homegamint}
\begin{split}
H(\omega_m)
&= - \frac{2}{\pi} \int_0^{\infty} d \sigma \, \sigma \, \theta(\sigma) \, \left( \dfrac{1}{\sigma^2 + \omega_m^2} - \dfrac{1}{\sigma^2 + \mathfrak w^2} \right) \\
&= \alpha (\omega_m - \mathfrak w) + \frac{2 \beta}{\pi} \ln \left( \frac{\omega_m}{\mathfrak w} \right) + O \left( \dfrac{1}{\omega_m} \right) \,.
\end{split}
\end{equation}
Thus, the leading contribution to $\exp H(\omega_m)$ at large m is
\begin{equation} \label{LeadContribExpH}
\exp H(\omega_m)  \approx e^{\alpha \left(\omega_m - \mathfrak w\right)} \left( \frac{\omega_m}{\mathfrak w} \right)^{\frac{2 \beta}{\pi}} = \mathfrak{g} \, m^{\frac{2 \beta}{\pi}} \, e^{a \alpha m} \, ,
\end{equation}
where $  \mathfrak{g}  = e^{\alpha \left( a q - \mathfrak{w} \right)} \left( a / \mathfrak{w} \right)^{\frac{2\beta}{\pi}} $ and we have used
the asymptotic form of the QNMs in \eqref{qnmasymform}. Moving to infinite product factors of the residue $C_m$ in \eqref{residuefull}, we can substitute the asymptotic form \eqref{qnmasymform} into it and execute the infinite product. We find that the leading order expression in large $m$ is given by
\begin{equation}\label{prodllarge}
 \omega_m^2 \prod_{n\neq m} \left(1-\frac{\omega_m^2}{\omega_n^2}\right)^{-1}
\prod_l \left(1-\frac{\omega_m^2}{\left({\omega}_l^{^\star}\right)^2}\right)^{-1} \approx m^{p} e^{\eta a m} \, ,
\end{equation}
where
\begin{equation}\label{petavals}
 p = \textcolor{black}{3}+ 4 \, \text{Re}\left[ \,q \, \right] = 5 \,\text{,}~~~\eta = \frac{2 \pi}{\lvert a \rvert^2} \text{Im} \left[ \, a \, \right] = - 7 \pi / 18 = - \alpha
\end{equation}
Here again, we have used the exact values of $a$ and $q$ form \eqref{paravals}. Putting together \eqref{LeadContribExpH} and \eqref{prodllarge}, the residue coefficients $\mathrm C_m$ reduces to
\begin{equation}\label{Cmeval}
 \mathrm C_m \approx m^{p \, + \, \frac{2 \beta}{\pi} } \, e^{a m \left( \alpha + \eta \right)} \, .
\end{equation}

We can now use this result back into the sum \eqref{sumGp}. Note that since $\mathrm C_m$ in \eqref{Cmeval} is only valid for large $m$, and we make an error if we substitute \eqref{Cmeval} in all the terms
in \eqref{sumGp}. However, this error will be a finite sum of exponential functions, the upper limit being determined by the value of $m$ beyond which \eqref{Cmeval} is a good approximation. But a finite sum of exponentials cannot generate any singular behavior at any finite point on the complex $t$ plane. Hence, this error term is of little interest to us
in understanding the complex-time singularities of $G_R(t)$. Thus, ignoring the subleading pieces, \eqref{sumGp} can be simplified as { \footnote{ \textcolor{black}{The terms in \eqref{polylog} capture the leading singular behavior. The less singular pieces contain sub-leading singular and regular terms. This sub-leading error is still singular, although less singular. It arises from the contribution of $C_m \approx m^{p - 1 +2 \frac{\beta}{\pi}} e^{a m (\eta + \alpha)}$, for which we get the following result with $p$ replaced by $p-1$. } } }

\begin{equation}\label{polylog}
\begin{split}
 G^{(p)}_R(t ) &= \sum_{m=0}^\infty  \text{Re} \, \left[ \left( m^{p \, + \, \frac{2 \beta}{\pi} } \, e^{a m \left( \alpha + \eta \right)} \right) \ e^{- i \omega_m t} \right] + \text{less-singular} \\
 &= e^{-i a q t} \, \text{Li}_{-\left(p + \frac{2\beta}{\pi} \right)}(e^{ - i a t})
 + e^{i (a q)^\star t} \, \text{Li}_{-\left(p + \frac{2\beta}{\pi} \right)}(e^{ i a^\star t }) + \text{less-singular} \, ,
\end{split}
\end{equation}
where in the second line we have evaluated the infinite sum in terms of polylog functions, using \eqref{qnmasymform} for $\omega_n$ in the exponential functions. Also, the fact that $\eta = - \alpha$ provide further simplification in the algebra. Note that the order of the Polylog Li evaluates to a negative integer, i.e.
$ - \left( p + \frac{2\beta}{\pi} \right) = - \textcolor{black}{4}$, using \eqref{petavals} and \eqref{albetvals}.
In this result, the polylogarithm Li function develops a singularity when its argument is 1, i.e. when $a \, t = 2 \pi n$ and
$a^\star \, t = 2 \pi n$. For $n=1$, this implies that $G_R(t)$ has singularities at
\begin{equation}
 t = 2 \mathcal X_\pm \, ,
\end{equation}
where we have used \eqref{paravals} to express $a$ in terms of $\mathcal X_\pm$ which is the optical distance from the boundary to the singularity
\begin{equation}
 \mathcal X_\pm = \frac{\pi \sqrt{2}}{9} \pm i \frac{7 \pi}{36} \, ,
\end{equation}
which was evaluated in \eqref{odval} (see appendix \ref{app:asymexact} for further details). This is the idea of
`bouncing singularity' which has been extensively discussed in the recent literature.

\vspace{0.5cm}
\noindent
{\bf{Branch-cut Contribution:}} A similar singular behaviour related to the optical distance is also observed in the branch cut contribution $G^{(c)} (t)$ reported in \eqref{cutcontribGt}, which is a novelty of the extremal case. In order to evaluate the integral \eqref{cutcontribGt} we need to quantify the discontinuity of  $\mathfrak G = \tfrac{2}{\omega} \text{Im} \, \mathcal G_R $ across the branch cut along the negative imaginary axis of the complex $\omega$ plane which we have parameterized by $\sigma$, i.e. $\omega = - i \sigma$ with $\sigma \geq 0$. Going back to \eqref{img}, let us recall that the discontinuity of $\mathfrak G$ is related to that of the Jost function $A(\omega)$ defined as the boundary
source in \eqref{fulbdyasym}. The cut-phase of the Jost function has been defined in \eqref{thetadef} as $\theta(\sigma)$, which provides the discontinuity of the phase of $\mathfrak G$.
Further, the magnitude of $A$ along the branch cut provides the magnitude of $\mathfrak G$ through \eqref{img}. Since we have already obtained the function $A(\omega)$ numerically for $k=1$, on both sides of the branch cut (see appendix \ref{app:discon}), we perform the following numerical fit
\begin{equation}\label{magfitfn}
 \ln \left( \frac{1}{\lvert A(\omega) A(-\omega)\rvert} \right)= - \ln \lvert A(\omega = - i \sigma)\rvert - \ln \lvert A(\omega = i \sigma)\rvert \approx c_0 + c_1 \sigma + c_2 \ln \sigma \, .
\end{equation}
An estimate of the fitting parameters is obtained using the range  $\sigma \in \left[ 40, 120 \right]$. We obtain
%
% Performing the numerical fitting in the range $\sigma \in \left[ 40, 120 \right]$, we obtain the following estimate of the fitting parameters
\begin{equation}\label{magfitpara}
 c_0 \approx - 1.500, ~~ c_1 \approx - 0.987 , ~~c_2 \approx 3.016
\end{equation}
In Fig.\ref{fig:magfit}, we have compared the numerical fitting with the original function. Our result \eqref{magfitfn} is particularly reliable for large $\sigma$. Hence,
we now have excellent control on the large $\sigma$ behaviour of $\mathfrak G (\sigma)$. Collecting the results for the discontinuity phase $\theta(\sigma)$ and the magnitude
of $A$ in \eqref{magfitfn}, we can make the following approximation for the discontinuity of $\mathfrak G (\sigma)$
\begin{equation}
\text{Disc} \,  \mathfrak G  \approx - \, \mathcal C_0 \, \sigma^{c_2} \, e^{c_1 \,\sigma} \left( e^{i \theta(\sigma)} -  e^{-i \theta(\sigma)}\right)
\end{equation}
We use the asymptotic form \eqref{thetapara} for $\theta(\sigma)$ and plug the large $\sigma$ approximation of the discontinuity $(\text{Disc} \,  \mathfrak G)$ back into \eqref{cutcontribGt}.
We obtain
{\footnote{Note that, just like the pole-contribution, the error conceded in using the asymptotic form of $\text{Disc} \,  \mathfrak G(\sigma)$, is a integral with a finite upper limit, however large. Such an integral can be bounded reliably and cannot lead to a singular behaviour of $G^{(c)} (t)$ for any finite complex $t$.}}
\begin{equation}
\begin{split}
 G^{(c)} (t) &\approx
  \frac{i \, \mathcal C_0}{\pi} \int_{0}^{\infty}  d \sigma \, \sigma^{c_2+1} \, e^{-\sigma (t - c_1)} \,  (e^{i (\alpha \sigma + \beta )} - e^{-i (\alpha \sigma + \beta )})
 + \text{regular} \, \\
 & =  \frac{i \, \mathcal C_0}{\pi} \int_{0}^{\infty}  d \sigma \, \sigma^{c_2+1} \,  \,  (e^{i  \beta }e^{-\sigma (t - \tau)} - e^{-i  \beta } e^{-\sigma (t - \tau^*)})
 + \text{regular} \, ,
\end{split}
\end{equation}
where $\tau = c_1 + i \alpha$. Recasting the integral in terms of gamma functions we obtain
\begin{equation} \label{cutsing}
\begin{split}
 G^{(c)} (t) &\approx   \frac{i \, \mathcal C_0}{\pi} \left( \frac{\Gamma(c_2+2)  \, e^{i  \beta} }{ (t-t_c)^{c_2+2}}
 - \frac{\Gamma(c_2+2)  \, e^{ - i  \beta} }{ (t-t_c^\star)^{c_2+2}} \right) + \text{Regular} \, .
\end{split}
\end{equation}
Finally, note that our numerical estimate of $c_1$ together with the exact form of $\alpha$ from \eqref{albetvals} elegantly assembles to give
\begin{equation}
 t_c = c_1 \pm i \alpha \approx - 2 \mathcal X_\mp  \, .
\end{equation}
where $\mathcal X_\pm$ is again the optical distance \eqref{odval}.
This implies that the branch cut contribution also becomes singular at $t = - 2\mathcal X_\pm$.

\begin{figure}[h!tbp]
  \centering
  \includegraphics[width=0.7\textwidth]{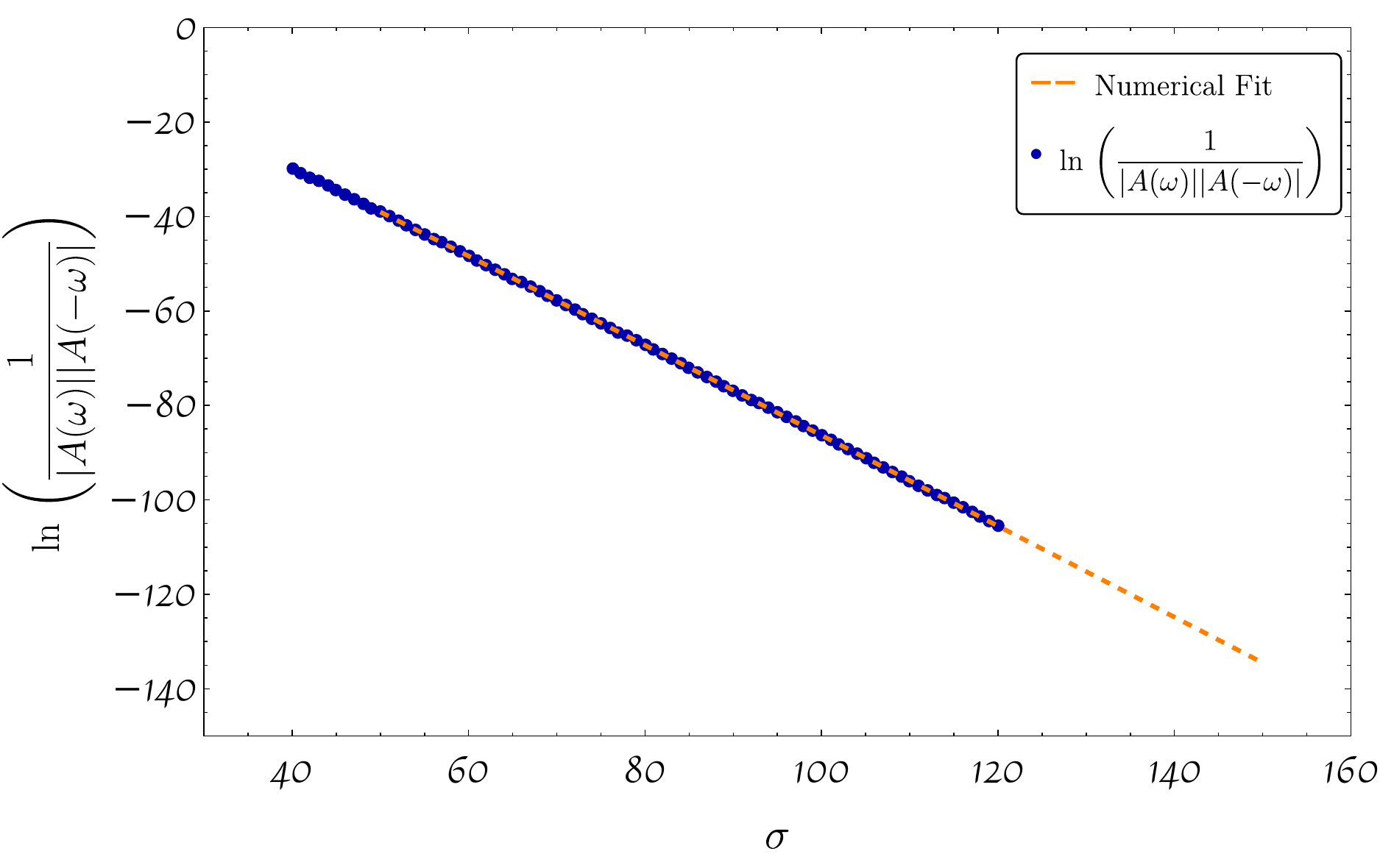}
  \caption{Plot of $\ln \left( \frac{1}{\lvert A(\omega) A(-\omega)\rvert} \right)$ for $k=1$. The solid blue line is the plot with
  the data generated using the procedure outlined in appendix \ref{app:discon}. This data was also used to obtain the phase discontinuity
  in Fig.\ref{fig:theta}. The orange dashed line is a numerical fit of the solid blue line with \eqref{magfitfn} using the parameter values given in \eqref{magfitpara}.}
  \label{fig:magfit}
\end{figure}

%****************************************
\section{Discussions}\label{sec:disco}
%****************************************

Our central result in this paper is the pole-cut formula \eqref{pole-cut} which explicitly demonstrates how the thermal spectral function 
is determined in terms of its QNM poles and its branch-cut discontinuity. For concreteness, in this paper we have focused only on the fluctuations 
of a neutral massless scalar in the background of  the extremal Reissner-Nordstr\"om black brane in AdS$_5$. However, we expect the pole-cut factorization 
structure of spectral functions to hold for more general fluctuations of extremal black branes. The main reason for this expectation is that \eqref{pole-cut}
follows straightforwardly from the thermal product formula which has been validated for more general fluctuations. When we have more than one coupled fluctuations
our formula may be suitably generalized following a similar situation for the thermal case as discussed in \cite{Bhattacharya:2025vyi}.

As a consequence of the pole-cut formula \eqref{pole-cut}, we are able to demonstrate that consistency with the OPE limit of the correlator enforces interesting relationships
between the two classes of spectral data, i.e. the QNMs and the branch-cut discontinuity function $\theta(\sigma)$. Particularly, the asymptotic form of the QNMs
becomes tied to the asymptotics of $\theta(\sigma)$. There are several questions which naturally arise
from this important observation. Moving to higher and higher orders in the large $\omega$ expansion we expect to produce more and more
relations between asymptotic expansions of QNMs and $\theta(\sigma)$. 
It would be interesting to find out whether this implies that the discontinuity function $\theta$ is 
completely determined in terms of the QNM data, upon a putative resummation of these asymptotic expansions. On an ambitious note, there is also a possibility that these relationships are so constraining that 
the entire bulk spectral data can be bootstrapped out of them (see also \cite{Barrat:2025nvu}). In a similar vein, moving to higher orders in the 
large-$\omega$ expansion, it may be possible to predict the boundary OPE coefficients appearing in \eqref{OPEfull} directly from the bulk spectral data. Finally, we would like to note that 
the bulk spectral constraints from the OPE limit arise due to the existence of dual boundary CFT. Hence it is tempting to wonder whether such a relationship 
between the QNMs and the branch cut data is an exclusive holographic signature. It may be also be interesting to search for a geometric justification in the bulk for such a relationship to exist.

Before concluding the paper let us appreciate the fact that the results here have provided us with important lessons to deal with scenarios where the spectral data consists of both
QNMs as well as branch cut discontinuities (see also \cite{Abbasi:2026rcc}). Similar significance of branch cut discontinuities can be expected for black holes in asymptotically flat space. In particular, 
if we consider a massless scalar field or metric fluctuations for Schwarzschild black holes in flat space, can we observe a structure similar to our pole-cut formula for asymptotic 
observables such as the grey-body factor?  We hope to answer this interesting question in the near future \cite{Bhattacharya:new}.

%~~~~~~~~~~~~~~~~~~~~~~~~~~~~~~~~~~~~~~~~~~~~~~
\acknowledgments 
%~~~~~~~~~~~~~~~~~~~~~~~~~~~~~~~~~~~~~~~~~~~~~~
We would like to thank Soumangsu Chakraborty, Diptarka Das, K. P. Yogendran for useful discussion. AS would like to acknowledge the support from
Prime Minister’s Research Fellowship (\href{https://www.pmrf.in/}{PMRF}), offered by the Government of India.

\vspace{0.2cm}
\noindent
{\bf{Statement on use of AI:}} OpenAI models ChatGPT Sol 5.6 and Astra 6.0, and Anthropic model Opus 4.8 were used to refine some of the analytical arguments
and other minor assistance with consistency checks and copy-editing. The numerical procedure used in appendix \ref{app:discon} was suggested by ChatGPT Sol 5.6, which was independently executed in Mathematica. The mathematical manipulation used to regulate the sum-rule leading to \eqref{c2lim} was suggested by ChatGPT Astra 6.0 which was also numerically verified independently by the authors
using Mathematica. All the ideas, arguments and writing in this manuscript originated from the authors. All the calculations and numerical computation were independently carried out by the authors in Mathematica.

%****************************************
\appendix
%****************************************

%********************************************************************************
\section{Imaginary part of the retarded correlator}\label{app:img}
%********************************************************************************

In this appendix, we provide the derivation of the formula \eqref{img}, beginning with the standard prescription \eqref{fullg} for the retarded holographic correlator \cite{Son:2002sd, Iqbal:2008by}.
Let us consider the fluctuation equation \eqref{phizeqnext}, which we denote by $E[\phi] =0$.
In this argument we will assume that $\omega$ and $k$ in \eqref{phizeqnext} are real parameters.
A solution $\phi (z)$ of \eqref{phizeqnext} is a complex function of the real variable $z$,  even for real $\omega$ and $k$. Through a simple algebraic manipulation, we can show that
\begin{equation} 
 \phi^\star \, E[\phi] - \phi \, E[\phi]^\star = z^3 \, \frac{\partial \mathcal F}{\partial z} \, , 
\end{equation} 
where 
\begin{equation} \label{Fconsv}
\mathcal F (z) = \frac{f}{z^3} \left( \phi^\star \,  \partial_z \phi(z) - \phi \,  \partial_z \phi(z)^\star \right) + \frac{2 i \omega}{z^3} \lvert \phi \rvert^2 \, . 
\end{equation} 
This implies that $\mathcal F(z)$ is radially conserved on-shell. 

Now, our solution to \eqref{phizeqnext} has the following behaviour at the end points 
\begin{equation} 
\phi(z) : ~ \phi_h (z \rightarrow 1) = 1 , ~ \phi_b (z \rightarrow 0) = A + \dots + B z^4 + \dots \, .
\end{equation} 
We can again normalize this solution dividing by $A$, which modifies the behaviour at end points in the following way 
\begin{equation} \label{varphi}
\varphi(z) = \frac{\phi(z)}{A} : ~ \varphi_h (z \rightarrow 1) = \frac{1}{A} , ~ \varphi_b (z \rightarrow 0) = 1 + \dots + \frac{B}{A} z^4 + \dots \, .
\end{equation} 
Since $\varphi(z)$ is a legitimate solution of \eqref{phizeqnext}, hence the quantity $\mathcal F (z)$ in \eqref{Fconsv} is also conserved for $\varphi(z)$. Evaluating 
$\mathcal F (z)$ at the ends points using \eqref{varphi} gives us 
\begin{equation} 
\mathcal F_b (z \rightarrow 0) = 8 i \, \text{Im} \left( \frac{B}{A} \right) , ~ \mathcal F_h (z \rightarrow 1) = 2 i \frac{\omega}{\lvert A \rvert^2} \, .  
\end{equation} 
Here, we have used $f(z \rightarrow 0) = 1$ and $f(z \rightarrow 1) = 0$. 
Now since $\mathcal F (z)$ is conserved along the radial direction $z$ we must have 
\begin{equation} 
 \text{Im} \left( \frac{B}{A} \right) = \frac{1}{4} \, \frac{\omega}{\lvert A \rvert^2}, ~~ \omega \in \mathbb R  \, .
\end{equation} 
This ensures that \eqref{img0} follows from \eqref{fullg}. It is important to note that this result is valid on the real axis of the complex $\omega$ plane.

%********************************************************************************
\section{Derivation of discontinuity in $H_T$}\label{app:HT}
%********************************************************************************
In this section, we will demonstrate that  $H_T(\omega)$ in \eqref{Hasint} has the discontinuity mentioned in \eqref{discHT}. 
The $\omega$ dependence and hence, the discontinuity of $H_T(\omega)$ arises only from the first term in \eqref{Hasint}, which is given by
\begin{equation} 
I (\omega) = 2 \int_0^\infty \sigma d \sigma \, \mathcal N(\sigma) ~\frac{1}{\sigma^2 + \omega^2} \, .
\end{equation} 
The discontinuity is computed as follows 
\begin{equation} \label{diseval}
\begin{split}
\text{Disc} \, H_T (\omega) =& \text{Disc} \, I(\omega) = \lim_{\epsilon \rightarrow 0} \big[ I ( - i \lambda + \epsilon) -  I (- i \lambda - \epsilon) \big] \\
=& \lim_{\tilde \epsilon \rightarrow 0}  ~2 \int_0^\infty \sigma d \sigma \, \mathcal N(\sigma) ~\left( \frac{1}{\left( \sigma^2 - \lambda^2 \right) - i \tilde \epsilon}
- \frac{1}{\left( \sigma^2 - \lambda^2 \right) + i \tilde \epsilon} \right) \, , 
\end{split}
\end{equation} 
where we have suitably redefined the infinitesimal parameter to $\tilde \epsilon$ in the final expression. 
Further, using the identity 
\begin{equation} 
\frac{1}{z + i \epsilon} = \text{P}.\text{V}. \, \left( \frac{1}{z}\right) - i \pi \delta\left( z \right)  \, ,
\end{equation} 
for both the terms within the bracket in \eqref{diseval}, we find that the principal values cancel between them and we are left with
\begin{equation} \label{findisc}
\text{Disc} \, H_T = 
 2 \int_0^\infty \sigma d \sigma \, \mathcal N(\sigma) ~\left( 2 \pi i \, \delta \left( \sigma^2 - \lambda^2 \right) \right) \, = 2 \pi i \,  \mathcal N (\lambda) , 
\end{equation} 
where we have used the fact that only the root $\sigma = + \lambda$ lies within the integration domain $[0, \infty)$. 
This is the same result we concluded in \eqref{discHT}, using the cumulative discontinuity of the individual Logarithms. Also, note that 
this argument does not use any special property of the function $\mathcal N(\sigma)$. Hence, any function with a form \eqref{Hasint} for any generic $\mathcal N(\sigma)$
will have a discontinuity \eqref{findisc}. 

%********************************************************************************
\section{QNMs in AdS$_2$}\label{app:ads2}
%********************************************************************************
In the extremal limit, an AdS$_2$ factor emerges in the near horizon geometry of the AdS black brane.
In this limit, the $f(z)$ in \eqref{backextsol} develops a double zero at the horizon $z=1$, which gives rise to this AdS$_2$ factor.
The importance of this AdS$_2$ and its role in the non-analyticity of boundary correlation functions is well studied in the
literature \cite{Faulkner:2009wj,Iqbal:2011in,Edalati:2010hk,Edalati:2010pn}.
Working in the near extremal limit, with a small but finite temperature
{\footnote{By small temperature, we mean $T \ll \mu$. See the discussion below \eqref{backextsol} and footnote \ref{fn:unit}.}}
, we can expand the function $f(z)$ in \eqref{backsol} in horizon adapted coordinate  $\rho = 1-z$ in the following way
\begin{equation}\label{fzexp}
\begin{split}
 f\left(1-\rho \right) &= (4 - 2 Q^2) \rho + (9 Q^2 - 6) \rho^2 + \mathcal O \left( \rho^3 \right) \\
 &= 4 \pi T \, \rho + 12 \, \rho^2 \, + \mathcal O \left( \rho^3 \right) ,
\end{split}
\end{equation}
where, in the second line of \eqref{fzexp},  we have used \eqref{Tmu} i.e. $Q^2 = 2 - 2 \pi T$ and we have ignored a term proportional to $T \rho^2$  which will be subleading in the near-extremal limit. In the near horizon limit, ignoring $\mathcal O \left( \rho^3 \right)$ terms, we can rewrite this function
after completing the square in the following form 
\begin{equation}
\begin{split}
 f \left(1-\rho \right) \approx 12 \left( \left( \rho + \frac{\pi T}{6} \right)^2 - \left( \frac{\pi T}{6} \right)^2 \right) 
\end{split}
\end{equation}
Hence, in the near horizon limit $z \rightarrow 1$ or equivalently $\rho \rightarrow 0$, the metric \eqref{backmet} takes the following form 
\begin{equation}\label{rcoordads2}
\begin{split}
ds^2 = - \frac{r^2 - r_h^2}{L^2} \, dt^2 + \frac{L^2}{r^2 - r_h^2} \, dr^2 + d\vec x^2  \, , 
\end{split}
\end{equation}
where
\begin{equation}
r = \rho + \frac{\pi \, T}{6}, ~~ r_h = \frac{\pi \, T}{6}, ~~ L^2 = \frac{1}{12}  \, . 
\end{equation}
Clearly, the subspace spanned by the coordinates $\{t,r\}$ is a finite temperature AdS$_2$ black hole with AdS radius $L = 1/\sqrt{12}$. 
In this coordinate,  $r = r_h$ is the horizon, and $r \rightarrow \infty$ is the boundary of this near horizon region. 
Note that, in the strict extremal limit, the linear term in \eqref{fzexp} drops, consequently $r_h$ vanishes, and the AdS$_2$ black hole
reduces to pure AdS$_2$ characteristic of the extremal geometry. 
For computational convenience, we further move to a coordinate $\zeta = L^2 / r$ which goes to zero at the near horizon boundary. With this coordinate change 
the metric \eqref{rcoordads2} reduces to 
\begin{equation}\label{zetacoordads2}
\begin{split}
ds^2 = \frac{L^2}{\zeta^2} \left( - h(\zeta) dt^2 + \frac{1}{h(\zeta)} \, d\zeta^2  \right) + d\vec x^2  \, , 
\end{split}
\end{equation}
where 
\begin{equation} \label{zetadef}
h(\zeta) = 1 - \frac{\zeta^2}{ \zeta_0^2}, ~~\text{with} ~~ \zeta_0 = \frac{L^2}{r_h} = \frac{1}{2 \pi T} \, .
\end{equation}

Hence the near-extremal geometry has a near horizon region given by \eqref{zetacoordads2} and a far region
which asymptotes to AdS$_5$ in the UV. We shall solve the fluctuation equation \eqref{phizeqnext} separately in the two regions and
patch them up through boundary conditions. Through this procedure, it is possible to solve the fluctuation equations
analytically for small frequencies $\omega$ ranging between zero and $\mathcal O(1)$
{\footnote{Strictly speaking, we mean $\omega/ \mu_e$ is between zero and $\mathcal O(1)$. See footnote \ref{fn:unit}.}}, as we now explain.

\vspace{0.40cm}
\noindent
{\bf{Inner Region:}} At first, let us consider the near horizon region \eqref{zetacoordads2}. Moving to a rescaled near horizon coordinate
$\xi = \zeta / \zeta_0 = 2 \pi \, T \, \zeta$, where $\zeta$ is given in \eqref{zetadef}, the fluctuation equation  \eqref{phizeqnext}
at the leading order in the $T \rightarrow 0$ limit is given by
\begin{equation}\label{nheq}
(1- \xi^2) \partial_\xi^2 \, \phi + \left( -2 \xi + 2 i \widehat \omega \right) \partial_\xi \, \phi - \frac{k^2 L^2 }{\xi^2} \phi = 0.
\end{equation}
where $\widehat \omega = \zeta_0 \omega = \omega / \left( 2 \pi \, T \right)$. Here we take the $T \rightarrow 0$ limit holding $\widehat \omega$ fixed.
It is important to note that this implies we are solving this equation in the low frequency limit. This is because, as $\widehat \omega$ ranges from zero to infinity, 
our original unscaled $\omega = 2 \pi \, T \, \widehat \omega$ ranges between zero and an $\mathcal O (1)$ number, in the $T \rightarrow 0$ limit
\footnote{Note that, if we are interested in values of $\omega$ beyond $\mathcal O \left( 1 \right)$, then we must consider corrections to the IR geometry
which are proportional to higher powers of $T$ i.e. retain higher order terms in \eqref{nheq}. These corrections to the near horizon geometry would capture the departure from the AdS$_2$ factor. In turn, this will also lead to corrections to the solution \eqref{insol}.}.

In this limit, the equation \eqref{nheq} is actually the equation of a massive scalar inside the AdS$_2$ black hole with $m = k$.
Hence the IR dimension of this scalar field is given by 
\begin{equation}\label{nuk}
\Delta_{\text{IR}} = \frac{1}{2} + \nu_k, ~~ \nu_k = \sqrt{\frac{1}{4} + \frac{k^2}{12}}.
\end{equation} 
We can solve \eqref{nheq} analytically to obtain
near horizon IR solution as follows
\begin{equation}\label{insol}
 \phi_{\text{in}} = (1-\xi)^{i \widehat \omega} \left( \mathcal A_{\text{IR}} ~~\xi^{\frac{1}{2}- \nu_k} ~  {_2F_1} \left( a_{-},b_-; c_- ; \xi^2\right)
 + ~\mathcal B_{\text{IR}} ~~\xi^{\frac{1}{2} + \nu_k} ~  {_2F_1} \left( a_+ , b_+ ; c_+ ; \xi^2 \right) \right) \, ,
\end{equation}
where
\begin{equation}
 a_\pm = \frac{1}{4} \pm \frac{\nu_k}{2} + \frac{i \widehat \omega}{2}\, ,~ b_\pm = \frac{3}{4} \pm \frac{\nu_k}{2} + \frac{i \widehat \omega}{2} ,~ c_\pm = 1 \pm \nu_k \, .
\end{equation}
We determine $\mathcal A_{\text{IR}}$ and $\mathcal B_{\text{IR}}$ by ensuring that our solution is ingoing at the horizon and is appropriately normalized. Since, we have already
considered an ingoing ansatz \eqref{anza}, the appropriate boundary condition at the horizon is $ \phi_{\text{in}} \rvert_{\xi \rightarrow 1} = 1$. Imposing this boundary condition
fixes the solution uniquely and we get
\begin{equation}\label{ABir}
 \mathcal  A_{\text{IR}} = \frac{2^{-\frac{1}{2} \textcolor{black}{+} \nu_k } ~\Gamma \left( 1- i \, \widehat \omega \right) ~ \Gamma \left( \nu_k\right)}{\sqrt{\pi} ~\Gamma \left(  \frac{1}{2} + \nu_k -i \, \widehat \omega \right)}, ~~
 \mathcal B_{\text{IR}} = \frac{2^{-\frac{1}{2} - \nu_k } ~\Gamma \left( 1- i \, \widehat \omega \right) ~ \Gamma \left( - \nu_k\right)}{\sqrt{\pi} ~\Gamma \left(  \frac{1}{2} - \nu_k -i \, \widehat \omega \right)}\,.
\end{equation}
Interestingly, the zeros of $\mathcal  A_{\text{IR}}$ are the values of on-axis QNMs which we have used in \eqref{ads2qnm}. We will explain this connection in more detail below.
As we approach the boundary of the near horizon region, this solution has the following asymptotic  behaviour
\begin{equation}\label{IRbdy}
\begin{split}
 \phi_{in} \rvert_{\xi \rightarrow 0} &= \mathcal A_{\text{IR}} ~ \xi^{\frac{1}{2} - \nu_k} + \mathcal B_{\text{IR}} ~ \xi^{\frac{1}{2} + \nu_k}  + \dots \\
 &=  \left( \mathcal A_{\text{IR}} \, {\zeta_0}^{-\frac{1}{2}  + \nu_k} \right) ~ \zeta^{\frac{1}{2} - \nu_k} +
 \left( \mathcal B_{\text{IR}} \, {\zeta_0}^{-\frac{1}{2}  - \nu_k} \right) ~ \zeta^{\frac{1}{2} + \nu_k} \, .
\end{split}
\end{equation}
Consequently, the IR retarded two point function is given by
\begin{equation}\label{Gir}
 \mathcal G_{\text{IR}} = \left( \zeta_0 \right)^{- 2 \nu_k} \frac{\mathcal B_{\text{IR}} }{\mathcal A_{\text{IR}} } = \left( \pi T \right)^{2 \nu_k} \frac{\Gamma\left( - \nu_k \right) \, \Gamma \left( \frac{1}{2} + \nu_k - i \widehat \omega \right) }{\Gamma\left(\nu_k \right) \,  \Gamma \left(\frac{1}{2} - \nu_k - i \widehat \omega \right)} \, .
\end{equation}
Note that while calculating the IR-retarded correlator, we have used the dimensionful coordinate $\zeta$, which ensures that the IR correlator scales with the right power of temperature that follows from dimension analysis
{\footnote{The two point function of an operator with dimension $\Delta_{IR}$ must have dimension $2\Delta_{IR}$. Hence, its Fourier transform must have dimension $2\Delta_{IR}-1$. From \eqref{nuk}, it follows that $2\Delta_{IR}-1 = 2 \nu_k$, which explains the scaling with temperature in \eqref{Gir}.}}.
As we will see below, the additional factor involving $T^{2\nu_k}$ in \eqref{Gir} will naturally emerge when we match
this near horizon solution with a solution from the outer region.

\vspace{0.40cm}
\noindent
{\bf{Outer Region:}} As we have observed before, our solution \eqref{insol} in the near horizon geometry is valid for $\omega$ ranging from 0 to an $\mathcal O (1)$ number. This allows us to match it up with solution of \eqref{phizeqnext} in the outer region which is obtained in a power series small $\omega$ expansion. Hence, for the solution in the outer region, 
we make the ansatz 
\begin{equation} 
\phi_{\text{out}} (z) = \varphi_0 (z)  + \omega \varphi_1 (z) + \omega^2 \varphi_2 (z) \dots \, . 
\end{equation} 
Plugging this back into \eqref{phizeqnext}, and expanding in $\omega$, we get the following equation at every order in $\omega$
\begin{equation}  \label{outeq}
   \varphi_{n}''+
  \left(\frac{f'}{f} - \frac{3}{z} \right) \, \varphi_n'
  - \frac{k^2}{ \textcolor{black}{f} } \, \varphi_n = - \frac{2 i}{f}  \,  \varphi_{n-1}' + \frac{3 i}{z f} \, \varphi_{n-1} \, ,
\end{equation}
Note that these equations are independent of $\omega$. The structure of these equations are such that the homogeneous part of the
equation on the LHS is same at all orders, while the source on the right is determined by the solution at the previous order.
For the leading order $\varphi_0$, we have a homogeneous equation with no source
\begin{equation}  \label{outeq0ord}
   \varphi_{0}''+
  \left(\frac{f'}{f} - \frac{3}{z} \right) \, \varphi_0'
  - \frac{k^2}{ \textcolor{black}{f} }\, \varphi_0 = 0 \, .
\end{equation}
We must solve these equations using the appropriate IR boundary conditions borrowed from
the asymptotic behaviour of the inner solution obtained in \eqref{IRbdy}.

The inner and outer regions have an overlapping range which are specified in terms of the inner and outer coordinates respectively as follows
\begin{equation}
 0 \ll \xi \ll 1, ~~~T \ll 1-z \ll 1
\end{equation}
This overlapping region is the asymptotic region of the AdS$_2$ black hole and serves as the region where the inner and outer region are matched.
Within this overlapping region, the AdS$_2$ coordinate $\xi$ is related to outer radial coordinate $z$ as follows
\begin{equation}
 \xi = \frac{1}{12 \zeta_0 \left( 1 - z + \frac{\pi T}{6}\right)}
 \approx \frac{1}{12 \zeta_0 \left( 1 - z \right)} , ~~ \text{for} ~~ T \ll 1-z \, .
\end{equation}
Using this in \eqref{IRbdy}, near $z \rightarrow 1$ the IR boundary conditions for the solution in the outer region is given by
\begin{equation}\label{IRbdyforout}
 \phi_{\text{out}} =  \tilde A_{\text{IR}}  ~ \left( 1- z\right)^{-\frac{1}{2} + \nu_k}
 + \tilde B_{\text{IR}} ~ \left( 1 - z\right)^{-\frac{1}{2} - \nu_k}  + \dots \, ,
\end{equation}
where
\begin{equation}\label{tildeAB}
 \tilde A_{\text{IR}} = \mathcal A_{\text{IR}} ~\left( 12 \zeta_0 \right)^{-\frac{1}{2} + \nu_k}, ~~
  \tilde B_{\text{IR}} = \mathcal B_{\text{IR}} ~\left( 12 \zeta_0 \right)^{-\frac{1}{2} - \nu_k}
\end{equation}
Note that the relative factor of $\zeta_0^{-2\nu_k}$ in the ratio $\tilde B_{\text{IR}} / \tilde A_{\text{IR}}$ was anticipated in \eqref{Gir}.
On the other hand, the asymptotic form of the outer solution in the true UV ($z \rightarrow 0$) is given by \eqref{fulbdyasym}
\begin{equation}
 \phi_{\text{out}} = \mathcal A + \dots + \mathcal B z^4 + \dots \, .
\end{equation}
Here $\mathcal A$ and $\mathcal B$ can be obtained order by order in $\omega$, by solving \eqref{outeq} with the boundary condition
\eqref{IRbdyforout}.
The $\mathcal A$ and $\mathcal B$ so determined, must have the following structural form
\begin{equation}\label{ABfullform}
\begin{split}
 \mathcal A &= \tilde A_{\text{IR}} \left( a_+^{(0)} + \omega a_+^{(1)} + \dots \right)
 + \tilde B_{\text{IR}} \left( a_-^{(0)} + \omega a_-^{(1)} + \dots \right) \, , \\
 \mathcal B &= \tilde A_{\text{IR}} \left( b_+^{(0)} + \omega b_+^{(1)} + \dots \right)
 + \tilde B_{\text{IR}} \left( b_-^{(0)} + \omega b_-^{(1)} + \dots \right) \, .
\end{split}
\end{equation}

\vspace{0.40cm}
\noindent
{\bf{QNMs:}}
Now, since the QNMs are the zeros of $\mathcal A$, they are given by the roots of the following equation
\begin{equation}\label{qnmcond0}
 \frac{\tilde B_{\text{IR}}}{\tilde A_{\text{IR}}}
 = - \frac{\left( a_+^{(0)} + \omega a_+^{(1)}+ \dots \right)}{\left( a_-^{(0)} + \omega a_-^{(1)} + \dots \right)} \, .
\end{equation}
Now, since we have \eqref{tildeAB}, we can identify that the LHS of \eqref{qnmcond0} is proportional to $\mathcal G_{\text{IR}}$ obtained in \eqref{Gir}. Thus we can write
\begin{equation}\label{qnmcond}
 \mathcal G_{\text{IR}}  \equiv \left(\zeta_0 \right)^{- 2 \nu_k} \frac{\mathcal B_{\text{IR}} }{\mathcal A_{\text{IR}} } \equiv \left(12\right)^{2 \nu_k} \frac{\tilde B_{\text{IR}}}{\tilde A_{\text{IR}}}
 = - \left(12\right)^{2 \nu_k} \frac{ \left( a_+^{(0)} + \omega a_+^{(1)}+ \dots \right)}{\left( a_-^{(0)} + \omega a_-^{(1)} + \dots \right)} \, .
\end{equation}
If we use $\mathcal G_{\text{IR}}$ from \eqref{Gir}, then \eqref{qnmcond} becomes the equation whose solutions for $\omega$ give
the QNMs, both the on-axis ones as well as the off-axis ones. Note that something very interesting happens
when we hit zero of $\mathcal A_{\text{IR}}$. From \eqref{ABir}, the zeros of $\mathcal A_{\text{IR}}$ are given by $\omega_m$
reported in \eqref {ads2qnm}. These are the QNMs of the IR AdS$_2$ black hole and they all lie along the negative imaginary axis
in the complex $\omega$ plane. When $\omega$ is near such a $\omega_m$,
the condition \eqref{qnmcond} reduces to
\begin{equation}
 T^{1 + 2 \nu_k} \frac{1}{\omega - \omega_m} = \Lambda_0 \, ,
\end{equation}
where $\Lambda_0$ is an $\mathcal O (1)$ number arising from the RHS of \eqref{qnmcond} as well as the remaining $\Gamma$-functions in \eqref{Gir} when $\omega$ approaches $\omega_m$.
Also we have
used $\widehat \omega = \omega / 2 \pi \, T$. Solving this equation gives
\begin{equation}\label{QNMcor}
 \omega = \omega_m + \frac{ T^{ 1 + 2 \nu_k}}{\Lambda_0} \, .
\end{equation}
But, from \eqref{nuk}, it follows that $2 \nu_k \geq 1$. Consequently, as we approach the extremal limit, the QNMs of the
AdS$_2$ black hole become the on-axis QNMs of the full geometry up to corrections of $\mathcal O \left( T^{2 \nu_k} \right)$.
The other solutions of \eqref{qnmcond} constitute the off-axis QNMs of the full geometry, where the specific values of $a_{\pm}^{(i)}$,
which are obtained from the outer geometry, plays an important role. This establishes that in the extremal
limit the near horizon AdS$_2$ is solely responsible for the branch cut, while the outer geometry determines off-axis QNMs.

Before concluding this discussion, we would like to point out a subtle feature of \eqref{QNMcor} which is very important for our purpose. The structure of
\eqref{QNMcor} may naively suggest that in the strict extremal limit the coalescence of $\omega_m$ correctly reproduces the density of QNMs everywhere on the
imaginary axis. But this is not true. The density of QNMs obtained from the extremal limit of $\omega_m$ in \eqref{ads2qnm} is correct on the
negative imaginary axis only near the real line
of the complex $\omega$ plane,
i.e. only for the small values of $m$ in \eqref{ads2qnm}. This is because, for large overtones ($m \sim \mathcal O \left( 1 / T\right)$), the contribution of the
external geometry through $\Lambda_0$ becomes important in \eqref{QNMcor} affecting the QNM density significantly.
If we parameterize the negative imaginary axis in the complex $\omega$ plane as $\omega = - i \lambda$, then for any fixed non-zero $\lambda = \sigma_m$, with $\sigma_m$ being given by \eqref{ads2qnm},
the large overtones become important, even in the strict extremal limit.
Consequently, the tower \eqref{ads2qnm}, in the $T \rightarrow 0$ limit correctly reproduces
the QNM density only in the neighbourhood of $\lambda \rightarrow 0$, i.e. only near the real axis.

\vspace{0.40cm}
\noindent
{\bf{Discontinuity of the source coefficient in the extremal limit:}}
From this analysis, it is also possible to derive the discontinuity
of the source coefficient in the strict extremal limit. In the $T \rightarrow 0$ limit, we have
\begin{equation}\label{extAG}
 \tilde A_{\text{IR}} \equiv \mathcal A_{\text{IR}} ~\left( 12 \zeta_0 \right)^{-\frac{1}{2} + \nu_k} = A_0 \left( - i \omega \right)^{\frac{1}{2} - \nu_k}, ~~\mathcal G_{\text{IR}} = G_0 \left( - i \omega \right)^{2 \nu_k} \, .
\end{equation}
This result can be either obtained by taking a straightforward $T \rightarrow 0$ limit on \eqref{ABir} and \eqref{Gir},
% {\footnote{Note that, more precisely, a $T \rightarrow 0$ limit on \eqref{ABir} gives
% $$ A_{\text{IR}} = A_0 \, \exp \left( {- i \widehat \omega \ln 2}  \right) \, \left( - i \omega \right)^{\frac{1}{2} - \nu_k} \, , $$
% where the additional phase is a remnant of finite temperature which is absent in the direct computation using the extremal geometry.
% This phase drops out from physical quantities such as $\mathcal G _{\text{IR}}$ and being analytic, it does not contribute to the discontinuity which
% is our main focus here.}}
or they can be obtained directly from the extremal geometry, using a near horizon AdS$_2$ rather than an AdS$_2$ black hole \cite{Faulkner:2009wj,Iqbal:2011in,Edalati:2010hk,Edalati:2010pn}.

The source coefficient of the full geometry continue to have the form \eqref{ABfullform}, which we compactly write as
\begin{equation}
 \mathcal A = \tilde A_{\text{IR}} \, a_+ \left( \omega \right) + \tilde B_{\text{IR}} \, a_-\left( \omega \right)
 = \tilde A_{\text{IR}}  \big( a_+ + \textcolor{black}{ {12}^{-2\nu_k} } \mathcal G_{\text{IR}} \, a_- \big)  \, ,
\end{equation}
where $a_\pm \left( \omega \right)  = a^{(0)}_\pm + a^{(1)}_\pm \omega + \dots $ are analytic functions of $\omega$. The discontinuity in $A$ therefore arises entirely from the
discontinuity of the IR source function $\tilde A_{\text{IR}}$ as well as the IR two-point function $\mathcal G_{\text{IR}}$
which can be evaluated directly using \eqref{extAG}. Hence, from the definition \eqref{thetadef}, we have
\begin{equation}
 \theta \left( \lambda \right) = \frac{1}{2  i} \text{Disc} \ln \mathcal A \left( \omega = - i \lambda \right) = \frac{1}{2  i} \left( \text{Disc} \ln \tilde A_{\text{IR}}
 + \text{Disc} \ln \left(  a_+ +\textcolor{black}{ {12}^{-2\nu_k} }  \mathcal G_{\text{IR}} \, a_- \right) \right).
\end{equation}
Evaluating this discontinuity using \eqref{extAG} and expanding in small $\lambda$, we get
\begin{equation}\label{thetaexp}
 \theta \left( \lambda \right) = \pi \left( \nu_k - \frac{1}{2} \right) + \mathcal C_0 \lambda^{2 \nu_k} \dots \, ,
\end{equation}
where we have retained only the leading order term in the small $\lambda$ expansion of the second term containing $\mathcal G_{\text{IR}}$. The leading order term in this expansion $$\theta_{\text{IR}} = \pi \left( \nu_k - \frac{1}{2} \right)$$ arises
entirely from the IR source coefficient $\tilde A_{\text{IR}}$, and it
explains the discontinuity about the branch cut arising from the coalescence of AdS$_2$ QNMs in the thermal product formula as explained in \ref{ssec:AdS2}.

%**********************************************************************************************************************
\section{Source function discontinuity across the branch cut}\label{app:discon}
%**********************************************************************************************************************
In this appendix we shall outline the numerical procedure adopted to compute the discontinuity in the source function $A(\omega,k)$
which lead to the plot in Fig.\ref{fig:theta} {\footnote{We thank ChatGPT model Sol-5.6 for suggesting and outlining this procedure, which was later independently executed in Mathematica.}}.

For this discussion, it would be convenient to redefine the radial coordinate as $u = 1 - z$, so that $u = 0$ is the horizon and $u = 1$ is the boundary. After performing this variable change, the extremal fluctuation equations \eqref{phizeqnext} in the $u$-coordinate reduces to \footnote{ \textcolor{black}{In going from \eqref{phizeqnext} to \eqref{ueq}, we have first rescaled $\phi(z) \rightarrow z^4 \phi(z)$. Then, we substitute $z = 1- u$, and dropped an overall factor in the equation \eqref{ueq}. } }
\begin{equation}\label{ueq}
 P(u) \, \partial^2_u \phi(u) + Q(u) \, \partial_u \phi(u) + R(u) \, \phi(u) = 0,
\end{equation}
where the coefficient functions are given by
\begin{equation}\label{coefs}
 \begin{split}
  P(u) &= 12 u^2 - 40 u^3 + 55 u^4 - 39 u^5 + 14 u^6 - 2 u^7 \, ,\\
  Q(u) &= - 2 i \omega + (24 + 2 i \omega) u - 168 u^2 + 332 u^3 - 303 u^4 + 132 u^5 - 22 u^6  \, ,\\
  R(u) &= - k^2 + 5 i \omega + (k^2 - 96) u + 336 u^2 - 432 u^3 + 240 u^4 - 48 u^5\, .\\
 \end{split}
\end{equation}

The asymptotic form of the solution near the boundary takes the form of \eqref{fulbdyasym}, which in the $u$-coordinate becomes
\begin{equation}
 \phi(u) \overset{u \rightarrow 1}{=} \frac{1}{\textcolor{black}{(1-u)^4} }\left( A(\omega, k) + \dots + B(\omega, k) (1-u)^{4} + \dots \,  \right).
\end{equation}

We wish to numerically obtain the discontinuity in $A(\omega,k)$ as we move across the imaginary axis in the complex $\omega$ plane. In other words,
we would like to compute the following two quantities
\begin{equation} \label{ApAMdef}
 A_+(\sigma,k) = \lim_{\epsilon \rightarrow 0} A(\omega = -i \sigma + \epsilon, k) , ~~   A_-(\sigma,k) = \lim_{\epsilon \rightarrow 0} A(\omega = -i \sigma - \epsilon, k)
\end{equation}
At first sight, this may seem straightforward. The ingoing boundary condition has already been incorporated into \eqref{ueq} (see \eqref{anza}), so we may wish to solve the equation numerically shooting from the horizon, using the boundary condition $\phi(u \rightarrow 0) = 1$ (see \eqref{horbdyc}). We can try to numerically evolve the differential equation from near $u = 0$ to say $u = u_b \approx 1$. Once we have the numerical solution
we can read off $A(\omega, k)$ from the leading asymptotic behavior at $u = u_b$. We can perform this numerical exercise substituting $\omega = -i \sigma \pm \epsilon$ in \eqref{coefs}, which
is expected to yield $A_\pm$ respectively. However, this approach is too naive and does not work for a very interesting reason which we now explain.

Note that the differential equation \eqref{ueq} has an \textcolor{black}{irregular} singularity at $u=0$. Apart from the regular series solution
\begin{equation}\label{Frosol}
 \phi_1(u) = \sum_n a_n u^n \, ,
\end{equation}
\eqref{ueq} also admits a second solution. This other solution may be obtained as follows. Let us substitute $\phi(u) = \exp \left( S(u) \right)$ in \eqref{ueq} and retain the leading order terms near $u=0$ so that \eqref{ueq} reduces to
\begin{equation}\label{dombal}
 12 u^2 \left( \left( \partial_u S \right)^2 + \partial_u^2 S \right) - 2 i \omega \, \partial_u S + \left(  5 i \omega - k^2  \right)  = 0 \, .
\end{equation}
Further, looking for the solutions where $\partial_u S$ is dominant, this equation reduces to
{\footnote{Note that, for the solution $S(u) = - \tfrac{ i \omega }{6 u}$, $\partial_u^2S \sim 1/u^3$ while $(\partial_u S)^2 \sim 1/u^4$. This justifies why we neglected the second derivative term in \eqref{dombal}.}}
\begin{equation}\label{Ssol}
 12 u^2 \left( \partial_u S \right)^2 - 2 i \omega \, \partial_u S = 0 \,  , ~ \Rightarrow ~ S = - \frac{ i \omega }{6 u} \, .
\end{equation}
Thus apart from the Frobenius solution \eqref{Frosol} there exists another solution, which near the horizon behaves as
\begin{equation}\label{expsol}
 \phi_2 (u) \overset{u \rightarrow 0^+}{=} \exp \left(  - \frac{i \omega}{6 u} \right) \, .
\end{equation}
Now in order to find the discontinuity about $\omega = - i \sigma \pm \epsilon$, if we try to numerically shoot from near $u = 0$, this solution branch \eqref{expsol} will behave as
$\phi_2 (u) \sim \exp \left(  - \frac{\sigma \pm i \epsilon }{6 u} \right)$, which will be exponentially suppressed sufficiently near the horizon for any non-zero value of $\sigma$.
Hence, this solution branch will be impossible to capture reliably in a numerical evolution of \eqref{ueq} starting from $u=0$. On the other hand, this is the crucial solution branch which decides whether we have $A_+$ or $A_-$ when we read off the source from the asymptotic data. This is the reason why the naive numerical approach, mentioned in the previous paragraph, breaks down. This is an example of the classic Stokes problem, the complex $\omega$ negative imaginary axis, i.e. $\omega = - i \sigma$ being the Stokes line across which the coefficient of this recessive solution jumps discontinuously. This makes the naive numerical shooting an ill-conditioned problem which fails to capture the discontinuity.

In order to circumvent this problem we have used a lateral contour prescription, where the central idea is to analytically continue $u$ to a complex variable
{\footnote{For earlier use of this idea in the black hole context see \cite{Motl:2003cd,Berti:2009kk} and the related references therein.}}.
Let us fix $\omega$ completely on the negative imaginary axis, i.e. set $\omega = - i \sigma$. We will not move across this branch cut line in terms of $\omega$. Instead,
we can now start the numerical evolution from $u = \pm i \epsilon$,
offsetting the $u = 0$ point in the imaginary direction. The numerical evolution is now performed along a contour which curves back to the real line away from $u =0$, and from the point it touches the real line, we evolve to boundary at $u =1$ along the real axis (see Fig.\ref{fig:latcont}). More precisely, we perform the numerical evolution in two parts. We start
at $u = \pm i \epsilon$ and integrate along the following contour
\begin{equation}\label{contpart1}
u = \epsilon \exp \left( \pm i \frac{\pi}{2} \left( 1-s \right) \right), ~~ s \in \left[0,1\right] \, .
\end{equation}
Here $s=0$ corresponds to the starting point where we use the boundary condition $\phi (u = \pm i \epsilon ) = 1$. We evolve the solution up to $s=1$ and use the value
of $\phi(u=\epsilon)$ as an initial value for numerical evolution along a second contour which lies along the real axis and is parameterized by
\begin{equation}\label{contpart2}
 u = \epsilon + \left( u_b - \epsilon \right) s   \, , ~~ s \in \left[0,1\right] \, .
\end{equation}
This ensures that at $s=1$ we have reached the boundary $u = u_b \approx 1$.
These contours for numerical evolution have been shown in Fig.\ref{fig:latcont}.

\begin{figure}[h!tbp]
  \centering
  \includegraphics[width=0.7\textwidth]{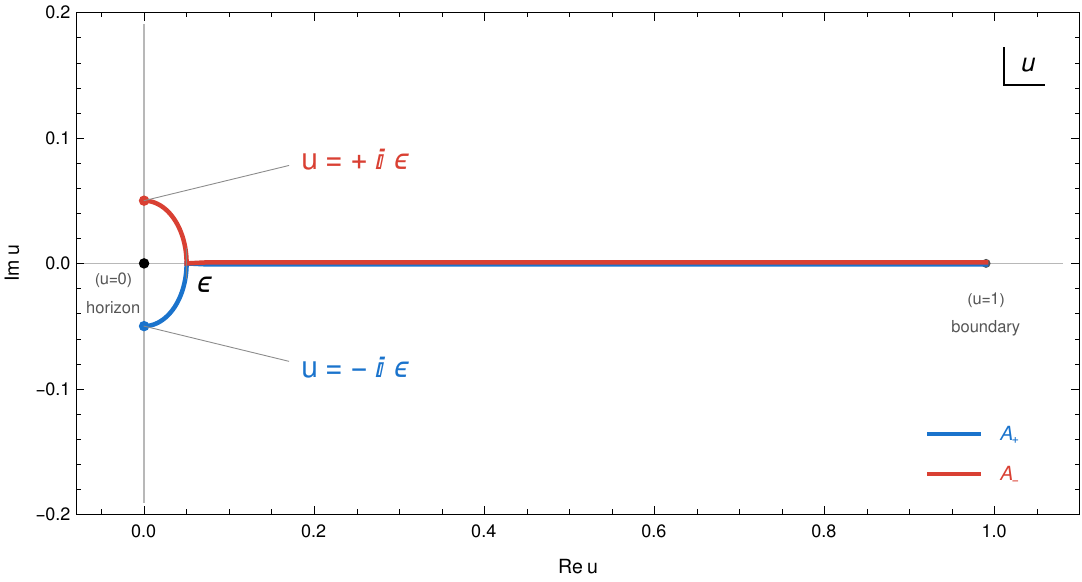}
  \caption{Lateral Contour Prescription.}
  \label{fig:latcont}
\end{figure}

In Fig.\ref{fig:latcont} the red and blue contours represent the starting points $u = \pm i \epsilon$ respectively.
Numerical integration is performed along a complex segment of each contour before joining the real axis at $u =\epsilon$.
On the complex arc, both the Frobenius solution $\phi_1$ \eqref{Frosol}, as well as the recessive solution \eqref{expsol} $\phi_2 \sim \exp \left(\pm i \sigma / 6 \epsilon \right)$
have comparable magnitudes. Hence, the initial condition supplied at $u = \epsilon$ for the segment of the contour on the real line, has the right admixture
of dominant and recessive solutions $\phi_1$ and $\phi_2$. The red and the blue contours with starting points $u = \pm i \epsilon$
lead to distinct coefficients of $\phi_2$ at $u=\epsilon$,
thus providing distinct solutions which corresponds to the left and right solutions across the branch cut $\omega = - i \sigma$.
We determine $A_\pm$ from the asymptotic behavior of these two solutions corresponding to the red and blue contours. Now let us determine
which of these two contours leads to $A_+$ and which one gives $A_-$.

Firstly, note that in \eqref{ApAMdef}, $A_\pm$ corresponds to
$\omega = - i \sigma \pm \epsilon$ respectively. If we put this form of $\omega$ in \eqref{Ssol}, we see that for $\omega = - i \sigma + \epsilon$,
we have $S= - \tfrac{\sigma}{6 u} - i \tfrac{\epsilon}{6 u}$, i.e. it corresponds to Im$\, S < 0$. Hence, we conclude that
Im$\, S < 0$ is associated with $A_+$, and similarly Im$\, S > 0$ is associated with $A_-$. But, in our numerical procedure we fix $\omega = - i \sigma$ and
consider $u = \pm i \epsilon$ as the starting point. When we take $u = - i \epsilon$, we have $S= - i \tfrac{\sigma}{6 \epsilon }$, i.e. Im$\, S < 0$. This means
that starting at $u = - i \epsilon$ leads to that recessive solution branch for which Im$\, S < 0$. Hence, integrating along the blue contour will
provide us with $A_+$. Similarly, integrating along the red contour starting from $u = + i \epsilon$ will yield $A_-$. Once we have $A_\pm$ for a definite numerical value
of $\sigma$, we can compute $\theta (\sigma)$ in \eqref{thetadef} using
\begin{equation} \label{app:thetadef}
 \theta (\sigma) = \frac{1}{2 i} \, \ln \left( \frac{A_+(\sigma, k)}{A_-(\sigma , k)}\right)
\end{equation}
This function $\theta(\sigma)$, thus determined numerically, has been plotted in Fig. \ref{fig:theta}.

One noteworthy feature of \eqref{ueq} together with the coefficient functions \eqref{coefs} is that, under a complex conjugation, it switches
from an equation for $\omega = -i \sigma + \epsilon$ to that corresponding to $\omega = -i \sigma - \epsilon$. Even if we fix $\omega = -i \sigma$, as we have done
in our numerical procedure, complex conjugation will exchange the red and the blue contours in the complex $u$-plane. This implies that the solutions across
the branch cut are complex conjugates of each other. Consequently, we must have
\begin{equation}
 A_-(\sigma, k) =  \left[ A_+(\sigma, k) \right]^{\star}
\end{equation}
This serves as an important cross-check of our numerical answer.

We would like to conclude this discussion by observing that our final numerical answer for $A_\pm$ does not depend on the chosen numerical value of $\epsilon$. This is because
the contour construction essentially ensures that we have the right coefficient of the recessive solution $\phi_2$ when we are back on the real axis and subsequently
numerical evolution occurs along real $u$. The exact point where we are back on the real axis does not have any impact on the asymptotic solution. Nevertheless, the choice of
the numerical value of $\epsilon$ must be done judiciously. It should be small enough so that the Frobenius series solution $\phi_1$ is reliable, i.e. $u = \pm i \epsilon$ must
lie within the radius of convergence of the series. On the other hand, it should be large enough so that, on the complex arc segment of the contours, the oscillation of the recessive solution $\phi_2 \sim \exp \left( - i \tfrac{\sigma}{6 \epsilon} \right)$ can be resolved by the step-size of the numerical evolution. These two constraints
do provide a stable range of values of $\epsilon$, which we have used in our numerics.

%********************************************************************************
\section{Asymptotic form of QNMs and branch cut discontinuity}\label{app:asymexact}
%********************************************************************************

%%%%%%%%%%%%%%%%%%%%%%%%%%%%%%%%%%%%%%%%%%%%%

The leading order coefficient of the QNMs in the large overtone limit \eqref{qnmasym} can be calculated analytically using a WKB analysis which we discuss in this
appendix. We will closely follow the excellent exposition of this analysis in \cite{Hartnoll:2026vhu} building up on earlier works \cite{Motl:2003cd,Cardoso:2004up,Natario:2004jd} (also see very recent works \cite{Jia:2026qvp,valach2026}).

For the WKB analysis, it would be convenient to move to the tortoise coordinate
\begin{equation}
 x (z) = \int_0^{z}  \ \frac{d \zeta}{f{(\zeta)}}
\end{equation}
where $f$ is given by \eqref{backextsol} in the extremal limit. Integrating, we get
\begin{equation}\label{tor}
 x(z) = \frac{z}{6 \left(1-z^2\right)}+\frac{7}{36} \log
   \left(\frac{z+1}{1-z}\right)+\frac{2}{9} \sqrt{2} \tan
   ^{-1}\left(\sqrt{2} z\right)
\end{equation}
Now rewriting the fluctuation equation \eqref{phizeqnext} in the tortoise coordinate
followed by the redefinition $\phi = z^{3/2} \exp \left( - i \omega x \right)\Psi$
{\footnote{Note that the factor of $\exp \left( - i \omega  x \right)$ removes the additional factor in \eqref{anza} introduced due to the definition of
$\omega$ with ingoing-time \eqref{intime}. This is necessary to see the Schr\"odinger equation type structure which is the most familiar setting to perform
the WKB analysis.}}, we obtain the following Schr\"odinger-type equation
\begin{equation}\label{scheq}
 \frac{d^2 \Psi}{dx^2} + \left( \omega^2 - U(x)\right) \, \Psi= 0\, ,
\end{equation}
where $U$ is the effective potential and is given by
\begin{equation}\label{Upot}
 U  = k^2 f + \frac{15 f^2}{4 z^2} - \frac{3 f f'}{2z} \, .
\end{equation}
Here $z$ is to be expressed in terms of $x$ by inverting \eqref{tor}.
% In Fig.\ref{fig:schpot} a plot of this effective potential has been shown.
For large overtones, we must find large $\omega$ solutions to \eqref{scheq}, which are ingoing at the horizon $z=1$ and normalizable at the boundary ($z \rightarrow 0$), i.e. satisfy the QNM boundary conditions. For large $\omega$, the equation \eqref{scheq} can be solved under the WKB approximation.
% A `complex turning point'
% {\footnote{Here, the term `complex turning point' implies the complex zeros of $\omega^2 - U$.}}
% is present near the singularity at $z \rightarrow \infty$, as shown schematically in Fig.\ref{fig:schpot}.
We will show that the `quantization condition' within the WKB approximation provides the asymptotic form of the QNMs.

% The behavior of the potential \eqref{Upot} near the singularity is given by
% \begin{equation}
%  U (z \rightarrow \infty) = -21 z^{10}+45 z^8 + \left(2 k^2-\frac{81}{4}\right) z^6 + \dots
% \end{equation}

% %%%%%%%%%%%%%%%%%%%%%%%%%%%%%%%%%%%%%%%%%%%%%%%%%%%%%%%%%%%%%%%%%%%%%%%%%%%%
% \begin{figure}[h!tbp]
%   \centering
%   \includegraphics[width=0.7\textwidth]{Figures/Upotplot.pdf}
%   \caption{Plot of the effective potential $U$ given in \eqref{Upot}.}
%   \label{fig:schpot}
% \end{figure}
% %%%%%%%%%%%%%%%%%%%%%%%%%%%%%%%%%%%%%%%%%%%%%%%%%%%%%%%%%%%%%%%%%%%%%%%%%%%%

\vspace{6pt}
\noindent
{\bf Optical distance between boundary and singularity:}
Before proceeding to evaluate the asymptotic form of the QNMs, it would be useful to evaluate the optical distance
from the boundary to the singularity. As we will see later, this quantity controls the large overtone coefficients.
For a radial null line, from the metric \eqref{backmet} we have
\begin{equation}
 0 = - f \ dt^2 +  \frac{dz^2}{f} \, .
\end{equation}
Integrating this we can obtain the analytically continued complex time from the boundary of AdS to the curvature singularity
\begin{equation}\label{opdis}
 \mathcal X = \int dt = \int_0^\infty \, \frac{dz}{f} = \int dx
\end{equation}
This is the optical distance between the boundary and singularity. In order to evaluate the integral
we notice that $f$ has a double zero at the horizon $z = 1$. In order to compute $\mathcal X$, we
must integrate past the horizon upto the singularity. Hence we need a pole prescription to deal with the
double pole at $z = 1$. For this we can deform the contour into a infinitesimal semi-circular arc of radius $\epsilon$ above or below the pole at
$z = 1$. This gives two distinct values of the integral which we denote by $\mathcal X_\pm$ respectively as shown in Fig.\ref{fig:torcont}.

%%%%%%%%%%%%%%%%%%%%%%%%%%%%%%%%%%%%%%%%%%%%%%%%%%%%%%%%%%%%%%%%%%%%%%%%%%%%
\begin{figure}[h!tbp]
  \centering
  \includegraphics[width=0.7\textwidth]{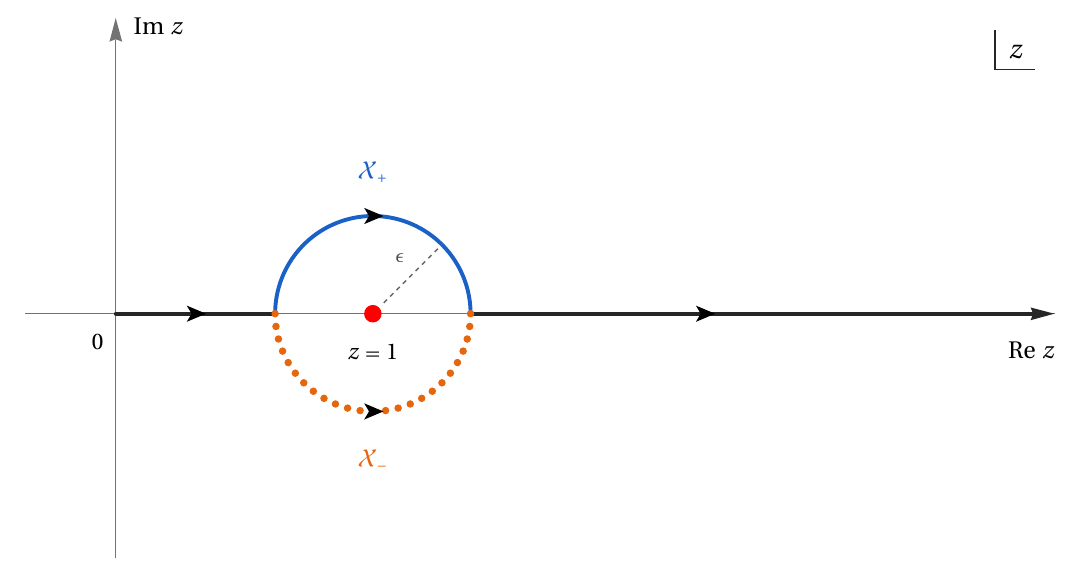}
  \caption{Plot of the contour prescription to obtain the optical distance $\mathcal X_\pm$ defined through the integral in \eqref{opdis}.}
  \label{fig:torcont}
\end{figure}
%%%%%%%%%%%%%%%%%%%%%%%%%%%%%%%%%%%%%%%%%%%%%%%%%%%%%%%%%%%%%%%%%%%%%%%%%%%%

The integral along the deformed contour has three segments
\begin{equation}\label{cont3terms}
 \mathcal X = \lim_{\epsilon \rightarrow 0} \left( \ \int_0^{1-\epsilon} \, \frac{dz}{f} + \int_{1+ \epsilon}^\infty \, \frac{dz}{f} + \int_{\text{arc}} \, \frac{dz}{f} \ \right) \, .
\end{equation}
The integral on the semicircular arc can be performed by considering the following expansion of $f(z)$ near the horizon
\begin{equation}
 \frac{1}{f} = \frac{1}{12 s^2} - \frac{7}{36 s} + \dots \, , ~ s = z-1 \, .
\end{equation}
For the integration over the arc, we take $s = \epsilon e^{i \theta}$ with $\theta$ varying from $\pi$ to $0$ for the arc on top of the pole, while it varies
from $\pi$ to $2 \pi$ for the arc below the pole. Executing the integral, the arc contribution from these two contours are given by
\begin{equation}
\int_{\text{arc}} \, \frac{dz}{f} = - \frac{1}{6 \epsilon} \pm i \frac{7 \pi }{36} \, ,
\end{equation}
where the $\pm$ sign for the imaginary part represent the contours above and below respectively.
Putting this back to \eqref{cont3terms}, we get
\begin{equation}
  \mathcal X_{\pm} = \underbrace{\lim_{\epsilon \rightarrow 0} \left( \ \int_0^{1-\epsilon} \, \frac{dz}{f}
  + \int_{1+\epsilon}^\infty \, \frac{dz}{f}  - \frac{1}{6 \epsilon} \right)}_{\text{F.P.}  \int_0^{\tiny {\infty}} \frac{dz}{f}}
  \pm i \frac{7 \pi}{36} \, ,
\end{equation}
where the terms within the bracket constitute the finite part of the integral and is given by
\begin{equation}
 \text{F.P.} \int_0^{\infty} \frac{dz}{f} = \frac{\pi \sqrt{2}}{9}
\end{equation}
Putting everything together, we have
\begin{equation}\label{odval}
 \mathcal X_\pm = \frac{\pi \sqrt{2}}{9} \pm i \frac{7 \pi}{36} \, .
\end{equation}
Note that, $\mathcal X_+$ will turn out to be the relevant optical distance for our QNMs with a negative
imaginary part.

\vspace{0.5cm}
\noindent
{\bf WKB procedure:}
The equation \eqref{scheq} admits legitimate solutions only for discrete values of
$\omega$ which are the values of QNMs we seek. For asymptotic values of QNMs, we seek solutions with large $\lvert \omega \rvert$
which justifies the applicability of the WKB procedure.
The leading order solution of \eqref{scheq}, in the WKB approximation is given by
\begin{equation}\label{wkbschsol}
 \Psi_\pm (x) \approx \frac{1}{(\omega^2 - U)^{1/4}} \exp \left( \pm \ i \int dx \ \sqrt{\omega^2 - U}  \right)  \, ,
\end{equation}
where $x$ is the tortoise coordinate which is related to the radial coordinate $z$ through \eqref{tor}.
Also the $\pm$ sign corresponds to the two independent solutions of the second order differential equation.
Since we seek complex QNMs, we analytically continue to the complex $z$ coordinate. The tortoise coordinate $x$ remains related to
complex $z$ by \eqref{tor} and is therefore also complexified. Our approximation breaks down when $U$ becomes comparable to $\omega^2$ where we have `turning points'.
Since $\lvert \omega \rvert$  is large by assumption, hence turning points
occur where the magnitude of $U$ is large. This can happen near the
boundary $z \rightarrow 0$ as well as near the singularity $z \rightarrow \infty$, as is evident from the
expansion of the effective potential $U$ near these points
\begin{equation}\label{Uexp}
 \begin{split}
  U (z \rightarrow 0) &= \frac{15}{4 z^2} + k^2 + \dots \\
  U (z \rightarrow \infty) &= -21 z^{10} + 45 z^8 + \left(2 k^2-\frac{81}{4}\right) z^6 + \dots
 \end{split}
\end{equation}
Near these turning points, \eqref{wkbschsol} is not valid and we must solve the equation \eqref{scheq} exactly in the local region, matching it with the solution  \eqref{wkbschsol}
away from the turning point.
Once we depart from the region near the turning points, we can ignore $U$ in comparison to $\omega^2$. Consequently, if we adhere to the anti-Stokes lines, the two solutions in \eqref{wkbschsol} remains completely oscillatory
\begin{equation}\label{oscsol}
  \Psi_\pm (x) \approx \exp \left( \pm  i \omega x \right)  \, .
\end{equation}
Our anti-Stokes lines are defined by the condition
\begin{equation}
 \text{Im} \left[ \int dx \ \sqrt{\omega^2 - U}   \right] = 0
\end{equation}
Away from turning points, our solution is a linear combination of the two solutions in \eqref{oscsol}.
The anti-stokes condition ensures that the two oscillatory solutions \eqref{oscsol} have equal magnitudes without any relative exponential damping or amplification.

%%%%%%%%%%%%%%%%%%%%%%%%%%%%%%%%%%%%%%%%%%%%%%%%%%%%%%%%%%%%%%%%%%%%%%%%%%%%
\begin{figure}[h!tbp]
  \centering
  \includegraphics[width=0.7\textwidth]{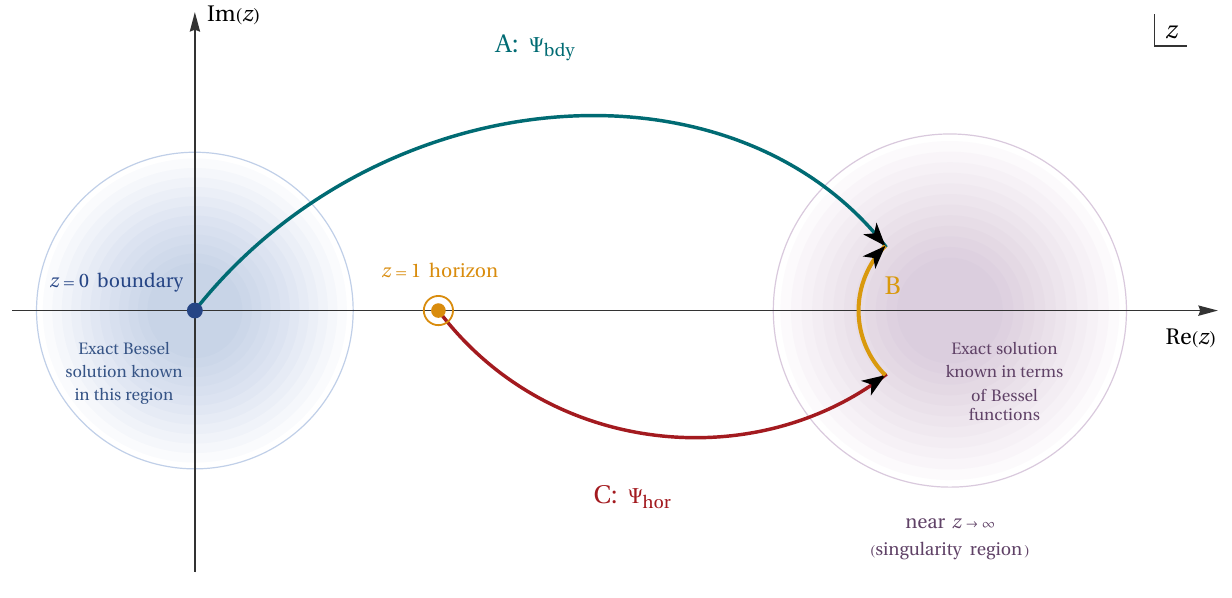}
  \caption{Antistokes contour for putting boundary condition.}
  \label{fig:AScont}
\end{figure}
%%%%%%%%%%%%%%%%%%%%%%%%%%%%%%%%%%%%%%%%%%%%%%%%%%%%%%%%%%%%%%%%%%%%%%%%%%%%

The right linear combination of the oscillatory solutions is determined by the QNM boundary conditions at the boundary $z=0$ and at the horizon $z=1$. So we need a solution which
passes through both these points to be able to impose both the boundary conditions simultaneously. However, the fact that we must lie on an anti-Stokes line poses
an interesting subtlety here. Firstly, note that the horizon  at $z=1$ is not a turning point of this problem. If we solve the equations
exactly near the boundary and move along the anti-stokes lines we can reach the horizon only after passing through the region near the singularity.
This fact has been schematically shown in Fig.\ref{fig:AScont}, where the contour segments A and C denote the anti-Stokes lines. We move
from A to C through the contour segment B near the singularity. The change in the solutions as we move along B is reliably captured
by using the exact solution near the singularity. We will now explain this procedure in more detail and demonstrate how the asymptotic QNMs are recovered.

Near the boundary we have $x \approx z$. Retaining the leading term in the potential \eqref{Uexp}, near $z \rightarrow 0$, the equation \eqref{scheq} reduces to
\begin{equation}
  \frac{d^2 \Psi}{dx^2} + \left( \omega^2 - \frac{15}{4 x^2} \right) \, \Psi= 0\, .
\end{equation}
The normalizable solution to this equation is expressed in terms of the Bessel function in the following way
\begin{equation}
 \Psi_{\text{bdy}} = A_0 \ \sqrt{\omega x} J_{2} \left( \omega x \right) \, .
\end{equation}
The region of validity of this solution has been schematically represented by the blue shaded region in Fig.\ref{fig:AScont}.
Recall that the behaviour of the Bessel function for large values of its argument is given by
\begin{equation}\label{bespro}
 J_\nu \left( \rho \right) = \sqrt{\frac{2}{\pi \rho}} \cos \left(- \rho + \left( 2 \nu + 1\right) \frac{\pi}{4} \right) + \dots
\end{equation}
This implies that away from the boundary $\rvert \omega x \rvert \gg 1$, our normalizable solution takes the form
\begin{equation}\label{bdysolexp}
 \Psi_{\text{bdy}} = A_0 \left( e^{- i 5 \pi / 4} \exp \left( i \omega x \right) + e^{ i 5 \pi / 4} \exp \left( - i \omega x \right) \right) \,
\end{equation}
Where $A_0$ is an overall normalization. In this way, normalizability has fixed a specific linear combination of the oscillatory solutions
as we move away from the boundary turning point.
We follow this solution along an anti-Stokes line A up to the singularity, as shown in Fig.\ref{fig:AScont}.

Near the singularity, we will again solve the equation \eqref{scheq} exactly. Here, it would be convenient to define a local coordinate
\begin{equation}\label{xidef}
 \xi = \int_z^{\infty} \frac{dz'}{f(z')} = \mathcal X_+ - x(z) \approx \frac{1}{10 z^5}
\end{equation}
where the relation to $z$ has been derived by performing the integral taking $f(z) \approx 2 z^6$ for large $z$, i.e. near the singularity.
Using this coordinate, together with leading order behaviour of the potential $U$ from \eqref{Uexp}, near the singularity ($z \rightarrow \infty$) \eqref{scheq}
reduces to
\begin{equation}
  \frac{d^2 \Psi}{d\xi^2} + \left( \omega^2 + \frac{21}{100 \ \xi^2} \right) \, \Psi= 0\, ,
\end{equation}
This equation admits the following exact solution in terms of Bessel functions
{\footnote{The two linearly independent solutions of the Bessel equation are generally considered to be $J_\nu$ and $Y_\nu$. However, since the order of the
Bessel equation is $\nu = 1/5$ which is not an integer, the $J_{1/5}$ and $J_{-{1/5}}$ are linearly independent functions which we choose to be
our basis for writing down the most general solution. Note that $Y_{1/5}$ can be expressed in terms of
these functions as follows
\begin{equation*}
 Y_{\frac{1}{5}} = \frac{\cos (\pi / 5) J_{\frac{1}{5}}-J_{-\frac{1}{5}}}{\sin (\pi / 5)}
\end{equation*}
}
\begin{equation}\label{singsol}
 \Psi_{\text{s}} =  B_+ \, \sqrt{ \omega \xi} \,  J_{\frac{1}{5}} \left( \omega \xi \right) + B_- \, \sqrt{ \omega \xi} \, J_{-\frac{1}{5}} \left( \omega \xi \right)
\end{equation}
The region of validity of this solution has been marked with the pink shaded region in Fig.\ref{fig:AScont}.
In the region $|\omega \xi| \gg 1$, using \eqref{bespro}, the solution \eqref{singsol} has the following behaviour
\begin{equation}\label{singsolexp}
\begin{split}
 \Psi_{\text{s}} &= \frac{1}{\sqrt{2 \pi}} \left(  \mathbb P \, \exp \left( i \omega \xi \right) + \mathbb Q \, \exp \left( - i \omega \xi \right) \right) \\
 &= \frac{1}{\sqrt{2 \pi}} \left(  \mathbb P \ e^{i \omega \mathcal X_+ } \,
 \exp \left(- i \omega x \right) + \mathbb Q \, e^{- i \omega \mathcal X_+ } \exp \left( i  \omega  x \right) \right) \,
\end{split}
\end{equation}
where
\begin{equation}
\begin{split}
 \mathbb P &= B_+ \exp\left( - i \frac{7 \pi}{20} \right) + B_- \exp\left( - i \frac{3 \pi}{20} \right) \, , \\
 \mathbb Q &= B_+ \exp\left( i \frac{7 \pi}{20} \right) + B_- \exp\left( i \frac{3 \pi}{20} \right) \, ,
 \end{split}
\end{equation}
and the second step in \eqref{singsolexp} is obtained using the definition of $\xi$ in \eqref{xidef}.
The solution \eqref{singsolexp} connects with the WKB solution on the stokes line A and comparing it to the boundary solution \eqref{bdysolexp}, we get
\begin{equation}\label{cond1}
 \frac{\mathbb P \ e^{i \omega \mathcal X_+ }}{\mathbb Q \, e^{- i \omega \mathcal X_+ }} = e^{ \left. i \frac{5 \pi}{2} \right. } \, .
\end{equation}

Now, we wish to go from the boundary anti-Stokes line A to the horizon anti-Stokes line C, through the arc B. This is achieved by rotating the local $\xi$ coordinate near the singularity in the following way
\begin{equation}\label{xirot}
 \xi_c = e^{i \pi} \xi \,
\end{equation}
where $\xi_c$ refers to the $\xi$ coordinate which continues on to the horizon anti-Stokes line C.
Since, $\xi$ is related to $z$ by \eqref{xidef}, this implies the rotation $z \rightarrow e^{-i \pi / 5} z$. This is moving to the adjacent anti-Stokes line in the $z$ coordinate
\footnote{The anti-Stokes directions in terms of $\arg z$ are given by $\theta_n = \left( \arg \omega - n \pi \right)/5$ with $n \in \mathbb Z$, which implies that the angle between adjacent anti-Stokes lines is $\pi/5$.}. This rotation is represented by the segment B in Fig.\ref{fig:AScont}. Note that the arc B is not an anti-Stokes line. But in this near
singularity region we are not relying on the oscillatory WKB plane-waves, rather we have the exact local Bessel solution. There is no prohibition against leaving the anti-Stokes line
since an exact local solution is available. The effect of the rotation on this local solution is given by
{\footnote{Here we have used the identity $J_\nu \left( e^{i \pi} u\right) = e^{i \pi \nu} J_\nu \left( u\right)$.}
\begin{equation}\label{horsol}
 \Psi_{\text{hor}} (\xi) = \Psi_s \left(e^{i \pi} \xi \right) =  B_+ e^{i \frac{7 \pi}{10}}  \, \sqrt{\omega \xi} \, J_{\frac{1}{5}} \left( \omega \xi \right)
  + B_- e^{i \frac{3 \pi}{10}}  \, \sqrt{\omega \xi} \, J_{-\frac{1}{5}} \left( \omega \xi \right)
\end{equation}
Expanding \eqref{horsol} for $\lvert \omega \xi \rvert \gg 1$ , we get
\begin{equation}\label{xihor1}
 \Psi_{\text{hor}} = \left( B_+ e^{ i \frac{7 \pi}{20}}  + B_- e^{ i \frac{3 \pi}{20}} \right) e^{i \omega \xi} +
 \left( B_+ e^{ i \frac{21 \pi}{20}}  + B_- e^{ i \frac{9 \pi}{20}} \right) e^{-i \omega \xi}
%  e^{i \omega \mathcal X_+} \left( B_+ e^{ i \frac{7 \pi}{20}}  + B_- e^{ i \frac{3 \pi}{20}} \right) e^{-i\omega x} +
%  e^{- i \omega \mathcal X_+} \left( B_+ e^{ i \frac{21 \pi}{20}}  + B_- e^{ i \frac{9 \pi}{20}} \right) e^{i\omega x}
\end{equation}
In order to identify the correct ingoing term at the horizon, let us recall that $\xi$ is the coordinate compatible with
contour A. The right coordinate compatible with contour C is $\xi_c = - \xi$, as implied by \eqref{xirot}. We can therefore
express the solution \eqref{xihor1} on the horizon anti-Stokes line C as
\begin{equation}\label{xihor2}
 \Psi_{\text{hor}} = \left( B_+ e^{ i \frac{7 \pi}{20}}  + B_- e^{ i \frac{3 \pi}{20}} \right) e^{-i \omega \xi_c} +
 \left( B_+ e^{ i \frac{21 \pi}{20}}  + B_- e^{ i \frac{9 \pi}{20}} \right) e^{i \omega \xi_c}
\end{equation}
Furthermore, the relation \eqref{xidef} continues to hold for $\xi_c$ as well. Consequently, we have
\begin{equation}\label{xihor1}
 \Psi_{\text{hor}} =
 e^{-i \omega \mathcal X_+} \left( B_+ e^{ i \frac{7 \pi}{20}}  + B_- e^{ i \frac{3 \pi}{20}} \right) e^{i\omega x} +
 e^{i \omega \mathcal X_+} \left( B_+ e^{ i \frac{21 \pi}{20}}  + B_- e^{ i \frac{9 \pi}{20}} \right) e^{-i\omega x}
\end{equation}
To ensure that this solution is in-going at the horizon, we must remove the out-going piece from this solutions, i.e. we should set the coefficient
of $e^{-i\omega x}$ to zero. This provides us yet another condition on $B_\pm$
\begin{equation}\label{cond2}
   B_+ \, e^{ i \frac{21 \pi}{20}}  + B_- \, e^{ i \frac{9 \pi}{20}} = 0
\end{equation}
Thus, the two QNM boundary conditions provide us with two conditions \eqref{cond1} and \eqref{cond2} on $B_\pm$. In order to have a legitimate solution for $B_\pm$ from these homogeneous coupled algebraic equations, the determinant of the coefficient matrix must vanish. This implies the following quantization condition
\begin{equation}
\exp \left(2 i \omega \mathcal X_+ \right)  = - \frac{1}{2 \cos \left( \pi / 5 \right)} \, .
\end{equation}
The values of $\omega$ for which this condition is satisfied can be obtained by taking Logarithm of this condition. We get
\begin{equation}\label{qnmfull}
 \omega_n = \frac{\pi}{\mathcal X_+} \left( n + \frac{1}{2} + \frac{i}{2 \pi} \ln \left( 2 \cos \frac{\pi}{5}\right)\right),
 ~~ n \in \mathbb Z _+ \, .
\end{equation}
These are our asymptotic QNMs. Note that for our WKB analysis to be valid, we need large values
of $n$, i.e. this formula is valid only for the large overtones. Writing these QNMs as
\begin{equation}\label{qnmasymform}
 \omega_n\simeq a\,(n+q) \, ,
\end{equation}
and using $\mathcal X_+$ from \eqref{odval} in to \eqref{qnmfull} leads us to the identification
%
% $$\theta(\sigma)\simeq\alpha\sigma+\beta$$
%
\begin{equation}\label{paravals}
a=\frac{\pi}{\mathcal X_+} = \frac{16\sqrt2-28i}{9},\qquad
q=\frac{1}{2}+\frac{i}{2\pi}\log\!\big(2\cos\tfrac\pi5\big),\qquad
% \alpha=\frac{7\pi}{18},\qquad \beta=-\frac\pi2 .
\end{equation}
We should emphasize that our analysis here is valid for arbitrary $k$. Consequently, these asymptotic QNM coefficients are independent of $k$.
Also note that, our use of the optical distance $\mathcal X_+$, as opposed to $\mathcal X_-$ leads to the QNMs with
negative imaginary part which are the right damped modes used in various analysis throughout this paper.

%%%%%%%%%%%%%%%%%%%%%%%%%%%%%
\bibliographystyle{JHEP}
\bibliography{Extremal_PCF}
\end{document}